\documentclass[a4paper,12pt]{iopart}
\usepackage{graphicx}
\def \bea {\begin{eqnarray}}
\def \eea {\end{eqnarray}}
\def \nno {\nonumber}
\begin{document}

\title{System size dependence of nuclear modification and
 azimuthal anisotropy of jet quenching}
\title[System size dependence....]{}

\author{Somnath De\footnote{somnathde@vecc.gov.in} and
 Dinesh K. Srivastava\footnote{dinesh@vecc.gov.in}}
\address{Variable Energy Cyclotron Centre, 1/AF Bidhan Nagar, Kolkata-700 064, India}

\begin{abstract}

We investigate the system size dependence of jet-quenching by analyzing transverse
 momentum spectra of neutral pions
in Au+Au and Cu+Cu collisions at $\sqrt{s_{\textrm{NN}}}$ =200 GeV for different
 centralities. The fast partons 
are assumed to lose energy by radiating gluons as they traverse the plasma 
and undergo multiple collisions.
The energy loss per collision, $\varepsilon$, is taken as proportional $ E$
 (where $E$ is
the energy of the parton), proportional to $\sqrt{E}$,
or a constant depending on whether the formation time of the gluon is less
 than the mean path, greater than the
mean free path but less than the path length, or greater than the path length 
of the partons, respectively.
 NLO pQCD is used to evaluate pion production by modifying the fragmentation function to account for the energy
loss. We reproduce the nuclear modification factor $R_\textrm{AA}$ by treating $\varepsilon$ as the only free 
parameter, depending on the centrality and the mechanism of energy loss. These values are seen to explain the
nuclear modification of prompt photons, caused by the energy lost by final state quarks before they fragment into
photons.  These also reproduce the  azimuthal asymmetry of transverse momentum distribution
 for pions within a factor of two and for prompt photons in a fair agreement
with experimental data.
\end{abstract}



\maketitle

\section{Introduction}

Exploration of the properties of quark gluon plasma (QGP) produced in relativistic heavy ion collisions represents 
a major theme of modern nuclear physics research. The formation of QGP is heralded by numerous signatures, e.g.,
jet-quenching~\cite{jetq_theo,jetq_exp}, elliptic flow of hadrons~\cite{flow_theo,flow_expt},
 quark number scaling of elliptic flow of hadrons\cite{recomb},
 radiation of thermal photons~\cite{phot_exp,fms,zakharov,phot_theory}, 
and suppressed production of $J/\psi$~\cite{jpsi_theo,jpsi_exp},
 etc. 
The celebrated
phenomenon of jet-quenching has its origin in energy loss suffered by high energy quarks and gluons as they 
traverse the QGP, colliding with other partons and radiating gluons. It leads to a suppressed production of hadrons
having large transverse momenta as compared to the case for NN collisions at the corresponding centre of mass energy
per nucleon. It is measured in terms of the nuclear modification factor $R_{AA}$, given by:
\begin{equation}
R_\textrm{AA}(p_T,b)=\frac{d^2N_\textrm{AA}(b)/dp_Tdy}{T_\textrm{AA}(b)(d^2\sigma_\textrm{NN}/dp_Tdy)} \nno\\
\label{raa}
\end{equation}
where $b$ is the impact parameter and $T_{AA}(b)$ is the nuclear overlap
function for impact parameter $b$. 

The jet-quenching has
several other manifestations. Thus, for example, in a non-central collision the jets moving in and out of the reaction plane
would cover differing distances inside the plasma, lose differing amounts of energy, and would be quenched to 
a different extent. This will lead to an azimuthal anisotropy of momentum distribution of hadrons which does not have
its origin in the so-called elliptic flow~\cite{jet_v2}. 

Next consider a high energy quark which is produced in a hard scattering. A photon
fragmented off this quark contributes to the prompt photon production at high transverse momenta. The quark may lose
energy before its fragmentation and thus this photon production may get suppressed~\cite{jeon,arleo}.
Following the same arguments as above, this may also lead to an azimuthal anisotropy 
in photon production for non-central collisions~\cite{v2_phot_highpt}. The jet-photon
conversion~\cite{fms} and jet-induced bremsstrahlung~\cite{zakharov} lead to 
 very
small contributions with opposite signs for the azimuthal
 anisotropy of the photon 
productions~\cite{v2_phot_highpt} and are not considered in the present work.

A new dimension to the azimuthal asymmetry and the jet-quenching has been added by the realization that the event
by event fluctuations of the initial conditions may affect the elliptic flow of the final state hadrons~\cite{v2h_event}. It is
not very clear these seriously affect the azimuthal  variation of the transverse momenta, 
which arises from jet-quenching~\cite{t.renk} as well.

It is expected that
the magnitude of all these effects will depend on the dynamics of the collision and the properties of the QGP. This
makes jet-quenching a powerful probe of the QGP.

While there are quite a few detailed studies on the jet-quenching at RHIC and LHC energies~\cite{jet_rev}, in the present work
we continue the study reported in Ref.~\cite{jeon,dks_jet}, where the Wang, 
Huang, and Sarcevic model~\cite{wang_xn1} was used to study 
the evolution of 
the mechanism of energy loss at RHIC and LHC energies in central collision of heavy nuclei. The model treats the
medium produced in the collision as a plasma of length $L$ through which the partons move, undergoing multiple
collisions, losing energy through radiated gluons, before fragmenting to form hadrons. The only parameters which then
enter the calculations are the path-length- $L$, the mean-free path- $\lambda$, and the energy loss per collision- 
$\varepsilon$. 

The energy loss per collision is then assumed to take a form as proportional to the energy of the parton- $E$, $\sqrt{E}$, or a constant;
depending upon the formation time of the radiated gluons, which is less than the mean free path, larger than the mean free
path but less than the path length $L$, and greater than the path length $L$, respectively.
 The three regimes are called the Bethe-Heitler regime, the LPM regime, and the complete
coherence regime (as the 
entire medium radiates as a whole)~\cite{baier}.  The formation 
time (or length) of the radiated
gluon is  $\approx \omega/k_T^2$, where $\omega$ is the energy of the gluon and $k_T$ is its transverse momentum.

In the present work we explore the limits of the above approach
by performing a detailed and systematic study of medium modification of production of neutral
pions and prompt photons having large transverse momenta as a function of centrality and system size, by analyzing
the corresponding data at 200 GeV/nucleon for collisions of gold nuclei and copper nuclei. We first seek an
accurate description
of the medium modification parameter $R_{AA}$ for different centralities and $p_T$, using the three approaches
for the energy loss mechanisms, by varying the coefficients of energy loss, while accounting for the variation of the
path length $L$. We find that the hadrons having low $p_T$ ($<$ 5 GeV/$c$)  fall into the Bethe-Heitler regime, 
those having intermediate
$p_T$ (5  GeV/$c$ $<$ $p_T$ $<$ 10 GeV/$c$) fall into the LPM regime, 
while those having large $p_T$ $>$ 8--10 GeV/$c$ are covered by the complete coherence regime. 
We find a systematic decrease of the energy loss per collision as we go to more peripheral collisions from central collisions,
as one might expect.

We then use the same parameters to determine the azimuthal anisotropy coefficient $v_2(p_T)$ for neutral pions having large $p_T$. We find that
in general, the experimentally obtained $v_2(p_T)$ is lower than those predicted by our calculations, though the
trend is generally reproduced. Part of this discrepancy may have its origin in the 
uniform densities for the
nuclei, used in the present work.

Next we calculate the prompt photon production as well as $v_2^\gamma (k_T)$. Again we get reasonable description
of the experimental data. Further improvements can perhaps be made by relaxing the condition of equality of
mean-free path and energy loss per collision for quarks and gluons assumed while obtaining 
a description for $R^{\pi^0}_\textrm{AA}$. This would, however, increase
 the number of parameters. 

We give our formulation for particle production, multiple scattering, and energy loss as applied to pion production
at $ \mathcal{O} (\alpha_s^3)$ in Sec.~2. The centrality dependence of nuclear modification of neutral pions and parton energy loss (dE/dx) for
Au-Au and Cu-Cu collisions at the highest RHIC energy are discussed in Sec.~3.
 The results for the azimuthal asymmetry of high $p_T$
neutral pions are shown in Sec.~4. In Sec.~5, we give the results for the nuclear suppression and  azimuthal anisotropy 
of prompt photons which arise due to the modification
of the fragmentation contribution. Finally we summarize our findings in Sec.~6.

\section{Formulation}
\subsection{Particle production in pp collisions}
In perturbative QCD, the inclusive cross-section for the production of a particle
 in proton-proton collisions is given by~\cite{P.aur1,P.aur2}
\bea
\frac{d\sigma^{AB\rightarrow C}}{d^2p_Tdy}
 =\sum_{a,b,c} \int \, dx_a \int \, dx_b \int \, \frac{dz}{z^2} & & F_{a/A}(x_a,Q^2) F_{b/B}(x_b,Q^2) \nno\\
& & D_{c/C}(z,Q^2_f)\frac{d\sigma_{ab \rightarrow c}}{d^2p_{cT}dy_c} ,
\eea
where $F_{a/A}(x,Q^2)$ is the parton distribution function (PDF) for the parton $a$  and $F_{b/B}(x,Q^2)$ is the PDF
for the parton $b$,  for the nucleons A and B respectively. $D_{c/C}$ gives the fragmentation probability of parton $c$ into
a particle $C$. In case of a hadron; $D_{c/h}$ is the fragmentation function evaluated at $z=p_h/p_c$, where z is the fraction 
 of the parton's momentum carried by the hadron. In case of photon production, $D_{c/\gamma}$ gives
the fragmentation probability of a photon fragmented off a quark with the momentum fraction $z=p_\gamma/p_c$. In addition,
we have an extra term where photon is directly produced in the hard collision ($c=\gamma$). In that case the fragmentation function
reduces to  $\delta(1-z)$.  
$\sigma_{ab \rightarrow c}$ is the parton-parton cross-section, calculated for the leading order processes $ \mathcal{O} (\alpha_s^2)$ 
such as:
\bea
 q+q \rightarrow q+q , \nonumber\\
q+\bar{q} \rightarrow q+\bar{q}, \nonumber\\
q+g \rightarrow q+g , \nonumber\\
g+g \rightarrow g+g , \nonumber\\
......                
\eea
At the next-to-leading order, $\mathcal{O} (\alpha_s^3)$ we include subprocesses like:
\bea
q+q \rightarrow q+q+g~,\nonumber\\
q+\bar{q} \rightarrow q+\bar{q}+g,\nonumber\\
q+q^{\prime} \rightarrow q+q^{\prime}+g, \nonumber\\
q+\bar{q} \rightarrow q^{\prime}+\bar{q}^{\prime}+g, \nonumber\\
g+g \rightarrow g + g+ g\nonumber\\
......
\eea
The running coupling constant $\alpha_s(\mu^2)$, is calculated at next-to-leading order
\begin{eqnarray}
\alpha_s(\mu^2)=\frac{12\pi}{(33-2N_f)\ln(\mu^2/\Lambda^2)}
 \left(1-\frac{6(153-19N_f)\ln\ln(\mu^2/\Lambda^2)}{(33-2N_{f})^2\ln(\mu^2/\Lambda^2)}\right) , \nno\\
\end{eqnarray}
where $\mu$ is the renormalization scale, $N_f$ is the number of flavours, and $\Lambda$ is the $\Lambda_{\bf{QCD}}$ scale.
We have used the \emph{CTEQ4M} structure function~\cite{H.lai} 
and the BKK fragmentation 
function~\cite{bkk-frag}. We set the factorization, renormalization, and fragmentation scales as equal to $p_T$, though we 
have checked scale dependence on particle production in pp collisions using $Q=0.5p_T$ and $Q=2.0p_T$ as well.

\begin{figure}[ht]
\begin{center}
\includegraphics[width=11.5cm, clip=true]{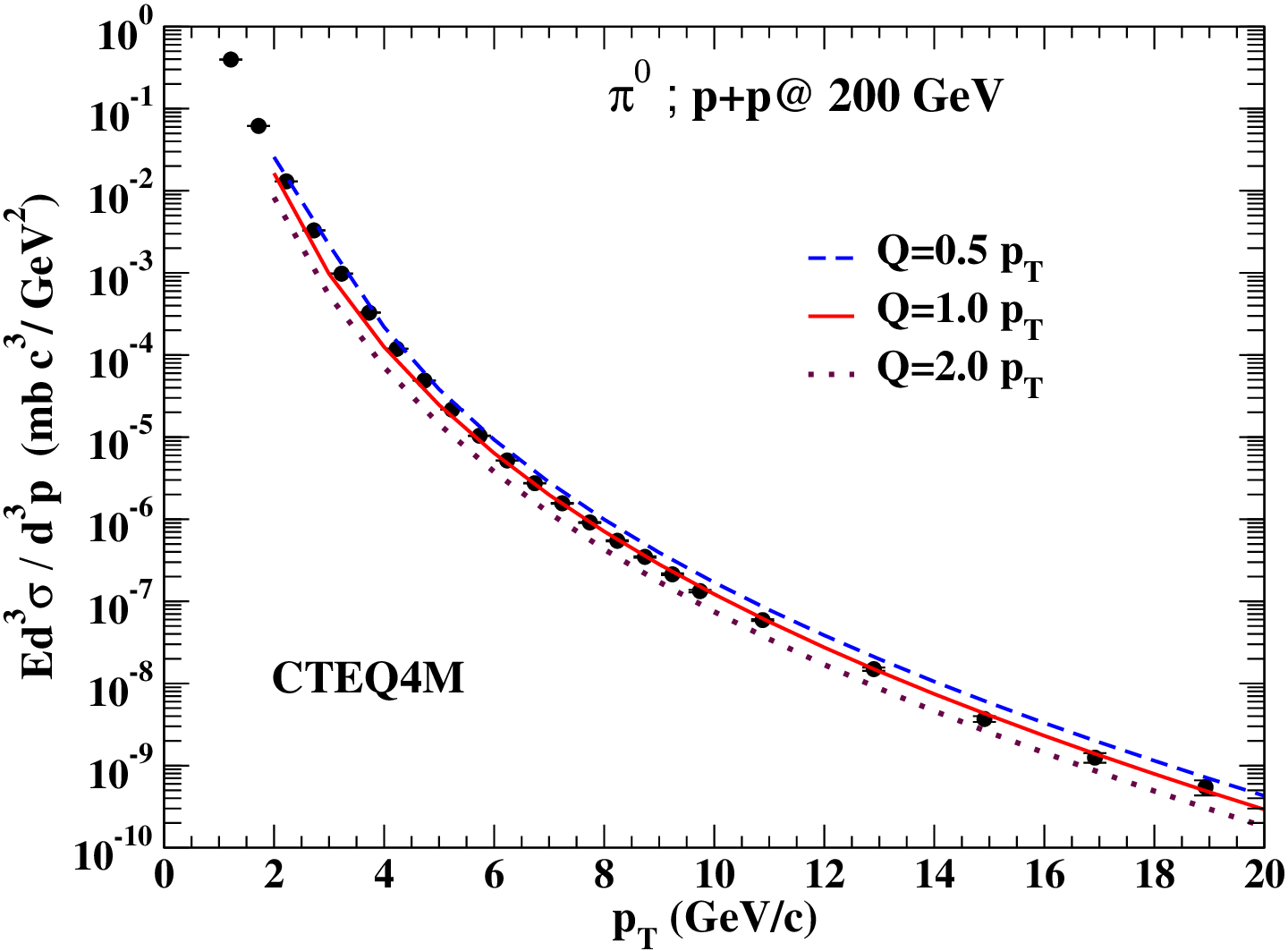}
\end{center}
\caption{(Colour on-line) A comparison of neutral pion yield in p+p collision
measured by the PHENIX collaboration~\cite{pi0_pp} at $\sqrt{s_\textrm{NN}}$ = 200 GeV
with NLO pQCD calculations.}
\label{fig1}
\end{figure}

We show the results of our calculations for the inclusive neutral pion yield for proton-proton collisions at 
the centre of mass energy of 200 GeV along with the data from the PHENIX measurements~\cite{pi0_pp} in 
~Fig.~\ref{fig1}. A very good agreement with the experimental data is seen, as in similar calculations by other authors
(see e.g., Ref.~\cite{vogelsang}). The quantitative description of the data from pp collisions provides us with a
reliable base-line to discuss nuclear modification of hadron production in nucleus-nucleus collisions.

\subsection{Model for parton energy loss}

 In order to calculate the inclusive pion spectra from nucleus-nucleus collisions
 we have used the effects
of shadowing and energy loss of partons.
The nuclear shadowing accounts for the modification of the parton distribution
function for nuclei compared to that for a free nucleon. The shadowing function is defined as:
\begin{equation}
S_{a/A}(x,Q^2)= \frac{F_{a/A}(x,Q^2)}{AF_{a/N}(x,Q^2)}\, ,
\end{equation}
where $F_{a/A}(x,Q^2)$ is the parton distribution for the nucleus A and $F_{a/N}(x,Q^2)$ is that for a nucleon.
 We have used the EKS98 parametrization
of nuclear shadowing obtained by Eskola, Kolhinen, and Salgado~\cite{EKS98}.

We use the simple phenomenological model,
first used by Jeon et al.~\cite{jeon} to obtain a description of 
preliminary $R_{AA}$ at RHIC energies, 
 to estimate parton energy loss.
The key features of this
treatment are: average energy loss per collision ($\varepsilon_a$) by the parton, the mean free path of the parton ($\lambda_a$), 
and the average path length of the parton in the medium (L). 
 The probability for a parton to scatter $n$ times before leaving the medium can be written as:
\begin{equation}
P_a(n,L)=\frac{(L/\lambda_a)^n}{n!}\,e^{-L/\lambda_a}~.
\end{equation}
The multiple scattering and energy loss suffered by the partons then necessitates a modification of the
fragmentation function $D_{c/h}(z,Q^2)$.  Following
the model of Wang, Huang, and Sarcevic~\cite{wang_xn1}, the modified
 fragmentation function can be written as:
\bea
zD_{c/h}(z,L,Q^2)&=&\frac{1}{C_N^a}\sum_{n=0}^N P_a(n,L) \, \times \nonumber\\
& &\left \lbrack z_n^a \, D_{c/h}^0(z_n^a,Q^2) +   
\sum_{m=1}^n z_m^a \, D_{g/h}^0(z_m^a,Q^2) \right \rbrack 
\label{mod_frag}
\eea
where $zE_T^a=z_n^aE_n^a=z_n^a(E_T^a-\sum_{i=0}^n\varepsilon_a^i)$, $z_m^a=zE_T^a/\epsilon_a^m$.
 N is the maximum number of collisions for which $z_n^a\leq 1$, $D_{c/h}^0$ is
the hadronic fragmentation function which gives the probability that quark or gluon would fragment into a hadron (in our case $\pi^0$)
and $C_N^a= \sum_{n=0}^N \, P_a(n,L)$~. The first term gives the fragmentation
 probability of the leading parton $a$ with a reduced energy 
$(E_T^a- \sum _{i=0}^n\varepsilon_a^i)$ and the second term comes from
the emitted gluon having energy $\epsilon_a^m$. \\

\begin{figure}[t]
\begin{center}
\includegraphics[width=7.2cm, clip=true]{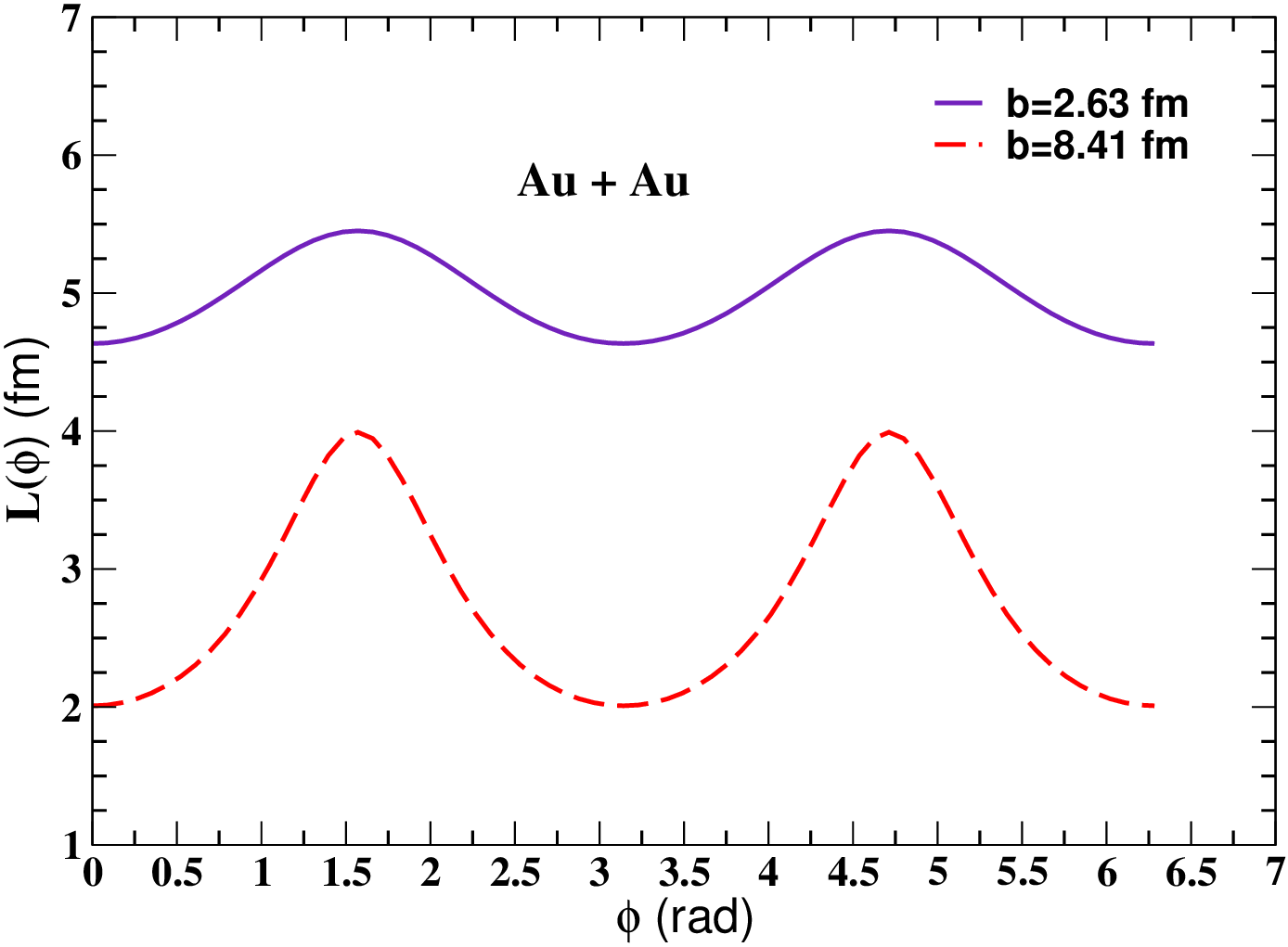}
\includegraphics[width=7.2cm, clip=true]{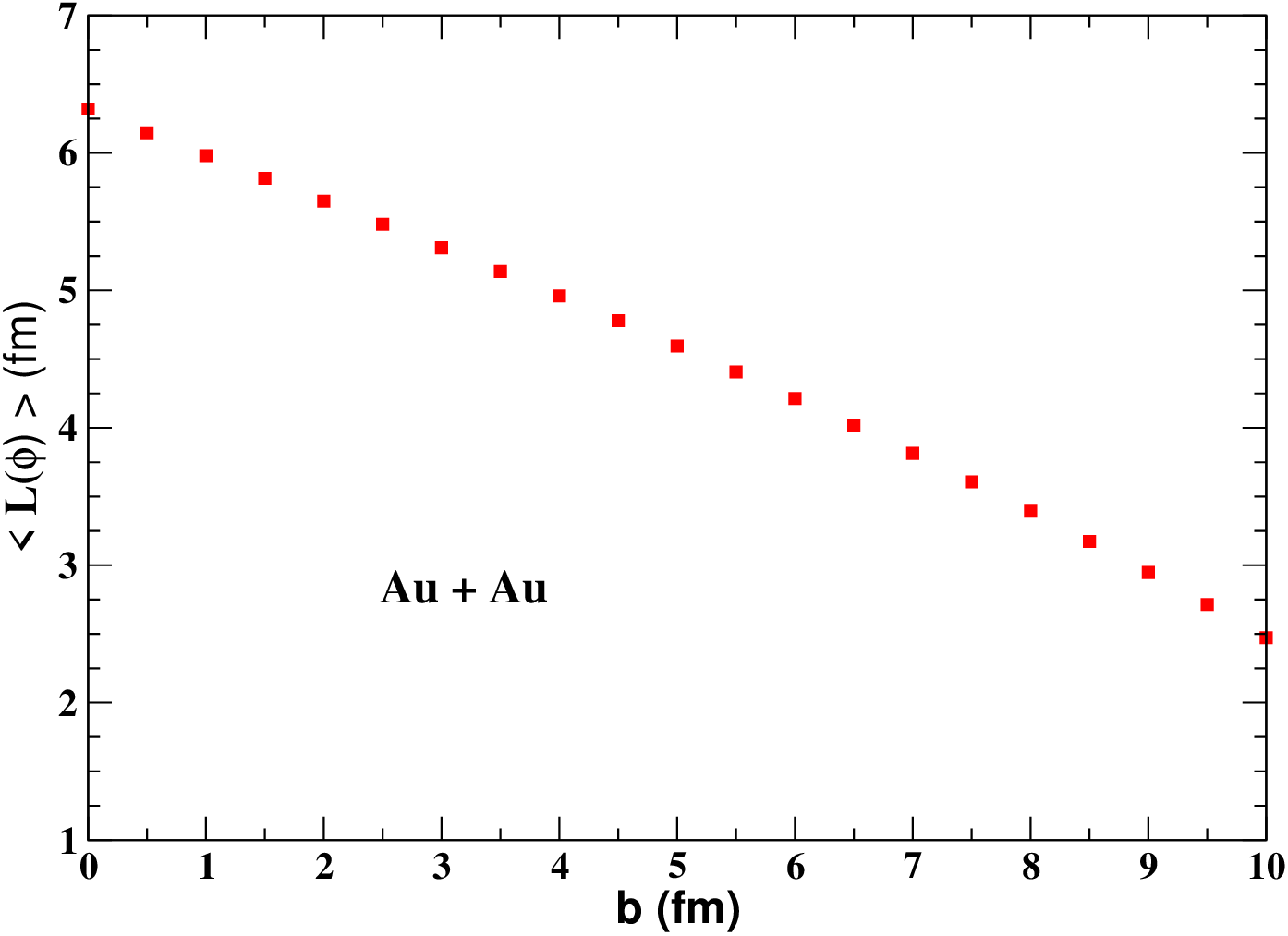}
\end{center}
\caption{(Colour on-line) (Left panel) Azimuthal variation of the average path length traversed by a parton in collision of gold nuclei.
 The impact parameter for the upper curve is an average for 0-10\% most central collisions and the lower one is for
 40-50\% centrality.
(Right panel) The average path length vs. impact parameter for Au+Au system.}
\label{fig2}
\end{figure}

 While one may easily ignore the azimuthal ($\phi$) dependence of the
average path length $L(b)$ for impact parameter $b$ for very central collisions, it needs to be considered for non-central collisions. 
We use a simple approach, based on Glauber model, to evaluate the dependence of the average path-length on the azimuthal angle
with respect to the
reaction plane. Assuming uniform densities for the colliding nuclei, the average path-length for an impact parameter $b$ and azimuthal angle
$\phi$ can be written as:

\begin{equation}
 L(\phi;b) = \frac{\int \! \int \ell(x, y, \phi, b) T_{AB}(x, y; b) \, dx \, dy}{\int \! \int T_{AB}(x, y; b) \, dx \,dy},
\end{equation}
where $x$ and $y$ are the transverse co-ordinates for the point where the partons scatter to produce the jet(s) which, travelling at
an angle $\phi$ with respect to the reaction plane, traverses the path length $\ell (x, y, \phi, b)$. 
  $T_{AB}(x, y; b)= t_A(x+b/2,y)t_B(x-b/2,y)$ is the nuclear overlap function and $t_A$ and $t_B$ are the transverse density profiles
of the two nuclei.
 An average of $L(\phi; b)$ over $\phi$  
(varying from zero to $2\pi$) gives the average path length $L(b)$ (see Fig.~\ref{fig2}).

We closely follow the excellent review on jet-quenching
by Baier et al~\cite{baier}, and assume that the collisional energy loss is quite small for light quarks and gluons compared
to the energy energy loss due to radiation of gluons. Now one can define the formation time of
the radiated gluon as:
\begin{equation}
t_\textrm{form}\simeq\frac{\omega}{k_T^2}~\\
\end{equation}
where $\omega$ is the energy of the  radiated gluon and $k_T$ is the transverse momentum. We would normally have
 $\omega \gg k_T$ where $k_T \approx \mu$ is a typical momentum
transfer in a partonic collision. The coherence length, $l_\textrm{coh}$, can now be written as:
\begin{equation}
l_{\textrm{coh}}\simeq \frac{\omega}{\langle k_T^2 \rangle_\textrm{coh}}\simeq \frac{\omega}{N_{\textrm{coh}}\langle k_T^2 \rangle},
\end{equation}
where $N_{\textrm{coh}}$ is the number of coherent scattering centers.
One can then write,
\begin{equation}
 N_{\textrm{coh}}= \frac{l_{\textrm{coh}}}{\lambda_a}\simeq \sqrt{\frac{\omega}{\lambda_a\mu^2}}\equiv 
\sqrt{\frac{\omega}{E_{\textrm{LPM}}}}
\end{equation}
where $E_\textrm{LPM}=\lambda_a\mu^2$ is the energy parameter introduced to separate the incoherent and coherent radiation of gluons. 

In the Bethe-Heitler (BH) regime, the energy of the radiated gluon is less than $E_{\textrm{LPM}}$, 
the coherence length $l_\textrm{coh}\leq \lambda_a$, leading to an incoherent radiation from $\textrm{L}/\lambda_a$ 
scattering centers. Thus 
energy loss per unit length in the soft $\omega$ limit  becomes: 
\begin{equation}
-\frac{dE}{dx}\approx\frac{\alpha_s}{\pi}N_c\frac{1}{\lambda_a}E~,
\label{BH}
\end{equation}
where $N_c=3$ and $E$ is the energy of the parton. We shall write $\varepsilon_a \approx k E$ for this case 
and determine $k$ from the measurement of $R_\textrm{AA}$. 

For $\omega > E_\textrm{LPM}$, we have a regime of coherent radiation, where the coherence length $l_\textrm{coh}$ is greater than $\lambda_a$
but less than average path length $L$. Thus, $N_{\textrm{coh}}(> 1)$ centres of scattering radiate coherently leading to
\begin{equation}
-\frac{dE}{dx}\approx \frac{\alpha_s}{\pi}
\frac{N_c}{\lambda_a} \sqrt{E_\textrm{LPM} E}~.
\label{LPM}
\end{equation}
We re-write this as
$\varepsilon_a \approx\sqrt{\alpha E}$
 and determine the values of $\alpha$ from the measurement of $R_\textrm{AA}$.

Finally, when the coherence length is larger than $L$ ($l_\textrm{coh} > L$) the entire medium acts as one coherent source
and energy loss per unit length becomes:
\begin{equation}
-\frac{dE}{dx}=\frac{\alpha_s}{\pi}N_c \frac{\langle k_T^2 \rangle}{\lambda_a} L~.
\label{Constant}
\end{equation}
This is denoted as constant energy loss ($\varepsilon_a=\kappa$) regime. 
We use the NLO code of Aurenche et al~\cite{P.aur1} adopted earlier by
 Jeon et al.~\cite{jeon} to account for the
energy loss of the partons, by modifying the fragmentation function,  for the
 calculations reported here.

A more detailed derivation of the energy loss for this case leads to:
\begin{equation}
-\frac{dE}{dx}=\frac{\alpha_s}{4}N_c \frac{\langle k_T^2 \rangle L}{\lambda_a}
\tilde{v}~.
\label{Constant2}
\end{equation}
where $\tilde{v}$ is Fourier transform 
of the normalized differential cross-section for parton-parton collisions
 for the appropriate momentum transfer scale (see Ref.~\cite{baier}). The
momentum transport coefficient $\widehat{q}$ is then defined as:
\begin{equation}
\widehat{q}= \frac{\langle k_T^2 \rangle}{\lambda_a} \tilde{v}~, 
\end{equation}
so that we can write:
\begin{equation}
-\frac{dE}{dx}=\frac{\alpha_s}{4}N_c  \widehat{q} L ~,
\label{qhat}
\end{equation}
which can be used to deduce the average momentum transport coefficient for any 
given centrality, for a comparison with other estimates for this.

\section{Results}
\subsection{Centrality dependence of $R_{AA}$ for Au-Au collisions at RHIC}
We have calculated the nuclear modification factor $R_{AA}$ for neutral pions
 for Au-Au collisions at the top RHIC energy for six centralities
using Bethe-Heitler (BH), LPM, and constant energy loss mechanisms, discussed above.
 We have kept the mean free path of gluons as
well as quarks fixed at 1 fm and also assumed that their energy loss
 per collision is identical. 
Thus once we have accounted for the variation of the average path length $L(b)$ with centrality,
 we are left 
with the energy loss per collision as the only adjustable parameter.

\begin{figure}
\begin{center}
\includegraphics[width=7.5cm, clip=true]{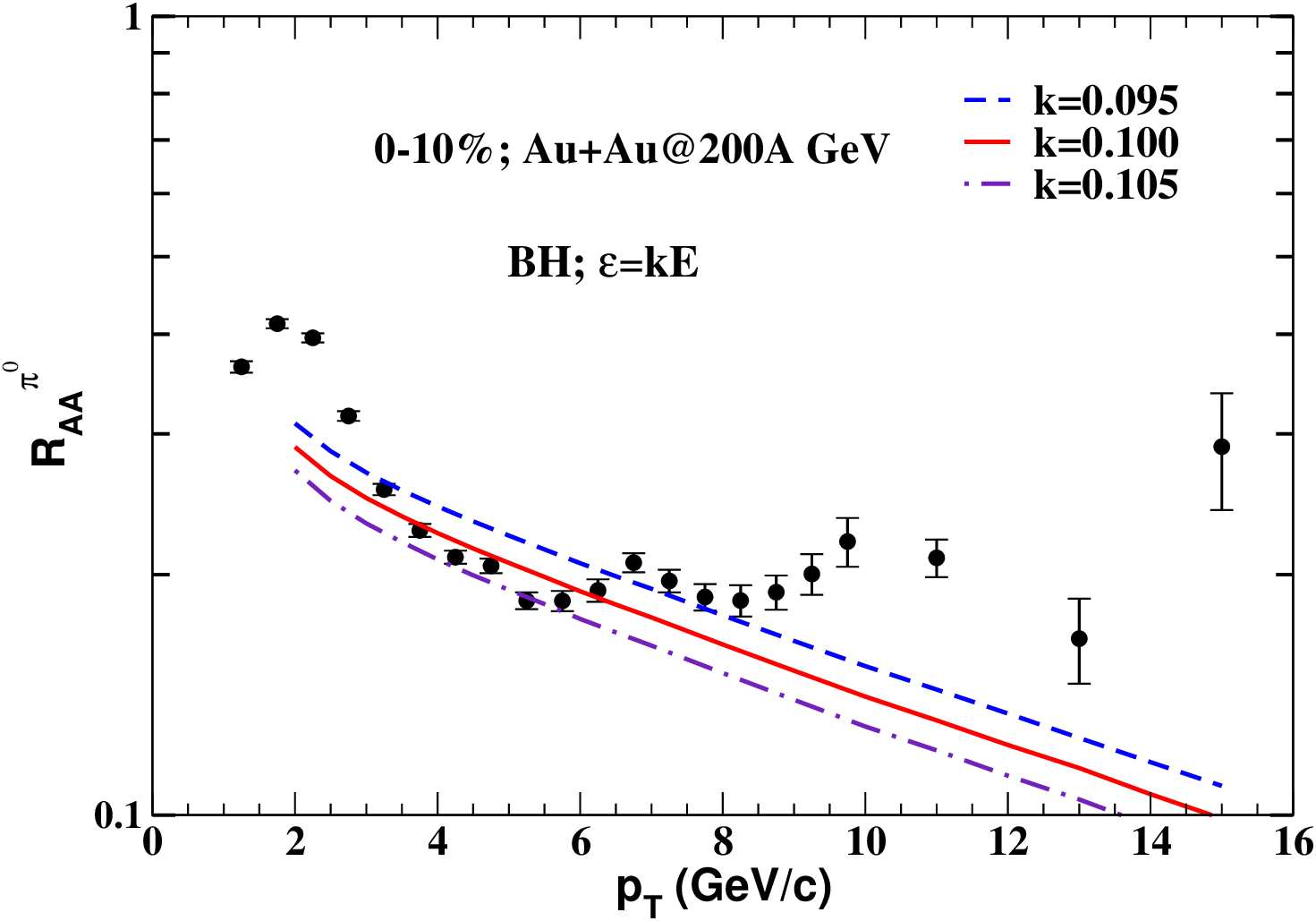}
\includegraphics[width=7.5cm, clip=true]{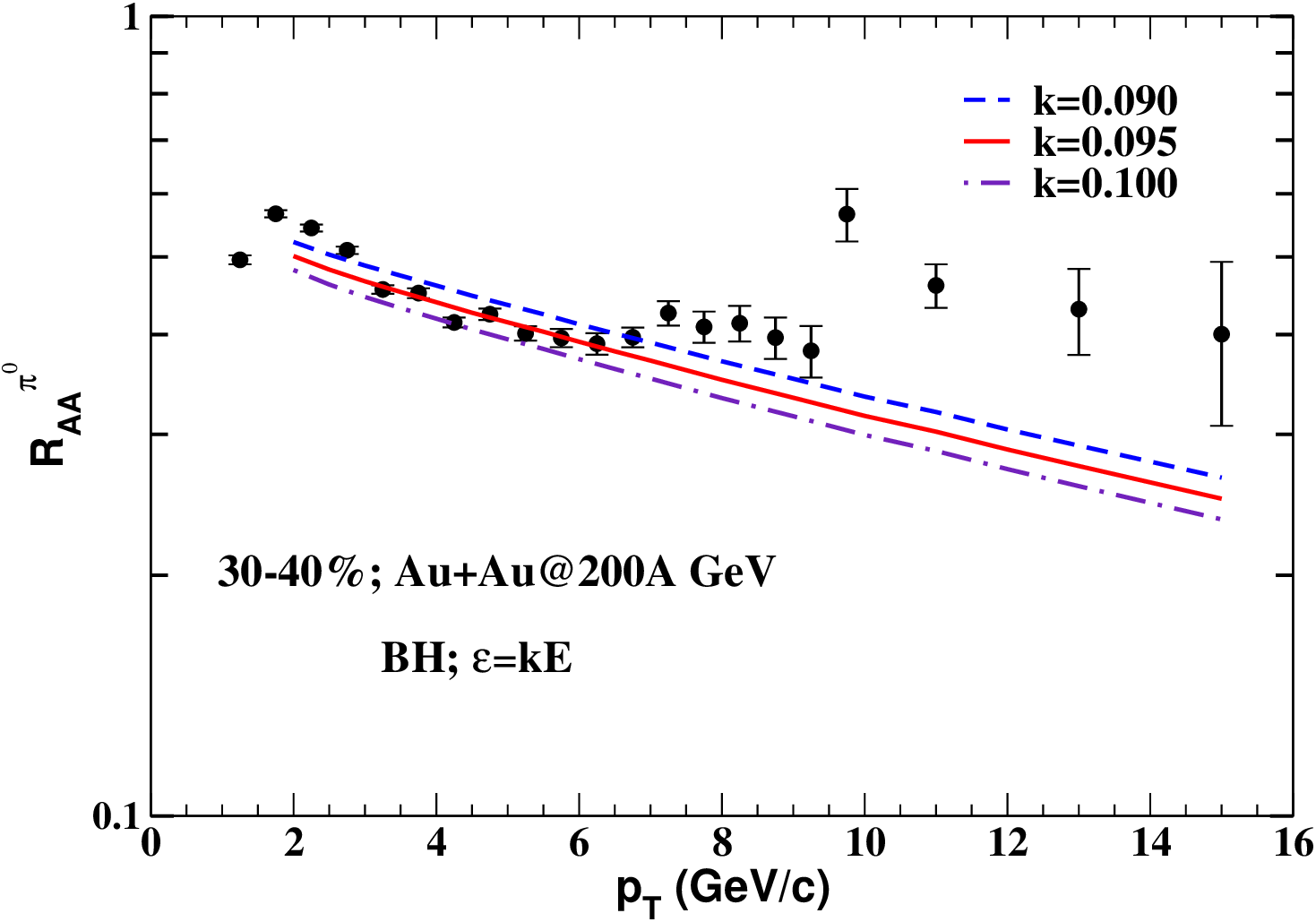}
\includegraphics[width=7.5cm, clip=true]{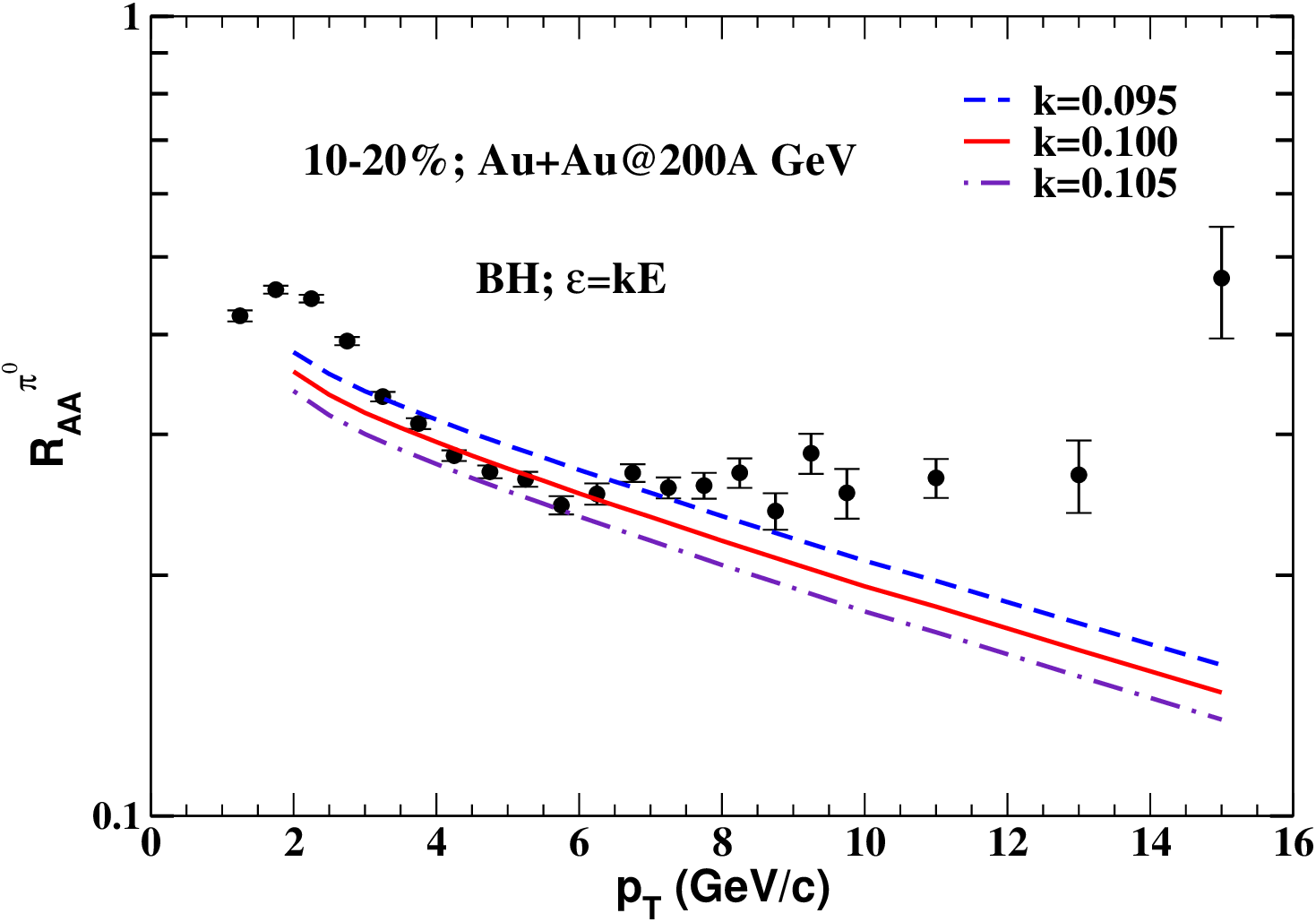}
\includegraphics[width=7.5cm, clip=true]{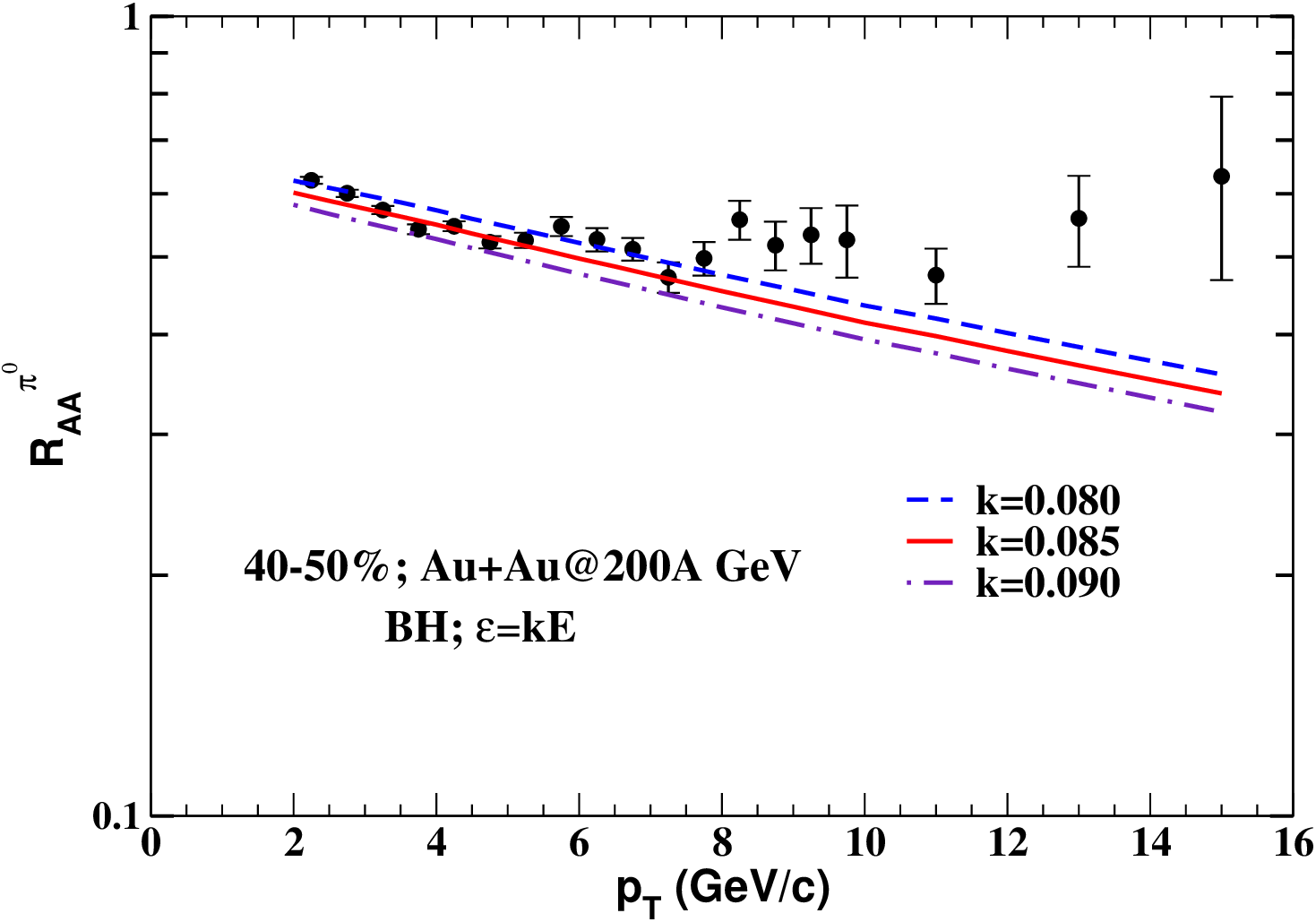}
\includegraphics[width=7.5cm, clip=true]{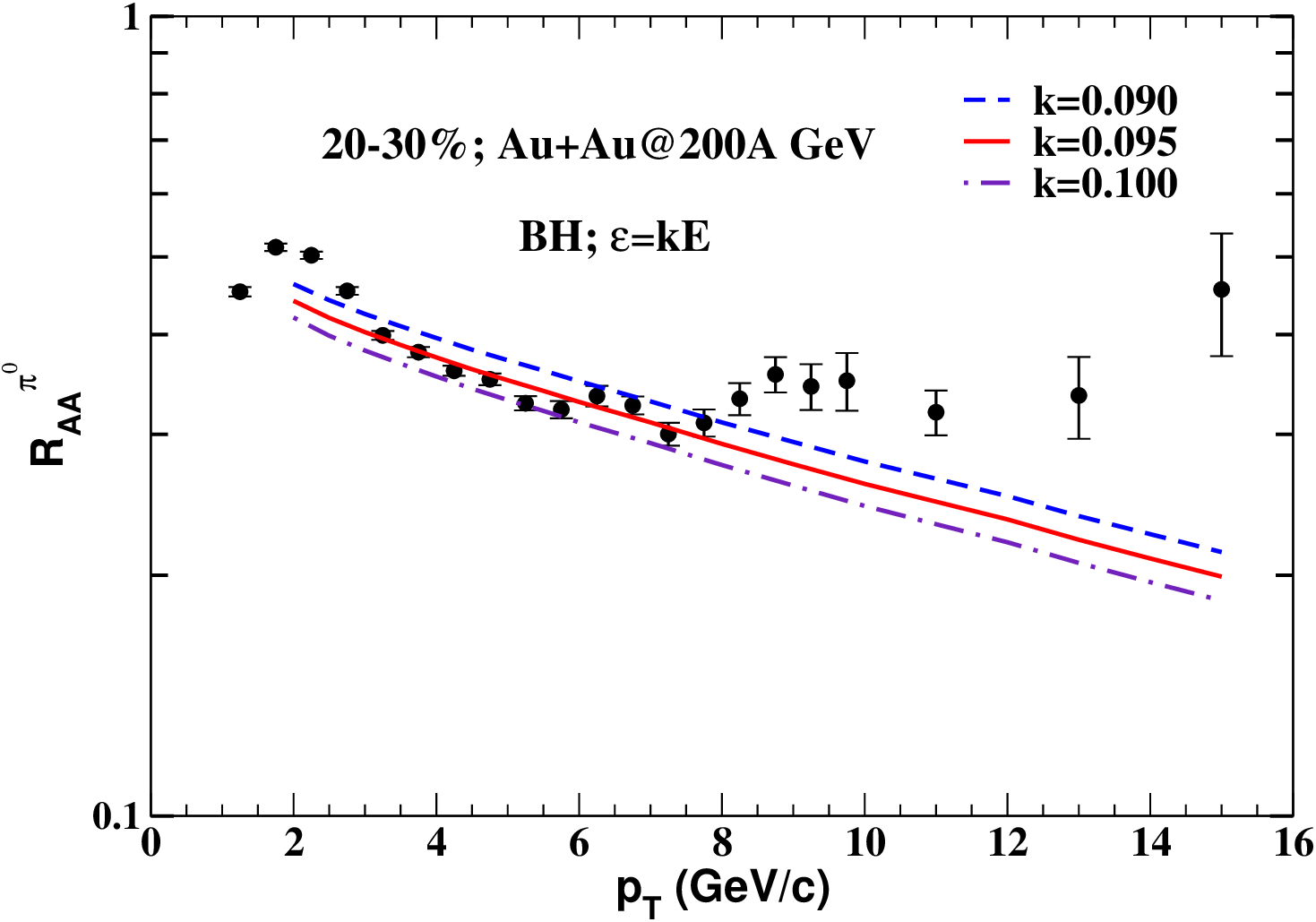}
\includegraphics[width=7.5cm, clip=true]{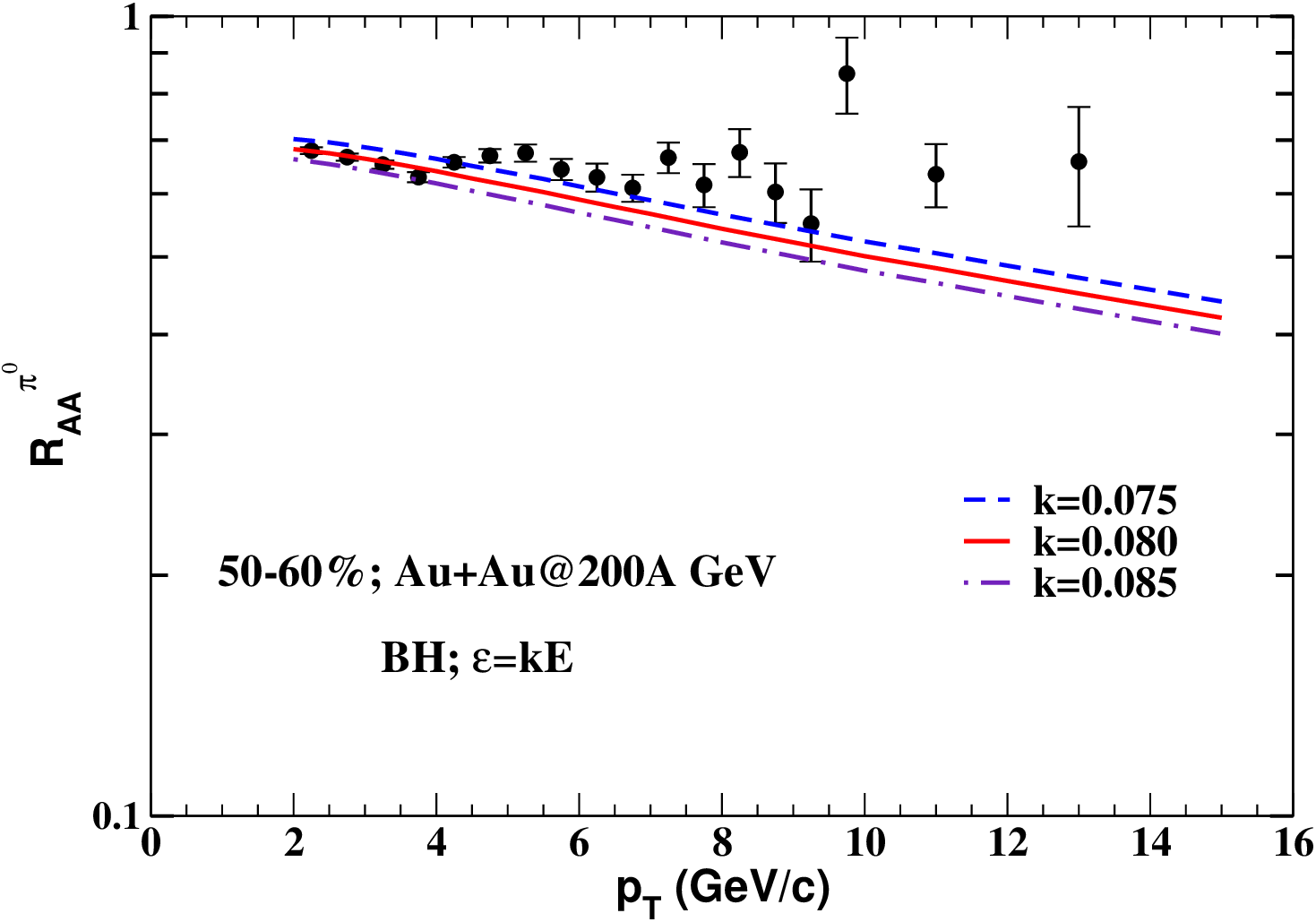}
\end{center}
\caption{(Colour on-line) 
Nuclear modification of neutral pion production for Au+Au collisions at $\sqrt{s_\textrm{NN}}$=200 GeV, 
using BH mechanism (see text). The experimental data are taken from Ref.~\cite{Raa_au}.}
\label{fig3}
\end{figure}

 The results for BH mechanism are shown in Fig.~\ref{fig3}
for near-central (0-10\%, 10-20\%, 20-30\%) and mid-central (30-40\%, 40-50\%, 50-60\%) collisions.  We see a good description
of the data for $p_T$ up to about 6 GeV/$c$. The necessary energy loss per collision is seen to
slowly drop from 10\% to about 8\% of the energy of the partons as we move from near central to mid-central collisions.
We also  note that the slope of $R_{AA}$ changes around $p_T$ equal to 5 GeV/$c$, which suggests a possible change in the
mechanism for energy loss for partons contributing to higher momenta~\cite{dks_jet}.

\begin{figure}
\begin{center}
\includegraphics[width=7.5cm, clip=true]{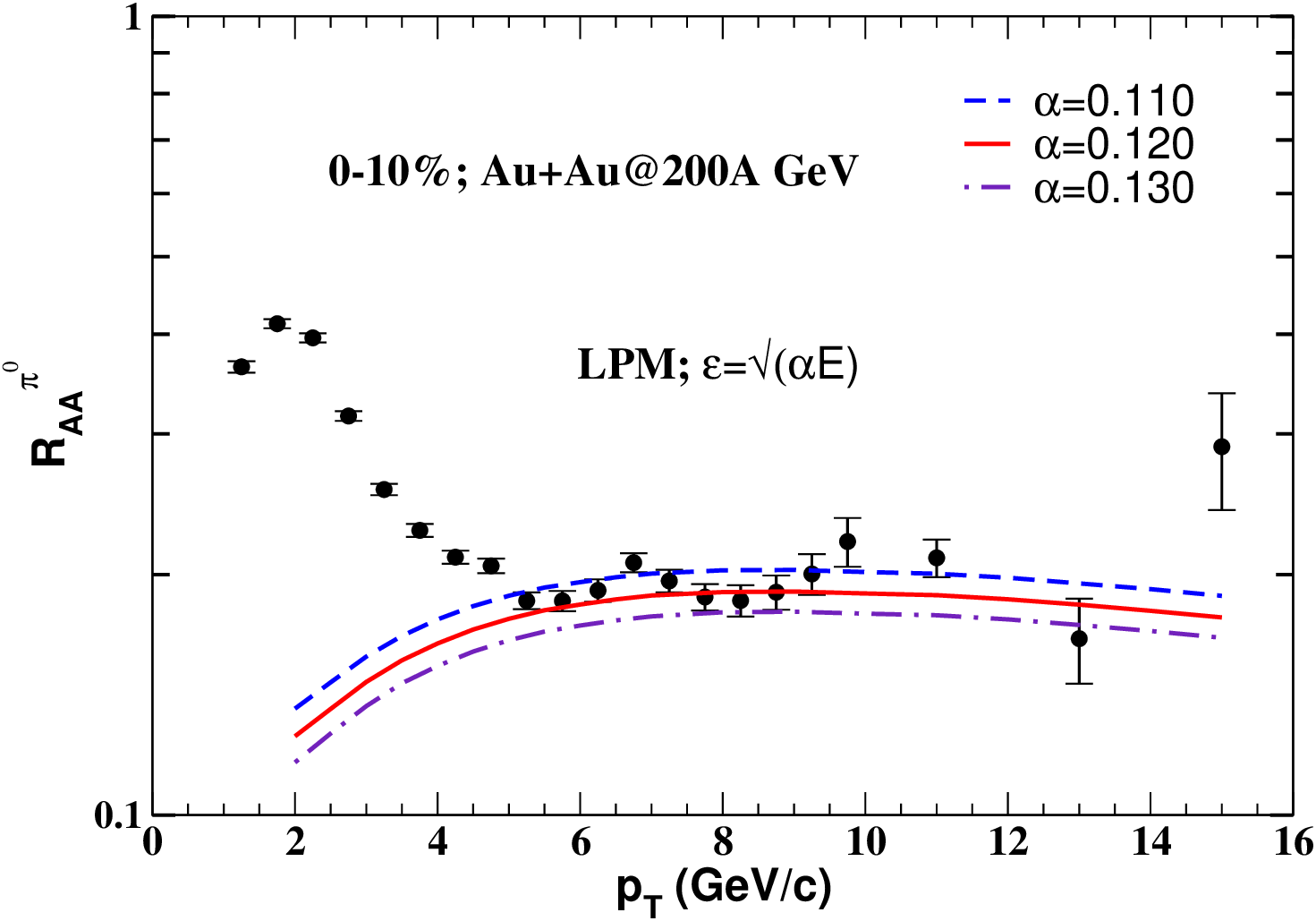}
\includegraphics[width=7.5cm, clip=true]{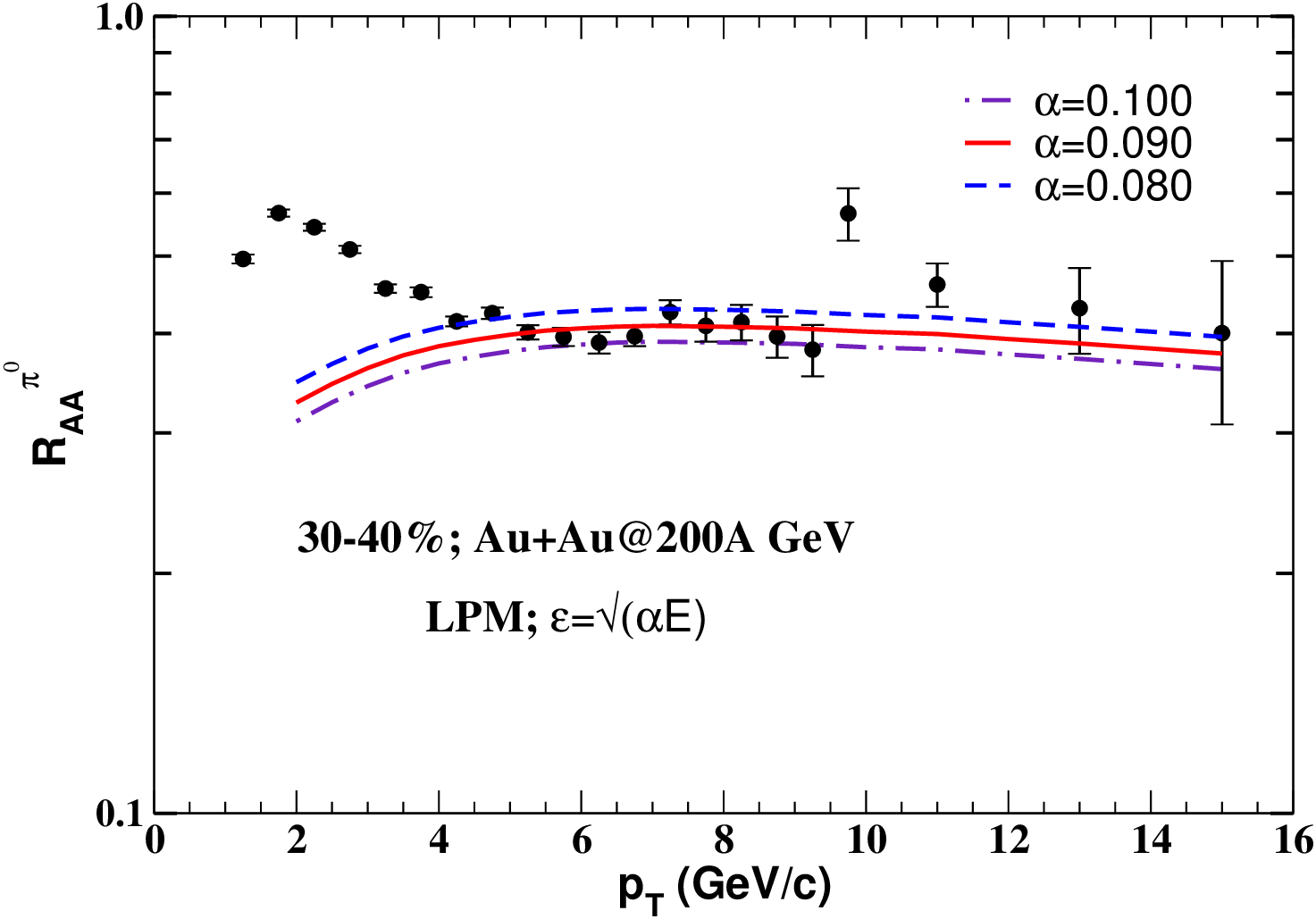}
\includegraphics[width=7.5cm, clip=true]{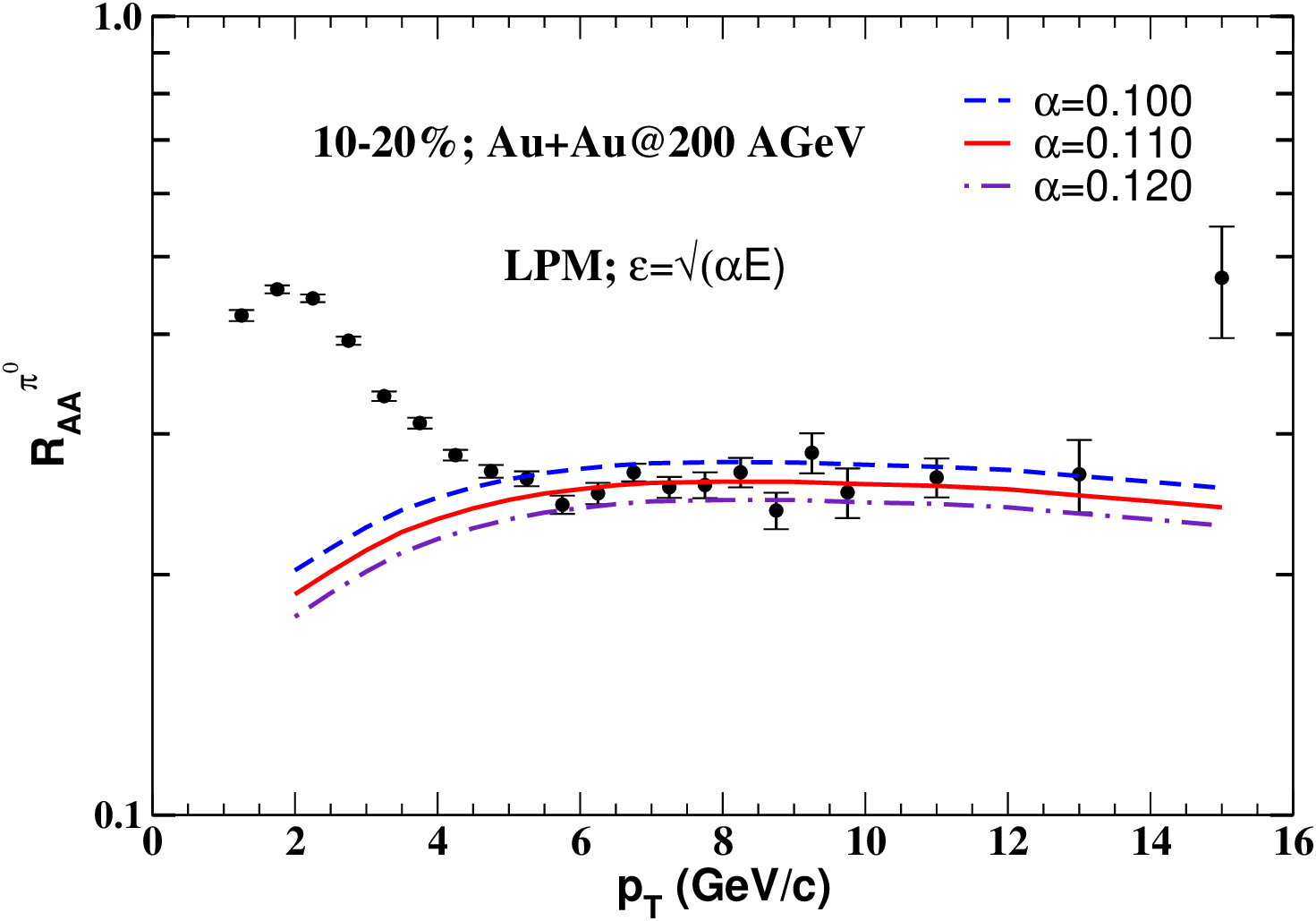}
\includegraphics[width=7.5cm, clip=true]{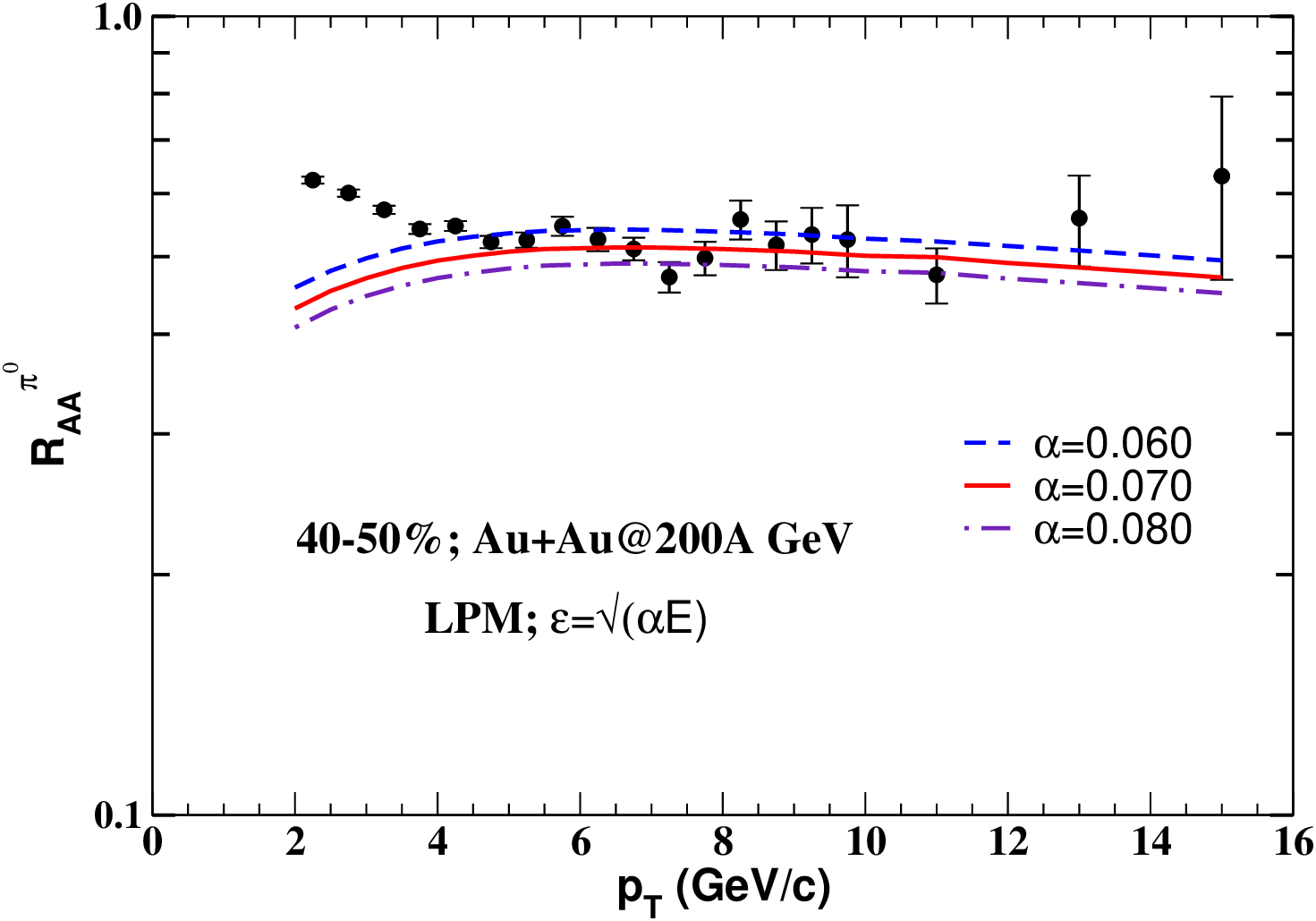}
\includegraphics[width=7.5cm, clip=true]{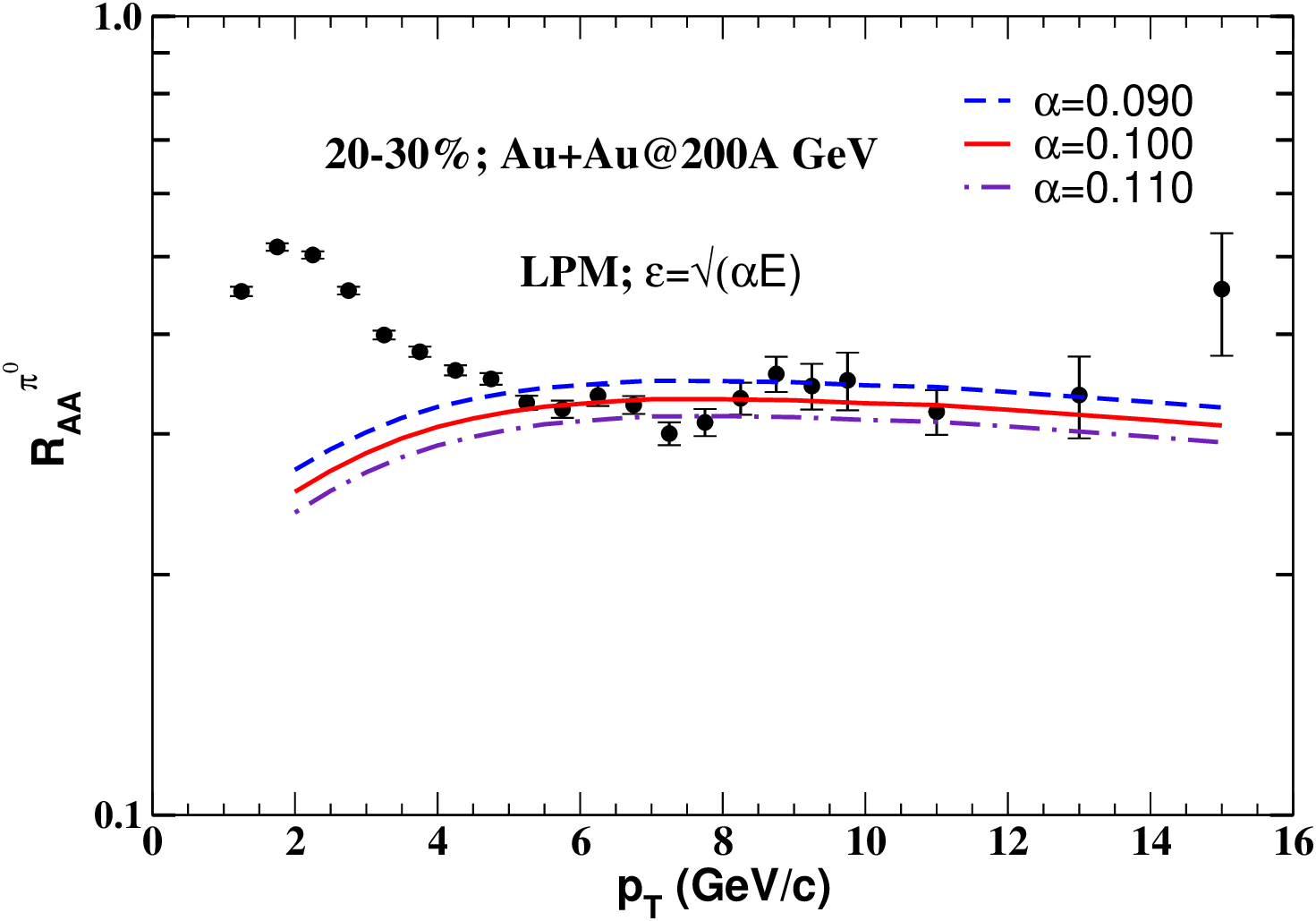}
\includegraphics[width=7.5cm, clip=true]{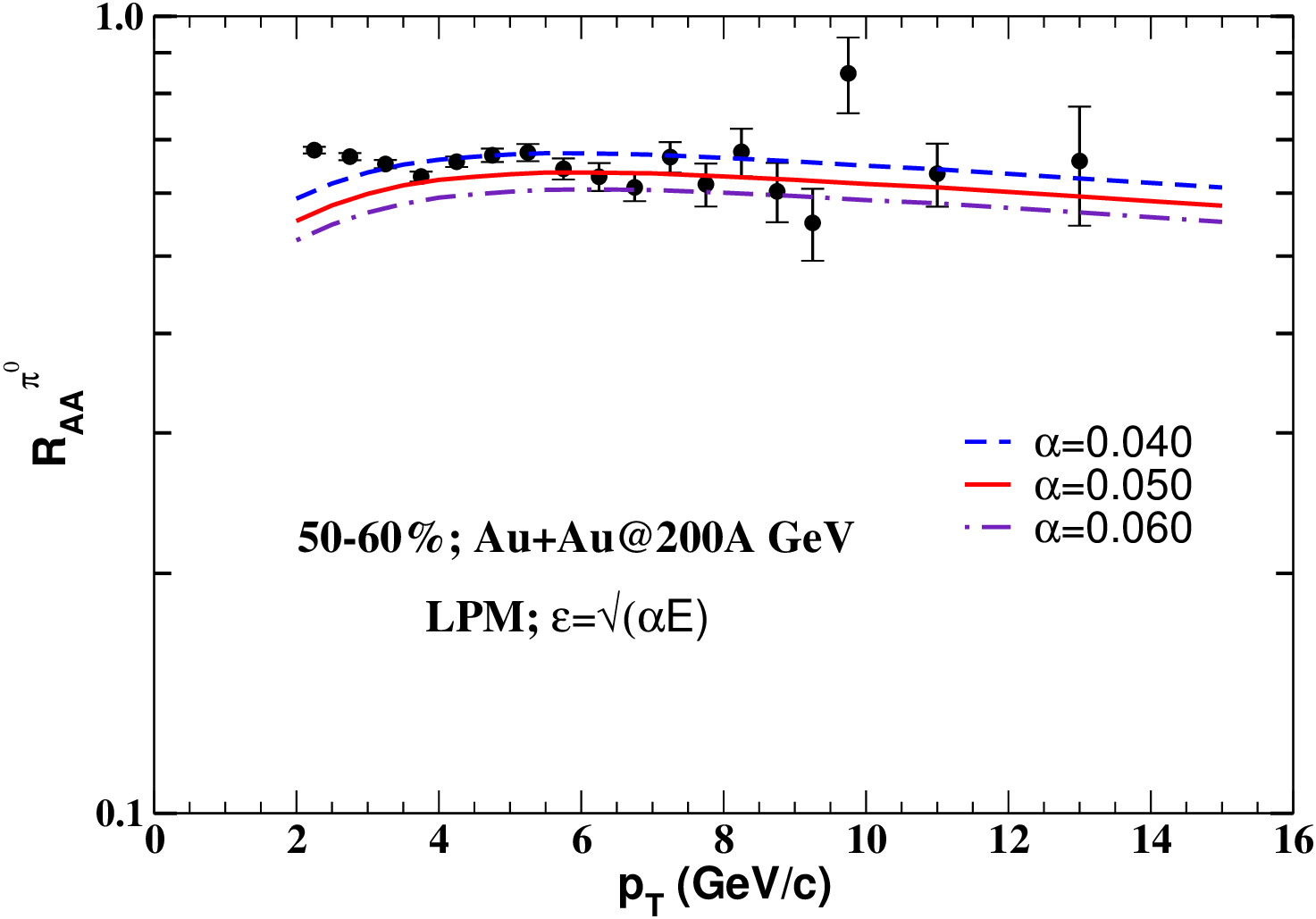}
\end{center}
\caption{(Colour on-line) 
Nuclear modification of neutral pion production for Au+Au collisions at $\sqrt{s_\textrm{NN}}$=200 GeV
using LPM mechanism. The experimental data are taken from Ref.~\cite{Raa_au}.}
\label{fig4}
\end{figure}

This expectation is confirmed by results shown in Fig.~\ref{fig4}, where the so-called LPM mechanism of energy loss
is seen to provide an accurate description of the nuclear modification function in the $p_T$ range of 6--10 GeV/$c$.
Once again we note that the energy loss coefficient $\alpha$ decreases systematically as the collisions become less central.
Thus a parton having an energy of 10 GeV is likely to lose about 1 GeV in the first collision in the most central
collision considered here and about 0.7 GeV in the collisions having a centrality of about 50--60\%.

\begin{figure}
\begin{center}
\includegraphics[width=7.5cm, clip=true]{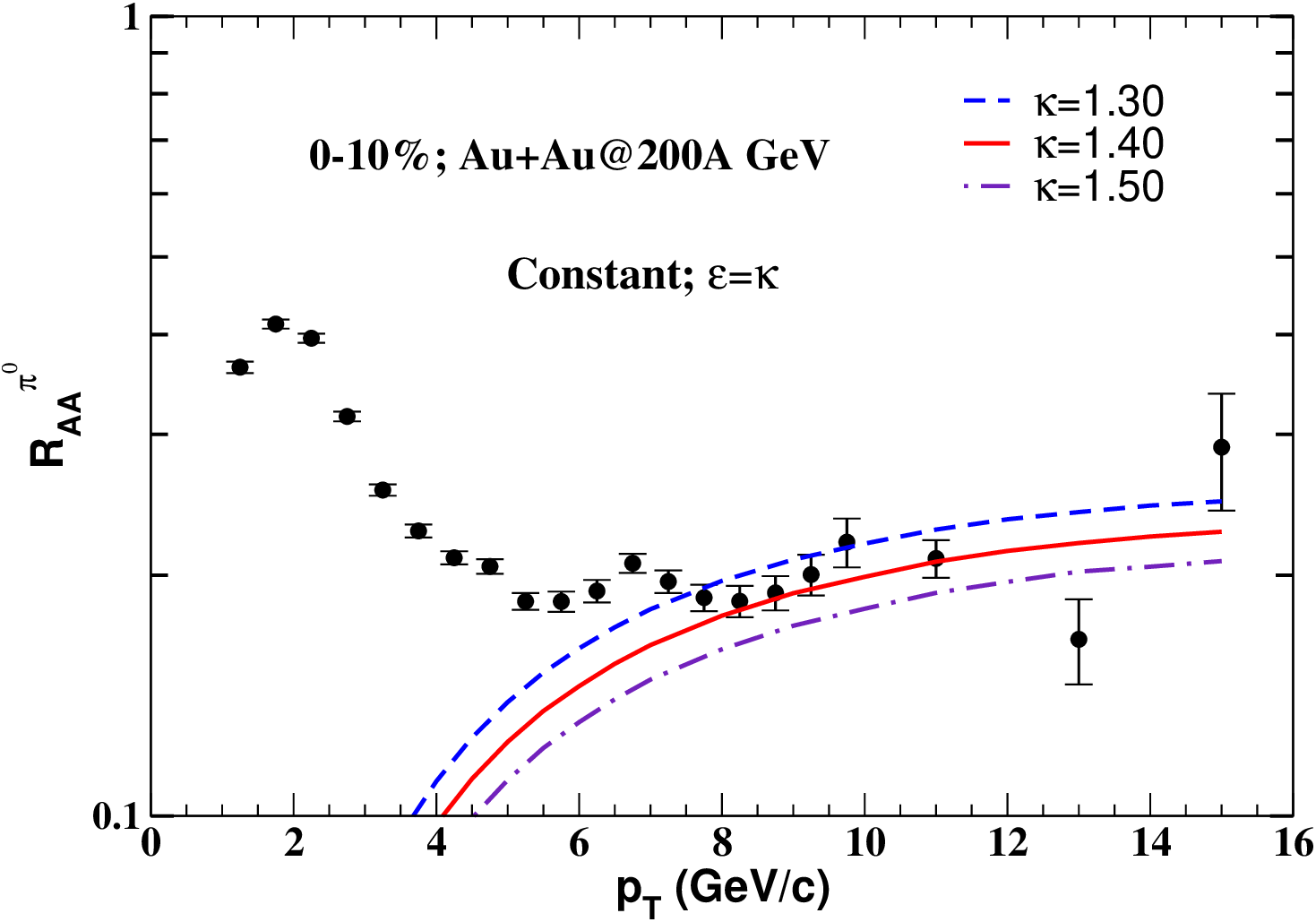}
\includegraphics[width=7.5cm, clip=true]{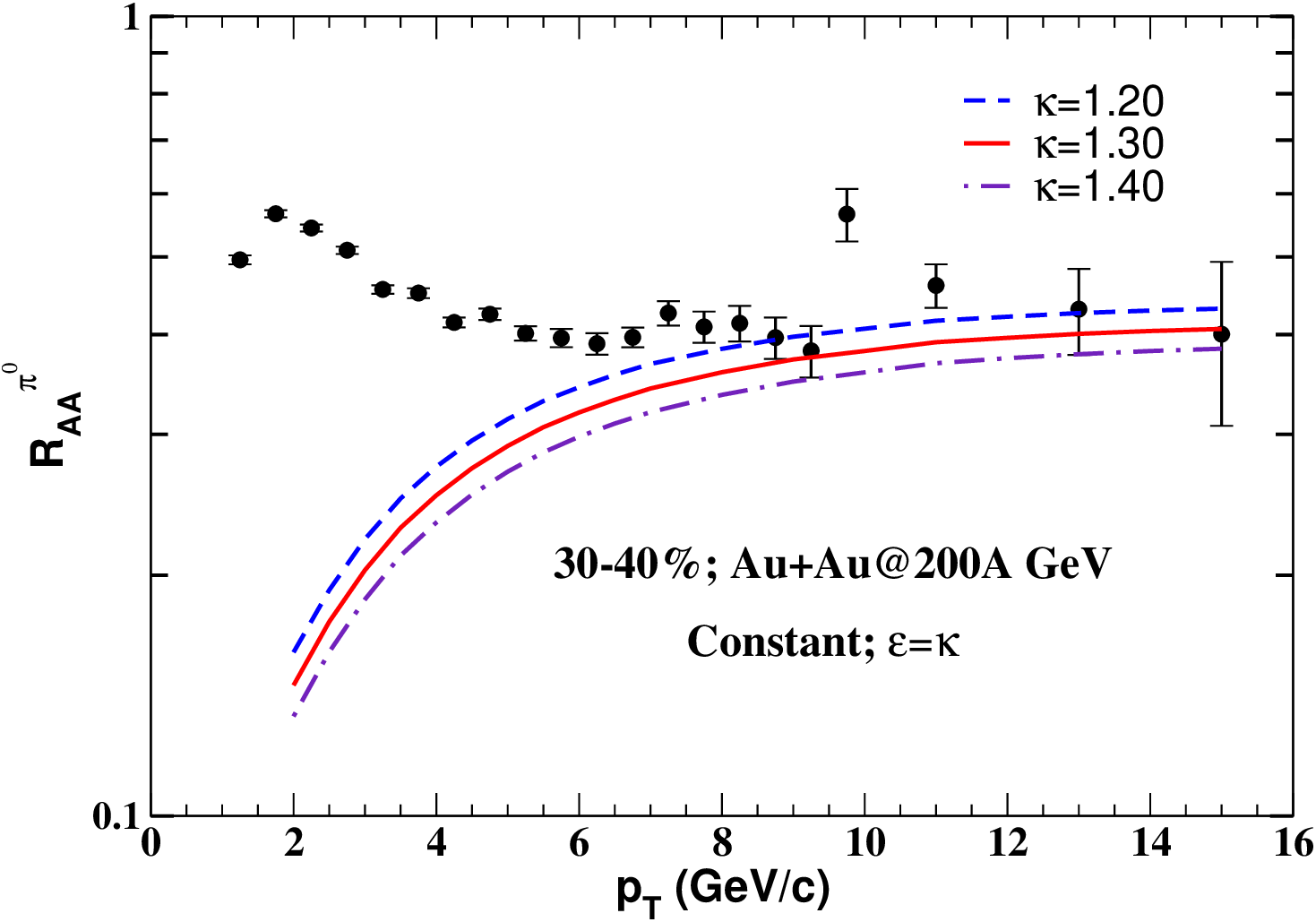}
\includegraphics[width=7.5cm, clip=true]{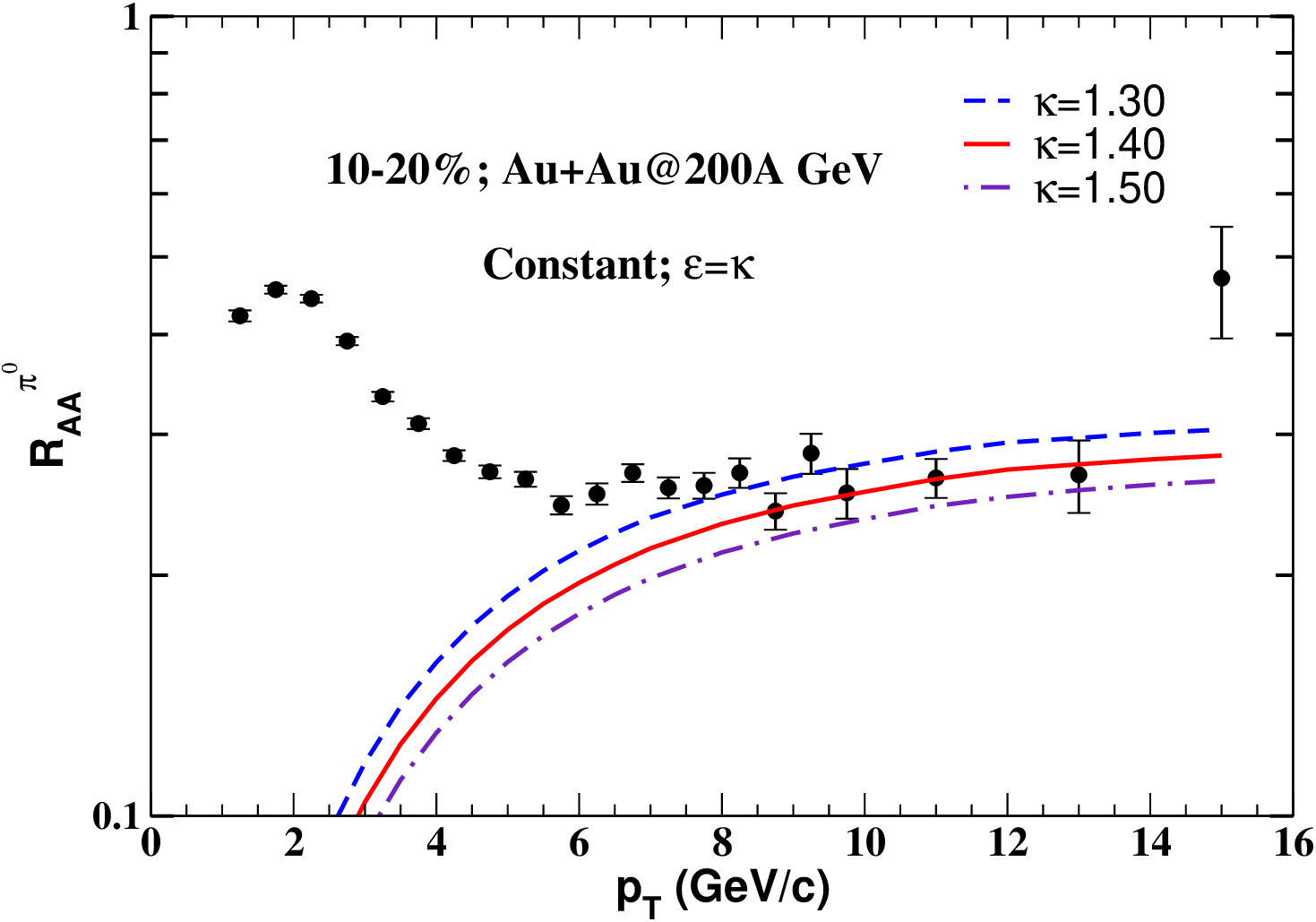}
\includegraphics[width=7.5cm, clip=true]{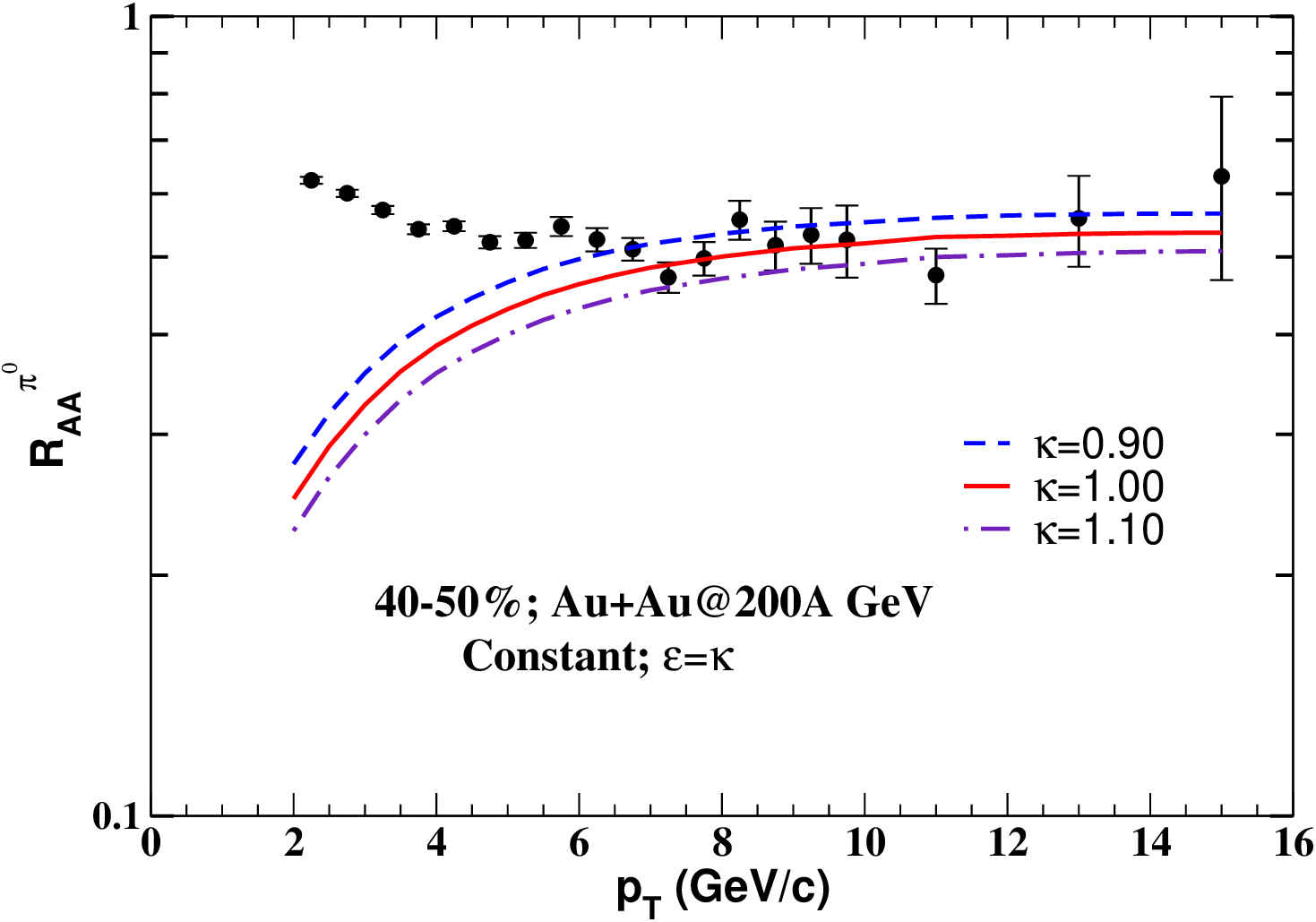}
\includegraphics[width=7.5cm, clip=true]{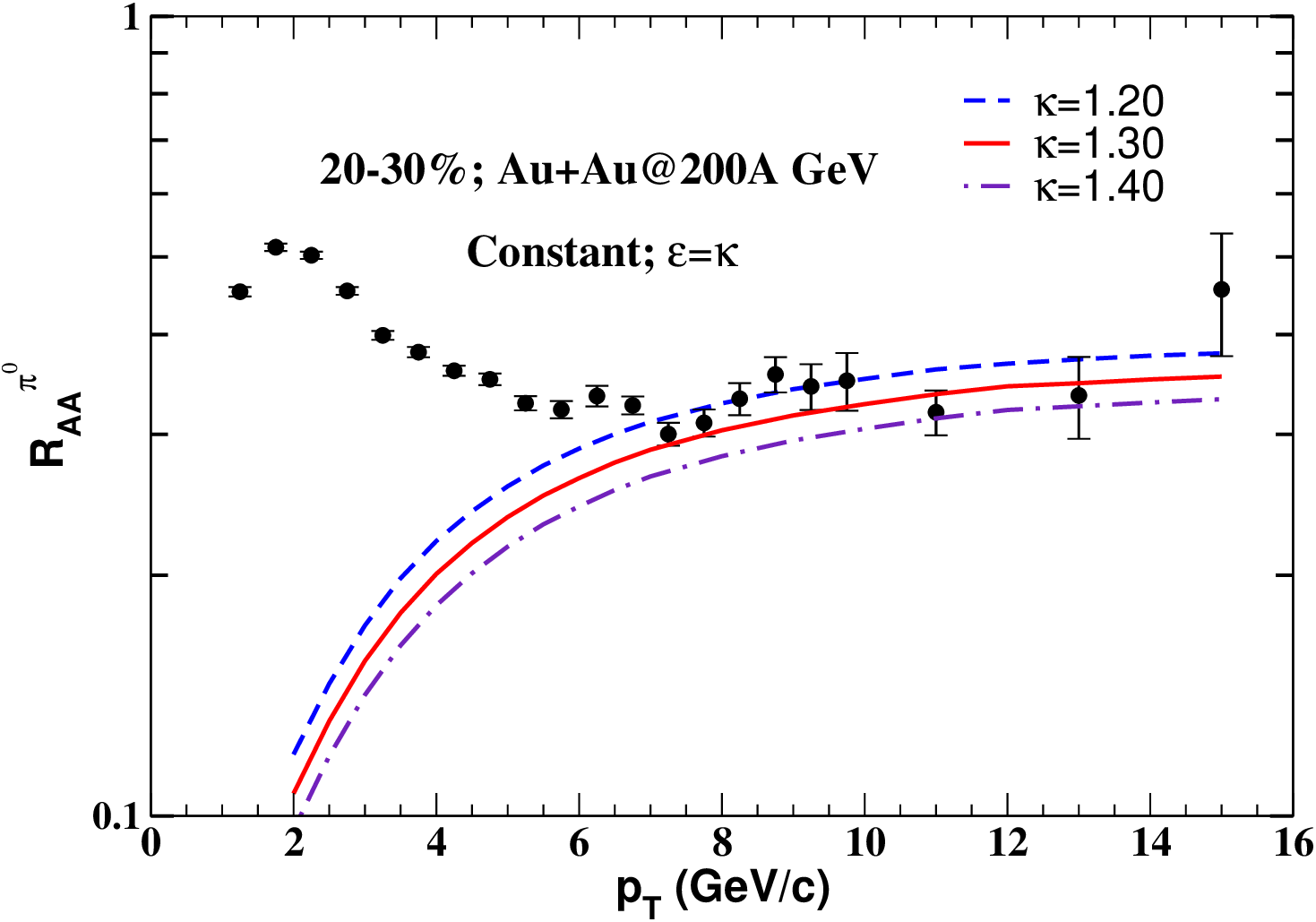}
\includegraphics[width=7.5cm, clip=true]{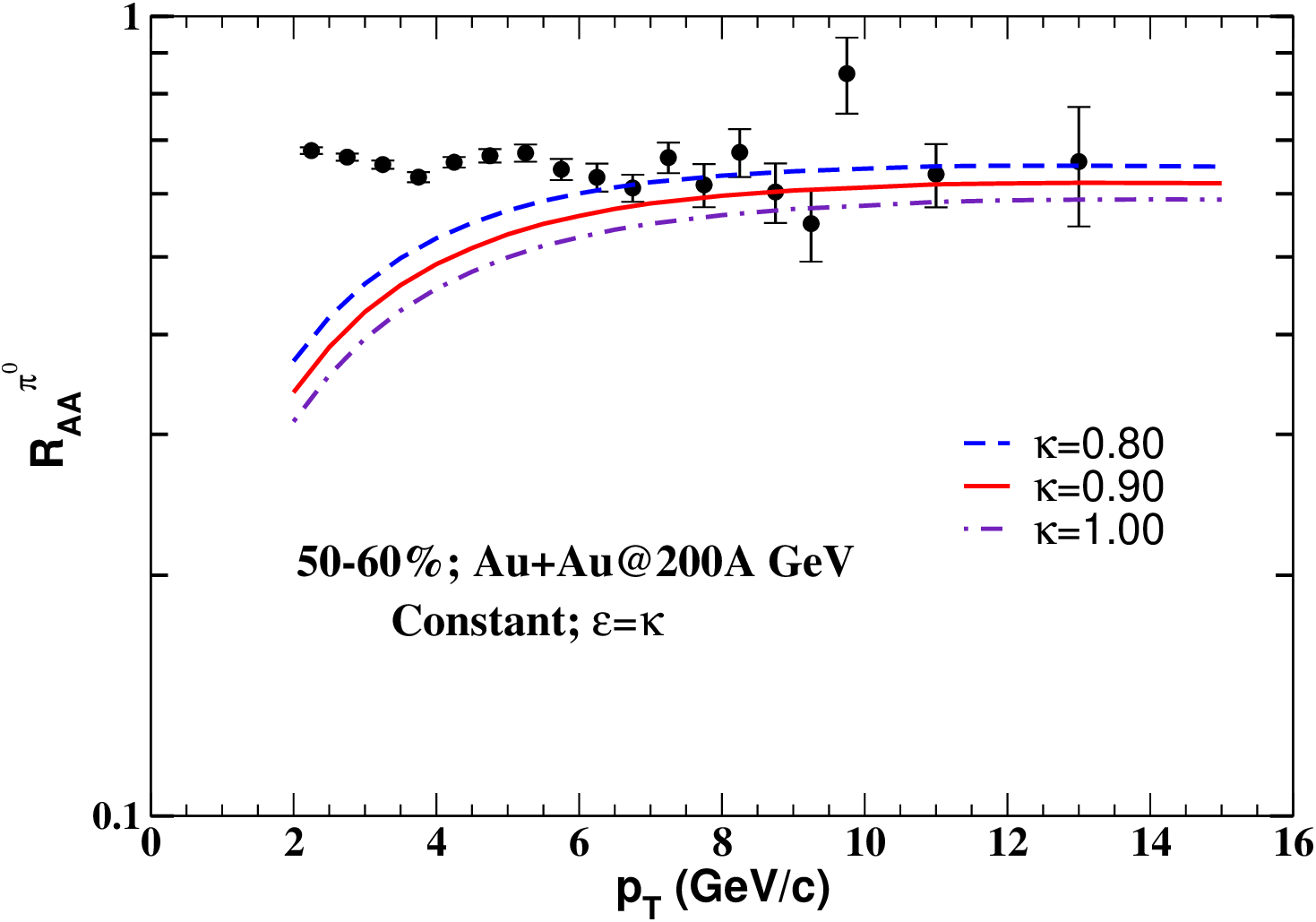}
\end{center}
\caption{(Colour on-line) 
Nuclear modification of neutral pion production for Au+Au collisions at $\sqrt{s_\textrm{NN}}$=200 GeV 
assuming a constant energy loss/collision. The experimental data are taken from Ref.~\cite{Raa_au}.}
\label{fig5}
\end{figure}

Finally, we see (Fig.~\ref{fig5}) that the medium modification function for hadrons having $p_T >$ 8 GeV/$c$ is best described
by the mechanism assuming a constant energy loss per collision. The energy loss per collision is seen to decrease from
about 1.4 GeV for the most central collisions to about 1 GeV for the collisions having a centrality of 50--60\%. 

The transition from the BH regime to the LPM seen above and noted
 earlier~\cite{dks_jet} at  $p_T \approx$  5 GeV/$c$ should not 
come as a surprise to us, since for the mean free-path $\lambda\sim$ 1 fm and
 $\langle k_T^2 \rangle \approx$ 1 (GeV/$c$)$^2$ per collision,
 $E_\textrm{LPM}$ is about 5 GeV, which
roughly separates the BH and LPM energy loss regimes for all centralities

\begin{figure}
\begin{center}
\includegraphics[width=7.5cm, clip=true]{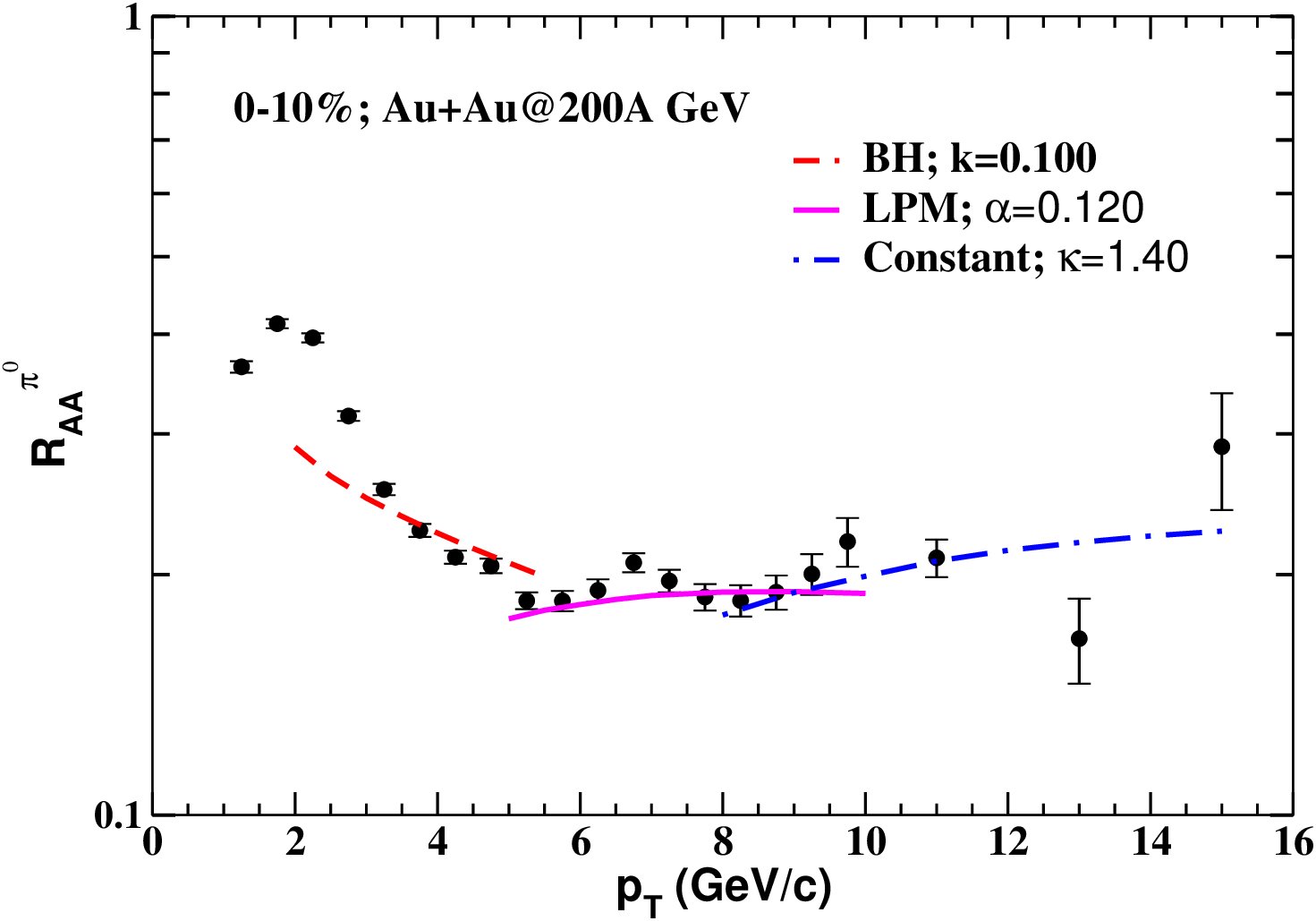}
\includegraphics[width=7.5cm, clip=true]{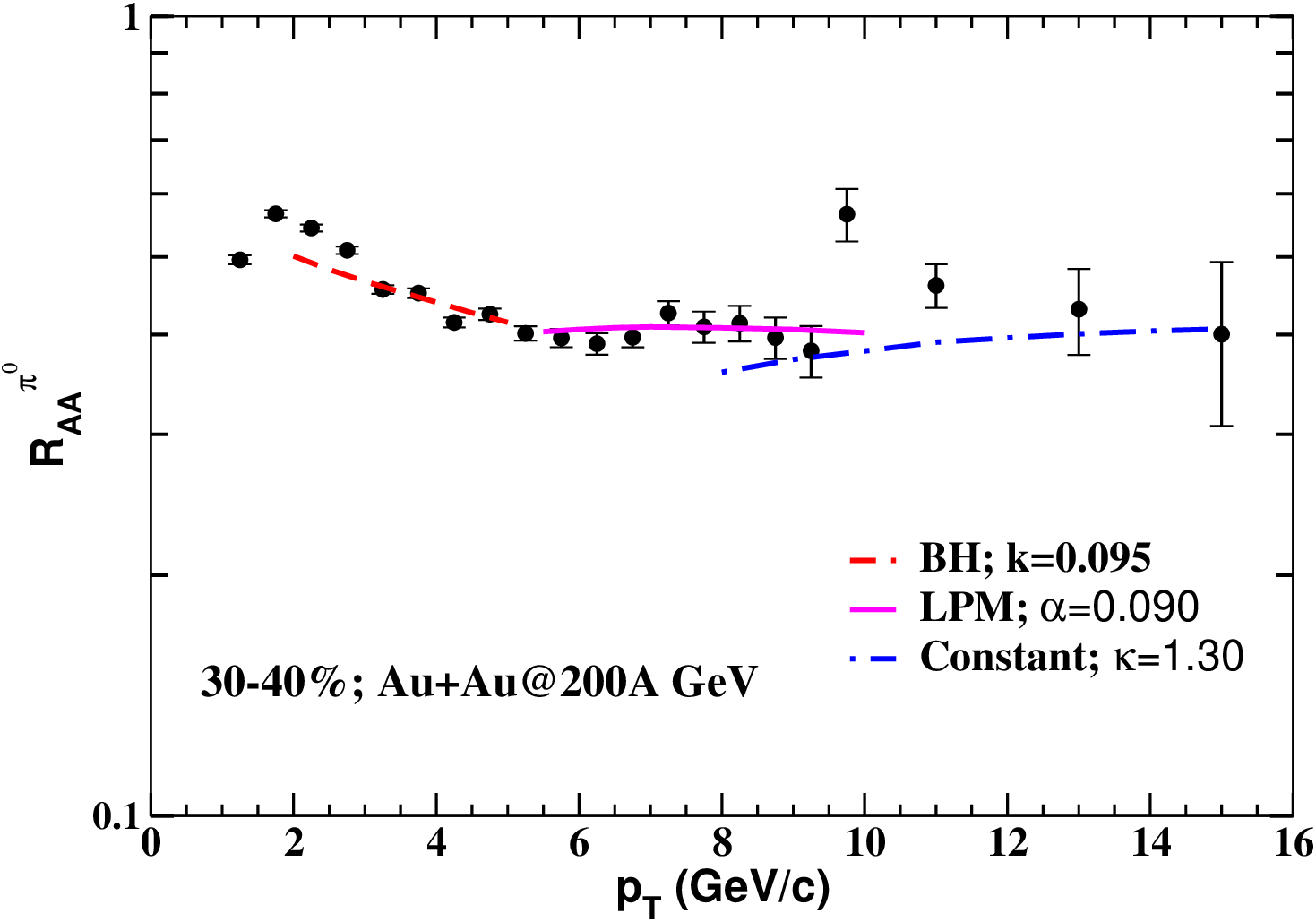}
\includegraphics[width=7.5cm, clip=true]{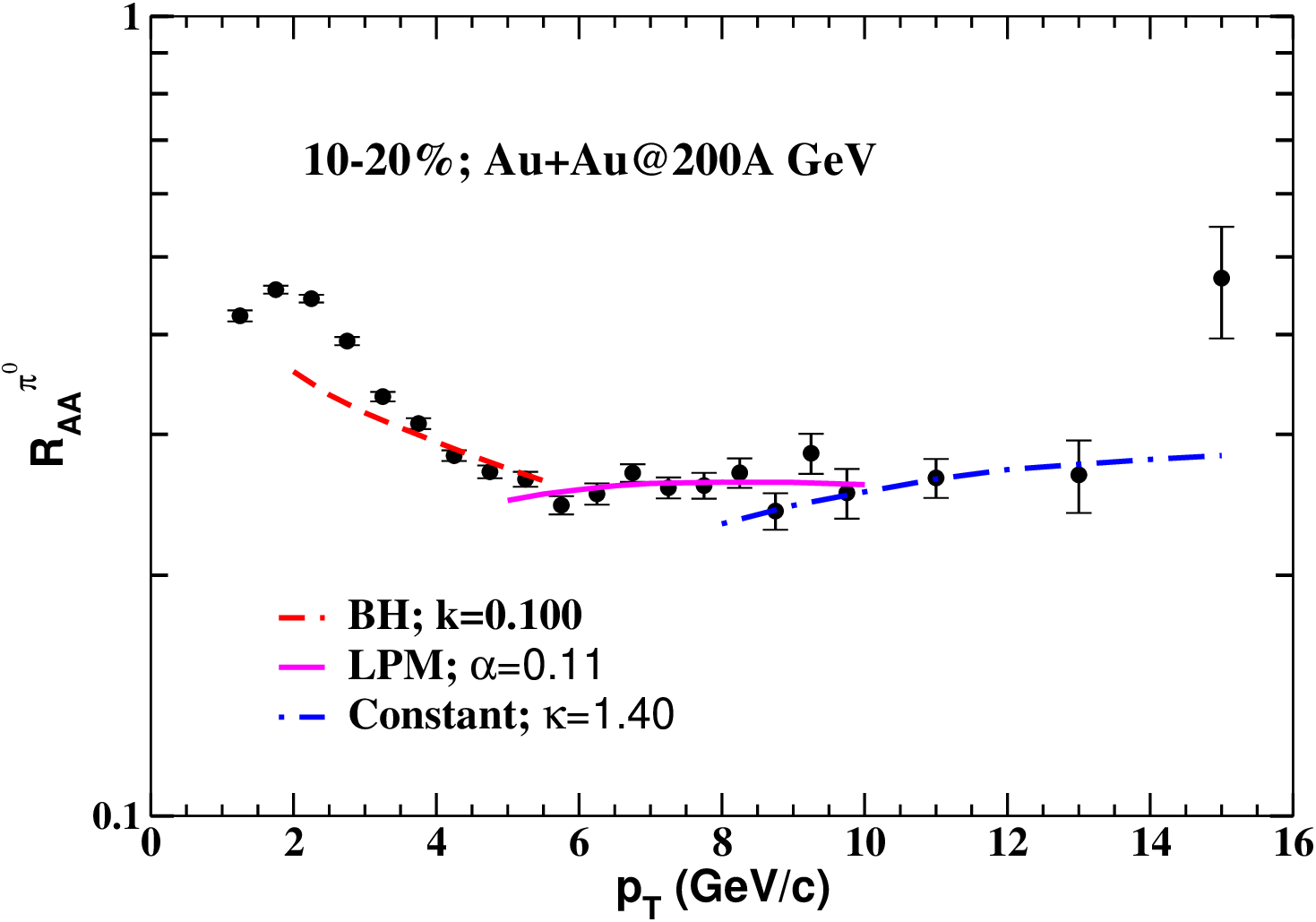}
\includegraphics[width=7.5cm, clip=true]{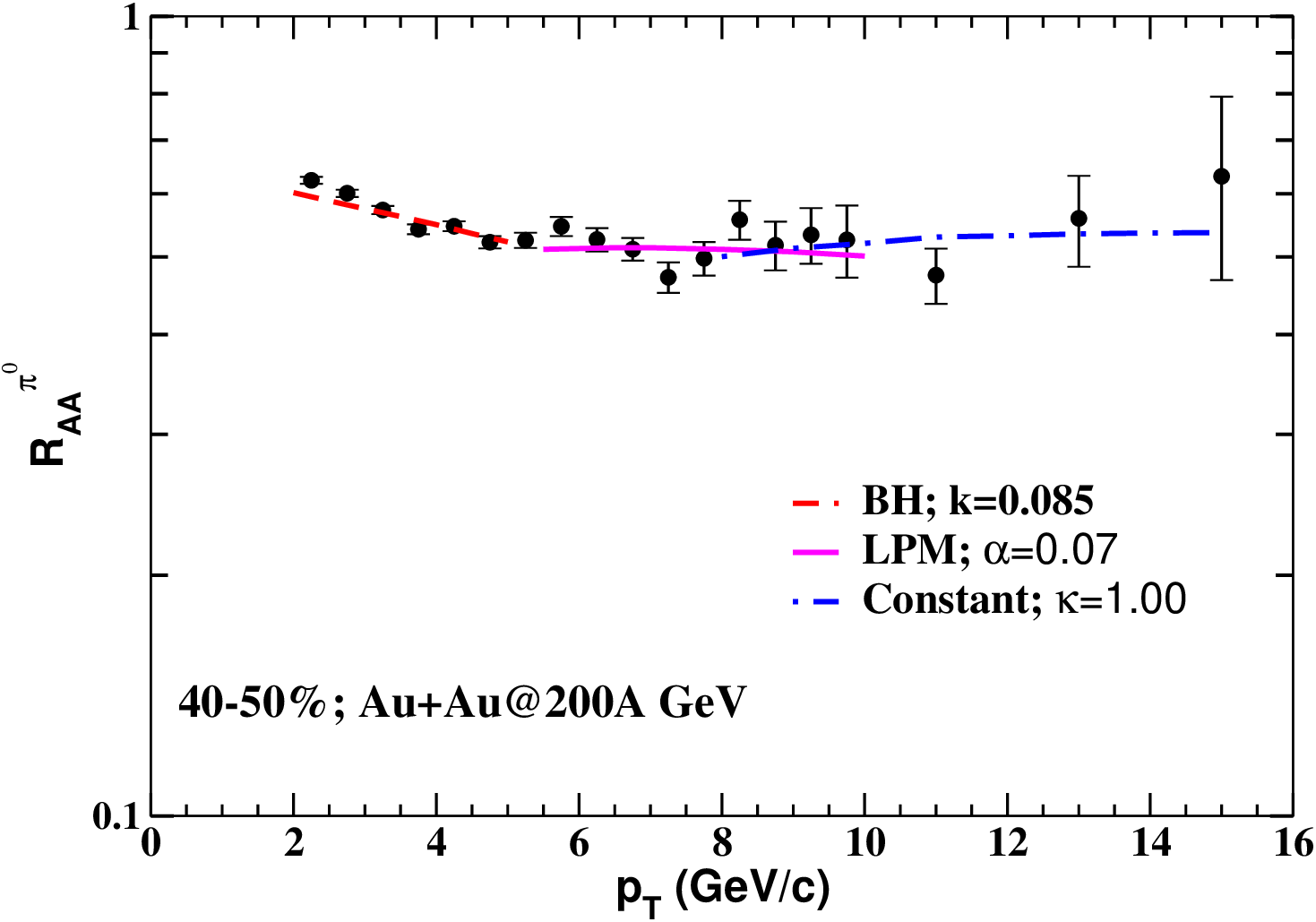}
\includegraphics[width=7.5cm, clip=true]{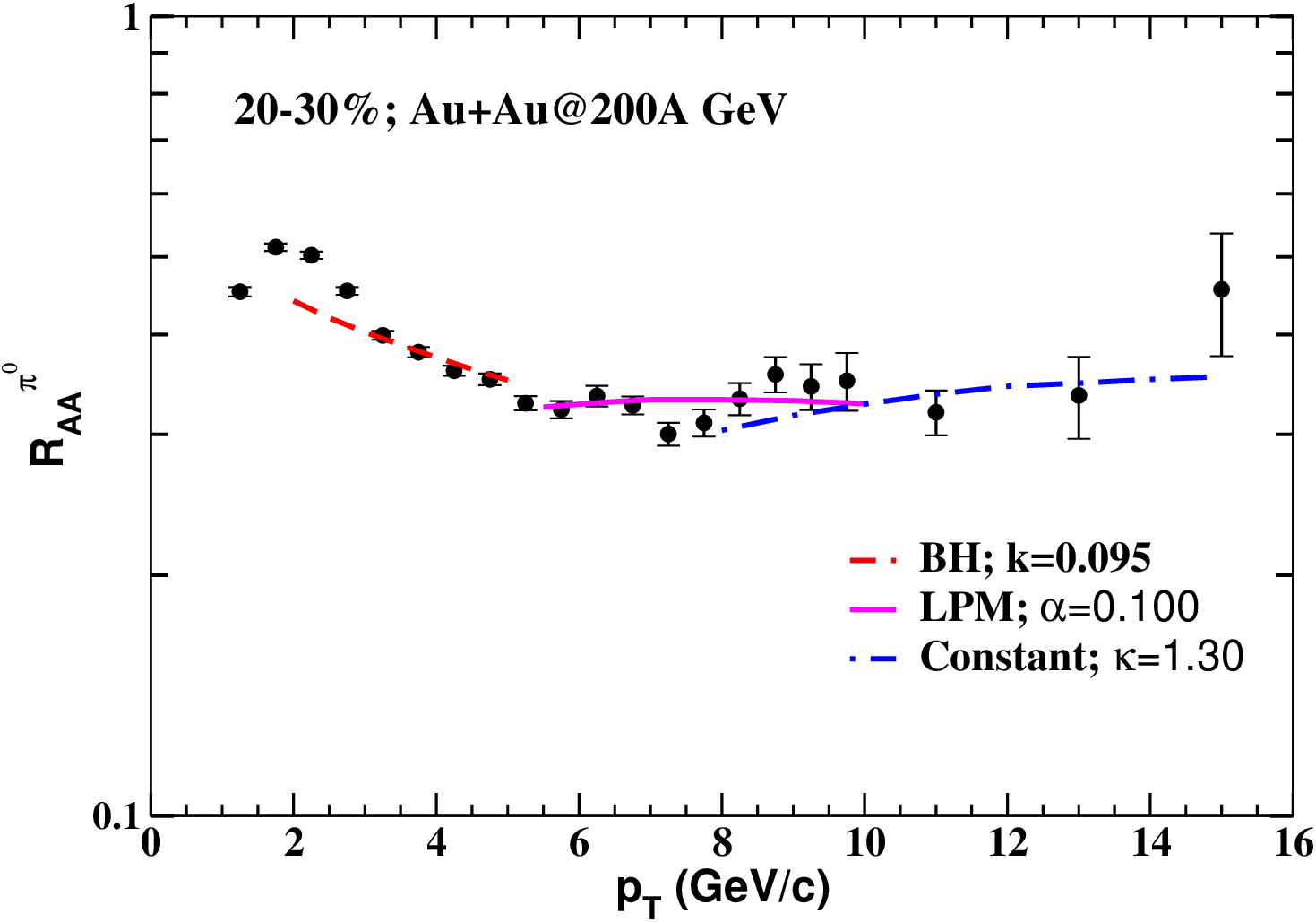}
\includegraphics[width=7.5cm, clip=true]{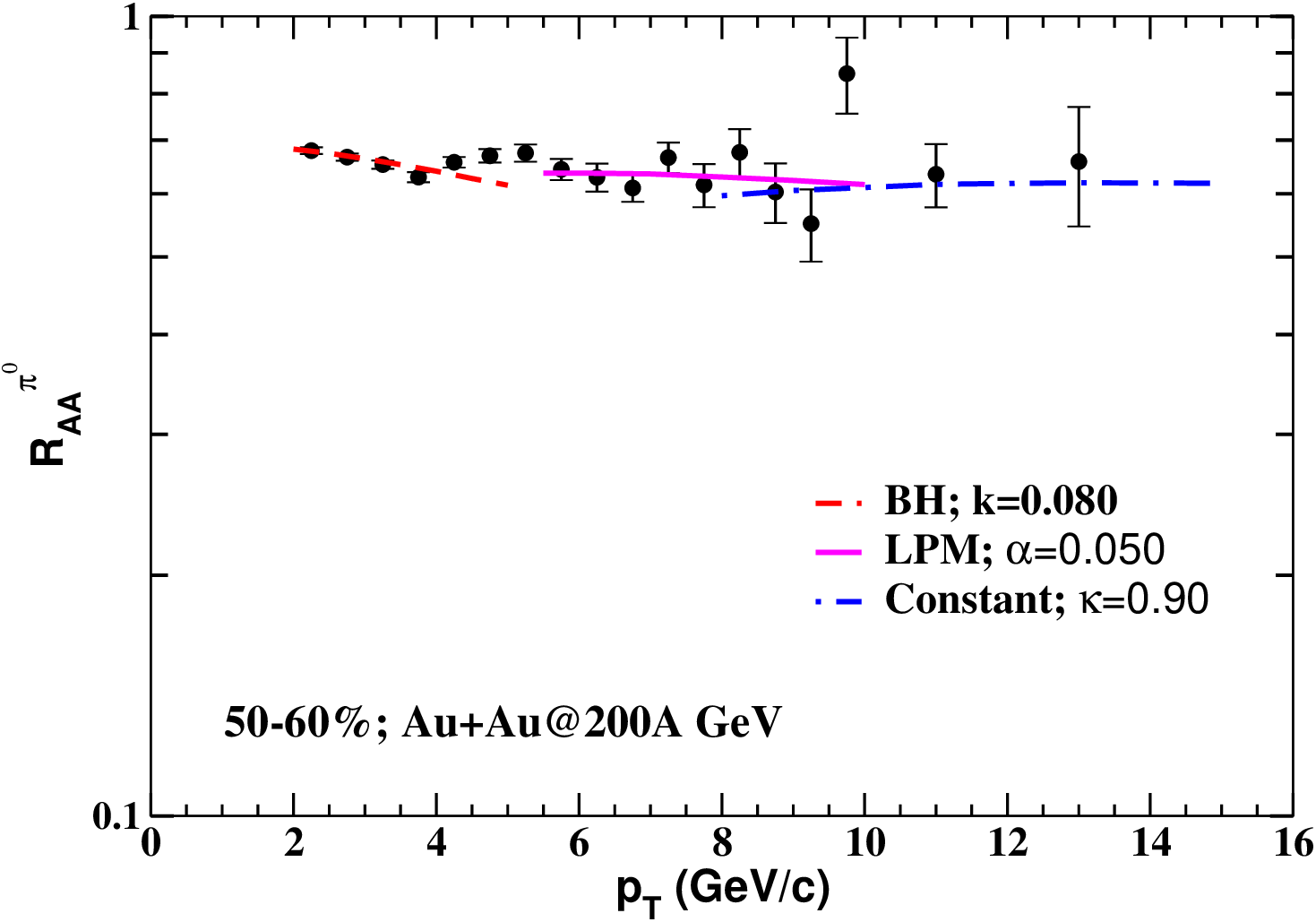}
\end{center}
\caption{(Colour on-line) 
Nuclear modification of neutral pion production for
 Au+Au collisions at $\sqrt{s_\textrm{NN}}$=200 GeV 
for the three energy loss schemes for different $p_T$ regimes.}
\label{fig5a}
\end{figure}

The results for these three figures are summarized in Fig.~\ref{fig5a}. 
Several interesting facts emerge.

 Let us first look at the results for the
BH regime operating for $p_T$ up to about 5--6 GeV/$c$. We note as we go to 
more peripheral collisions the data, even for lower $p_T$, are in a better 
agreement with our calculations. This, we feel, is due to the reducing importance
of radial flow (which gives a transverse kick to the partons/hadrons) in
peripheral collisions. 

The LPM regime operating over the $p_T$ window of 5--10 GeV/$c$ and the
complete coherence regime operating at higher $p_T$ are seen to correctly
describe the details of the changing curvature of the nuclear modification
factor. 

We add that a look at Fig.~\ref{fig4} may prompt one to conclude that the 
description using the LPM mechanism is also working reasonably well
till the largest $p_T$ considered here.
 However a closer look at the Figs.~\ref{fig4} and \ref{fig5} reveal that the
LPM description mostly misses the data for larger $p_T$ values while the
description using the constant energy loss per collision correctly reflects the
curvature of the data till the largest $p_T$.
 A more accurate data extending up to even larger $p_T$
would be very valuable to settle this question more firmly.

Even though  we have identified the regions of the applicability of the
 mechanisms of energy loss, purely on the basis of a good description of the experimental
values of the nuclear modification factor, it is heartening to note that the
separation of the BH and the LPM regimes is very close to 5 GeV/$c$ expected 
by us, as discussed above. We recall that the LPM and the complete coherence
regimes differ by the formation time of the gluons, which is larger than the mean
free path but smaller than the path-length for the former, while it is larger than
the path-length (and of-course the mean free path) for the later. This is
 reflected in a slight change in the curvature
of the nuclear modification factor, around $p_T \approx$ 8--10 GeV/$c$.

\begin{figure}[ht]
\begin{center}
\includegraphics[width=11.5cm, clip=true]{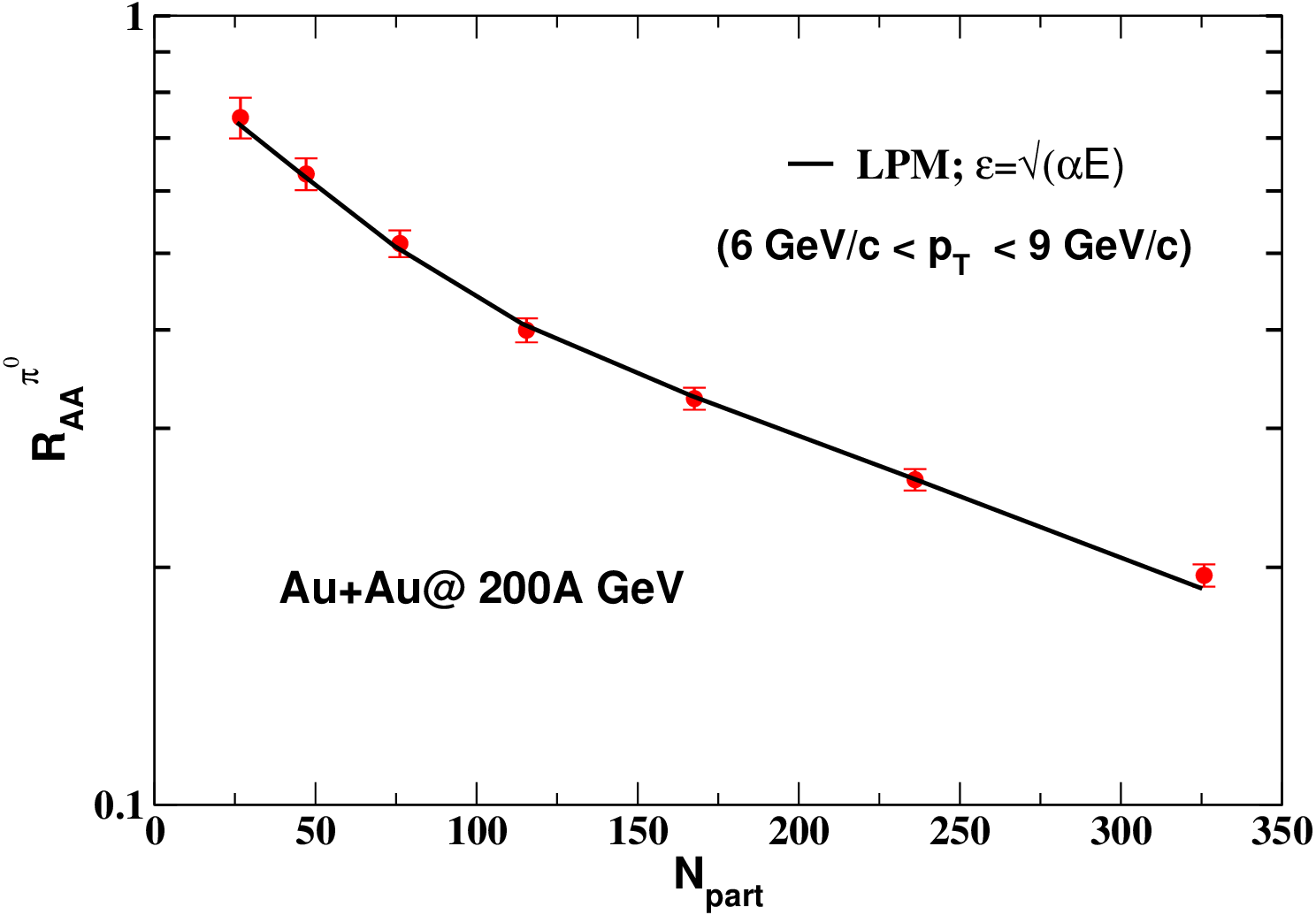}
\end{center}
\caption{(Colour on-line) Centrality dependence of $R_{AA}$ of neutral pions calculated for LPM mechanism of energy loss.
The data are from the PHENIX collaboration~\cite{pi0v2_phenix} for Au+Au collisions at $\sqrt{s_\textrm{NN}}$ = 200 GeV.}
\label{fig6}
\end{figure}

In brief, we have noted that the energy loss parameter in the 
Wang, Huang, and Sarcevic model for energy loss of partons in relativistic
heavy ion collisions at RHIC can be tuned to obtain an accurate description of the nuclear modification factor $R_{AA}$
for different ranges of the transverse momenta. This observation is further strengthened by Fig.~\ref{fig6} where the variation of
$R_{AA}$ integrated over $p_T$ ranging from 6 to 9 GeV/$c$ is shown as a function of number of participants and compared with
the data from PHENIX collaboration~\cite{pi0v2_phenix}.

\subsection{Centrality dependence of $R_{AA}$ for Cu-Cu collisions at RHIC}

In the next step we apply the above treatment to the collisions of copper nuclei at the same centre of mass energy of 200 GeV/nucleon.
These collisions present an interesting system for the study of jet-quenching which complements as well as
supplements the results from the collision of gold nuclei discussed earlier. It is well known that the more central collisions
of copper nuclei have  number of participants similar to those for mid-central collisions of gold nuclei. Thus a comparison
would give us results for two systems, involving similar number of participants,
one of which has nearly central collisions and the other one has a large eccentricity.

\begin{figure}
\begin{center}
\includegraphics[width=7.5cm, clip=true]{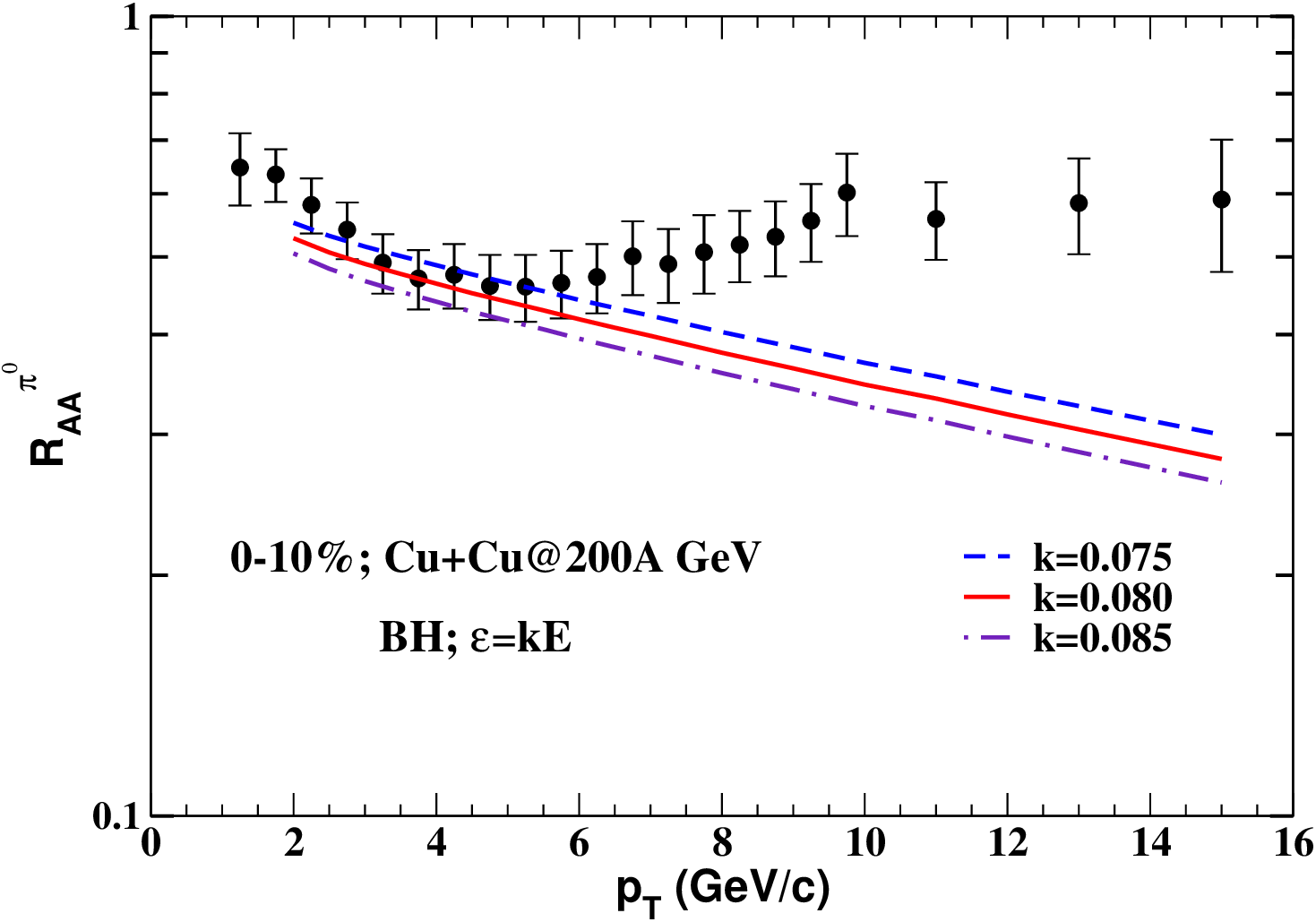}
\includegraphics[width=7.5cm, clip=true]{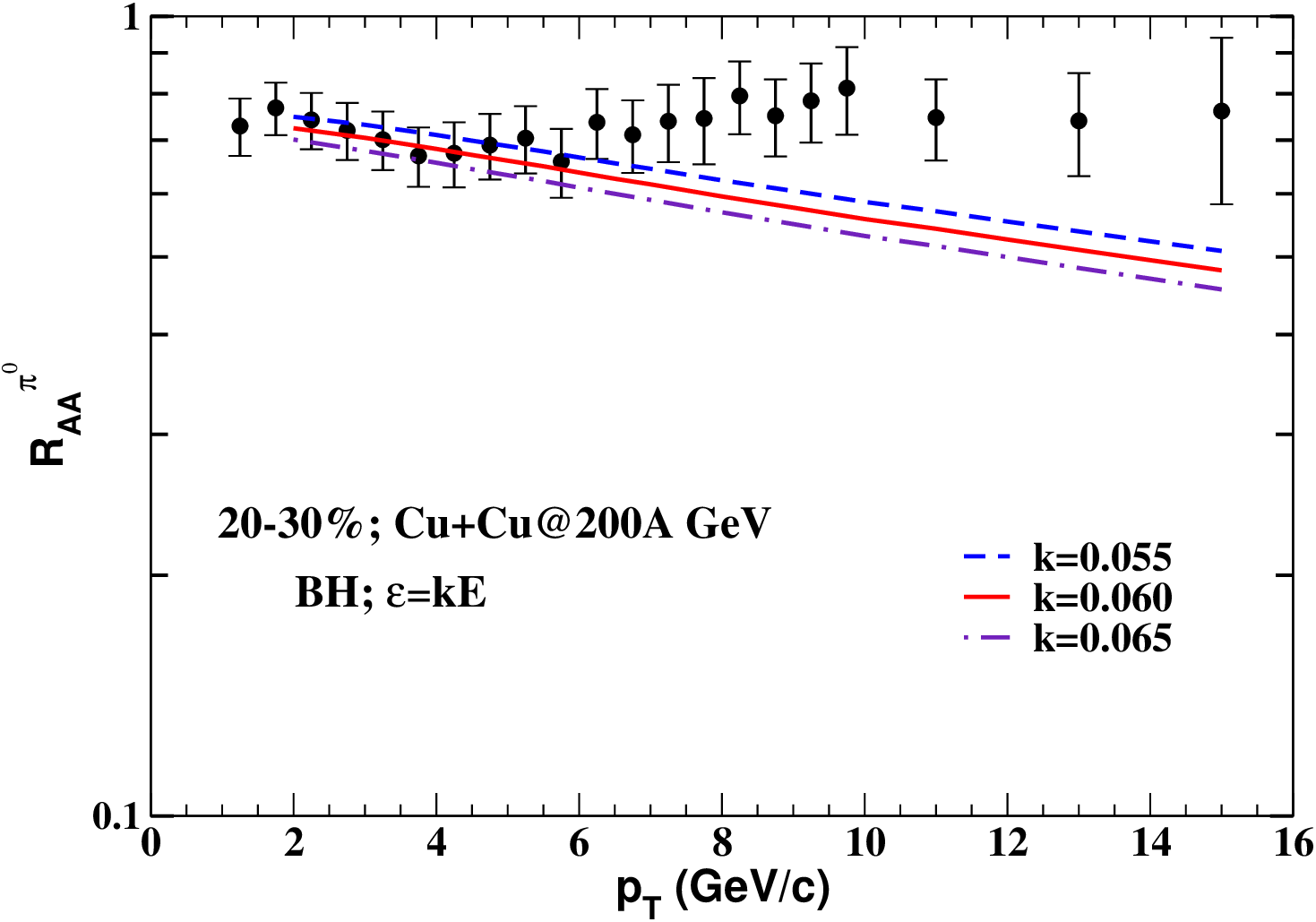}
\includegraphics[width=7.5cm, clip=true]{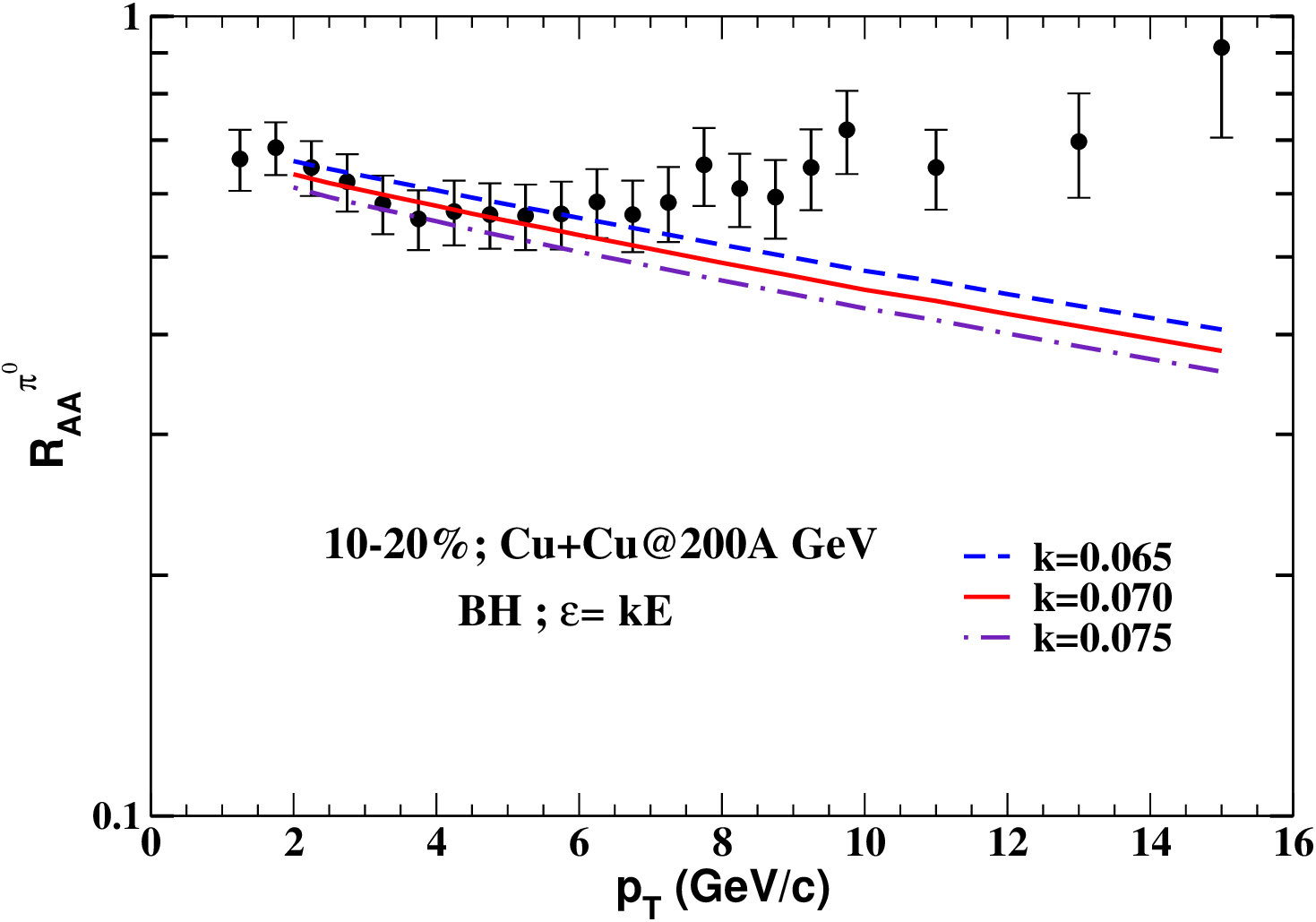}
\includegraphics[width=7.5cm, clip=true]{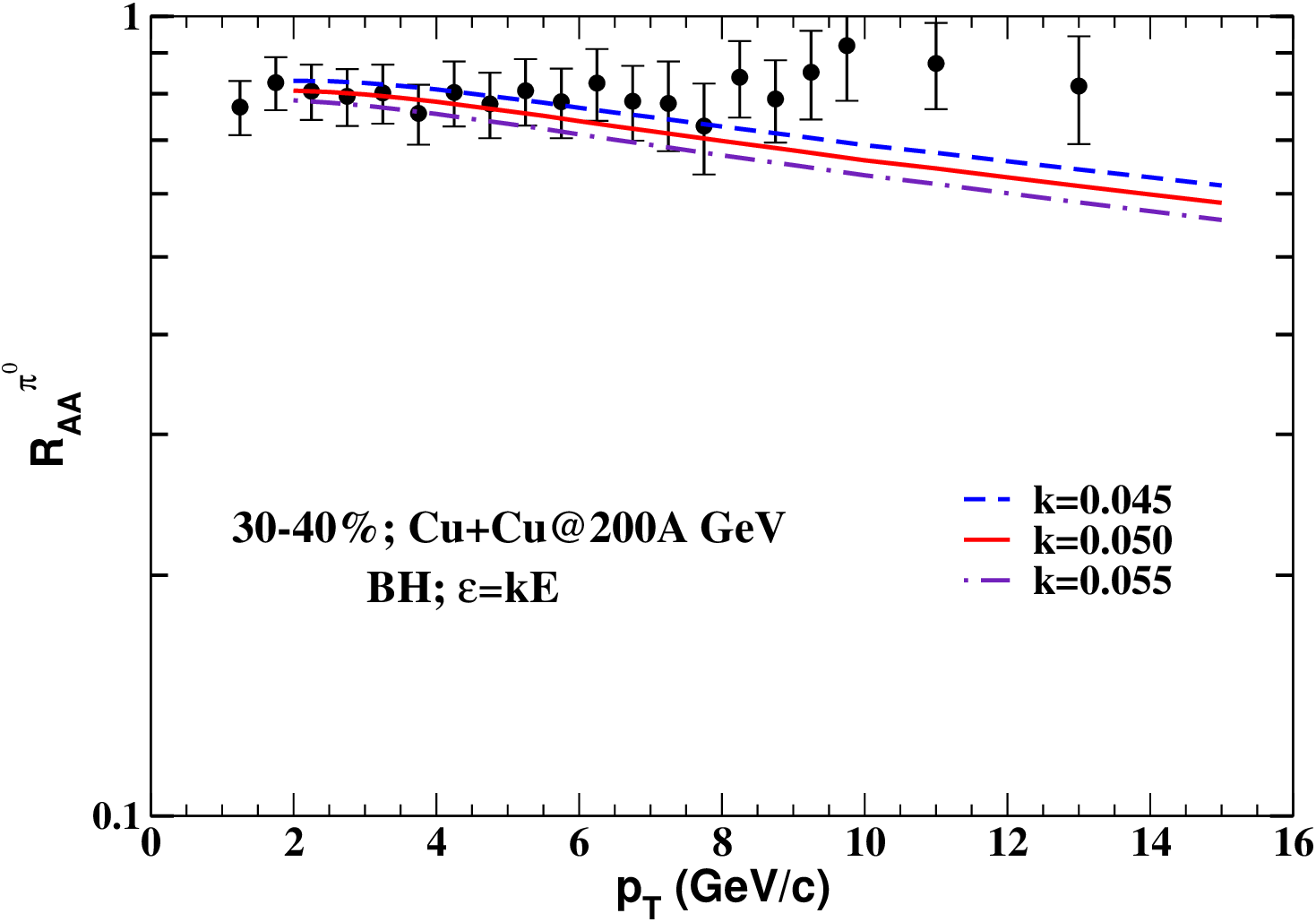}
\end{center}
\caption{(Colour on-line)
Nuclear modification of neutral pion production for Cu+Cu collisions at $\sqrt{s_\textrm{NN}}$=200 GeV for BH mechanism of energy loss
along with the results from the  PHENIX collaboration~\cite{phenix_cu}. }
\label{fig7}
\end{figure}

\begin{figure}
\begin{center}
\includegraphics[width=7.5cm, clip=true]{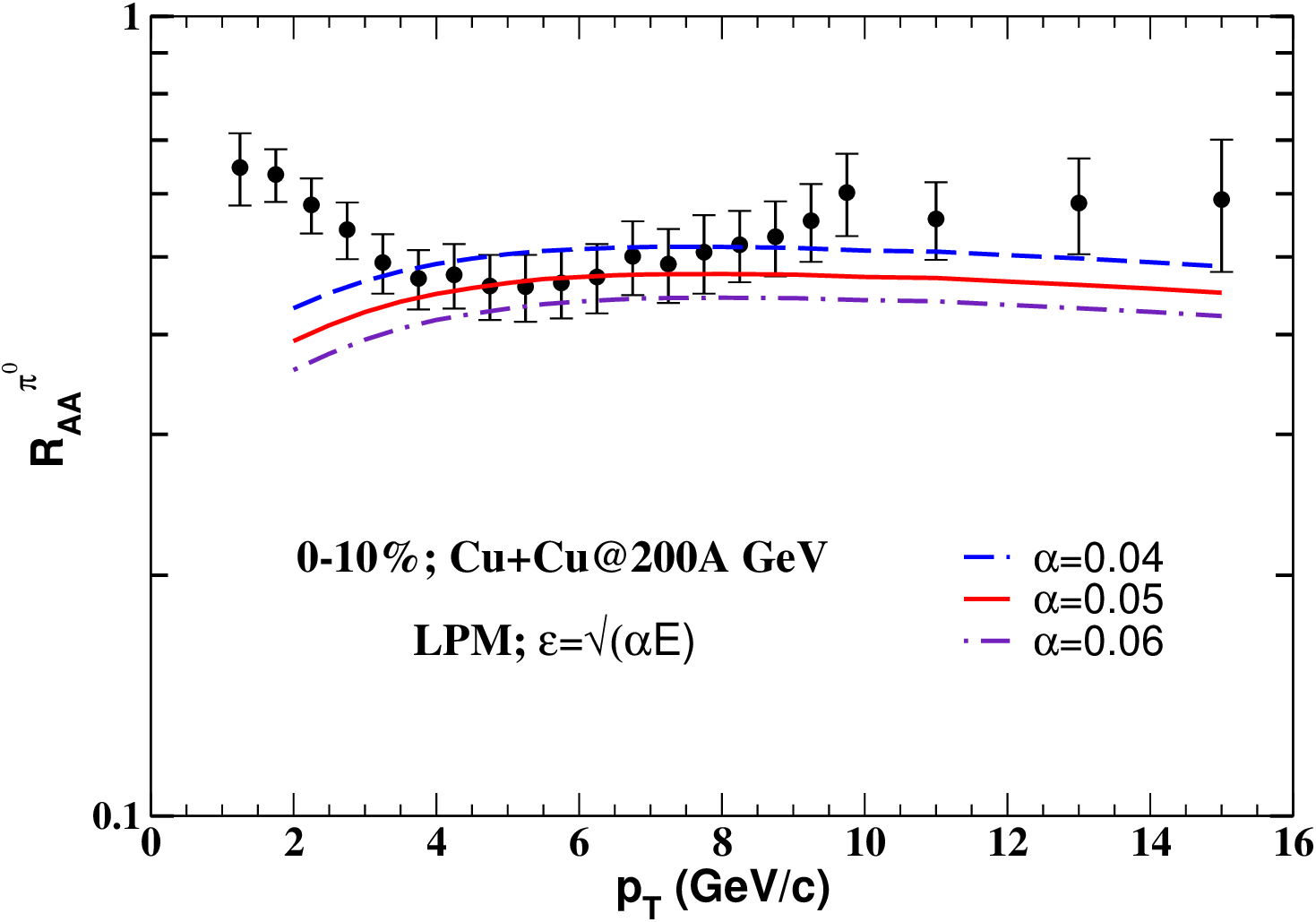}
\includegraphics[width=7.5cm, clip=true]{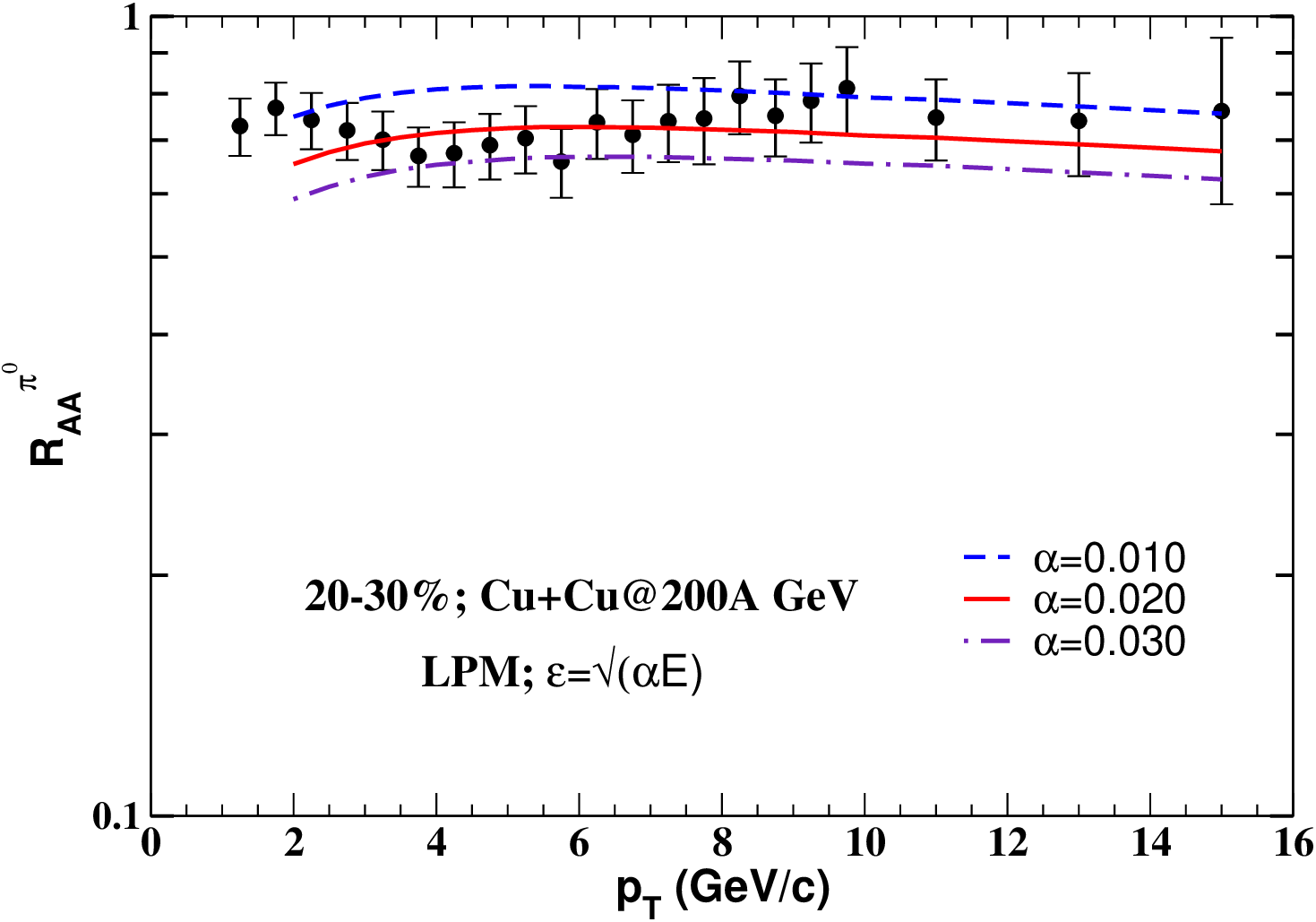}
\includegraphics[width=7.5cm, clip=true]{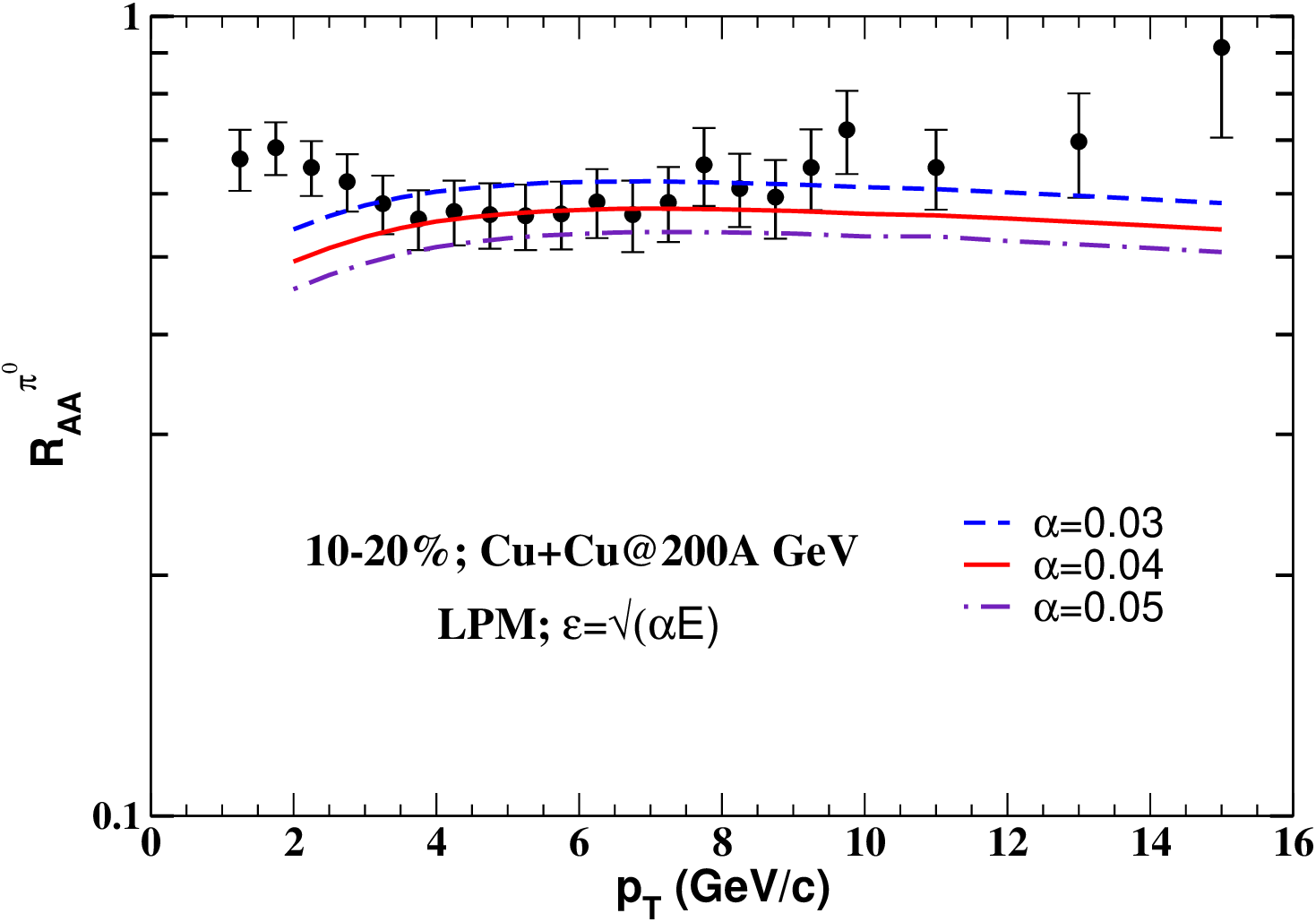}
\includegraphics[width=7.5cm, clip=true]{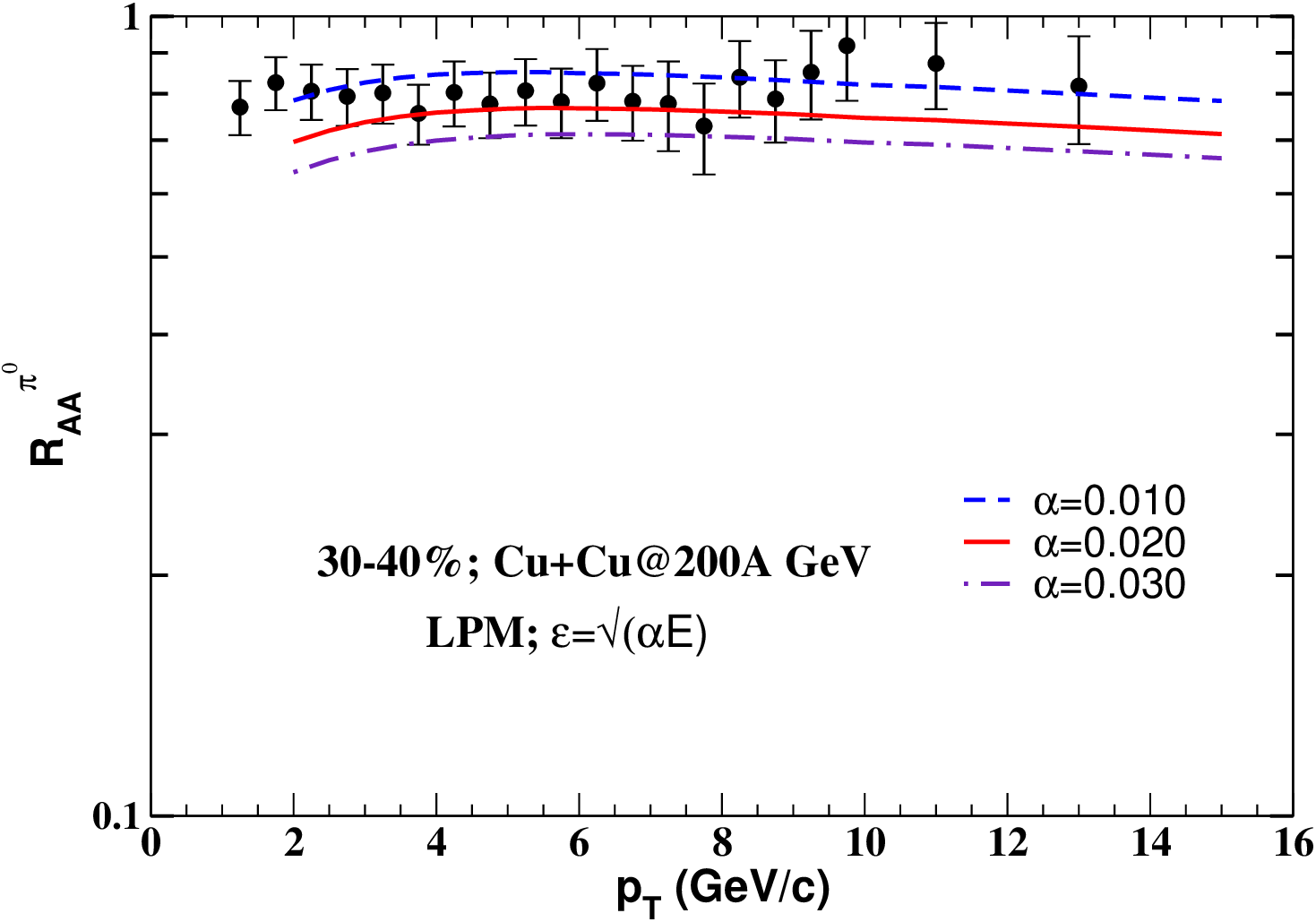}
\end{center}
\caption{(Colour on-line) 
Nuclear modification of neutral pion production for Cu+Cu collisions at $\sqrt{s_\textrm{NN}}$=200 GeV
for LPM mechanism of energy loss
and compared with PHENIX results~\cite{phenix_cu}.}
\label{fig8}
\end{figure}

\begin{figure}
\begin{center}
\includegraphics[width=7.5cm, clip=true]{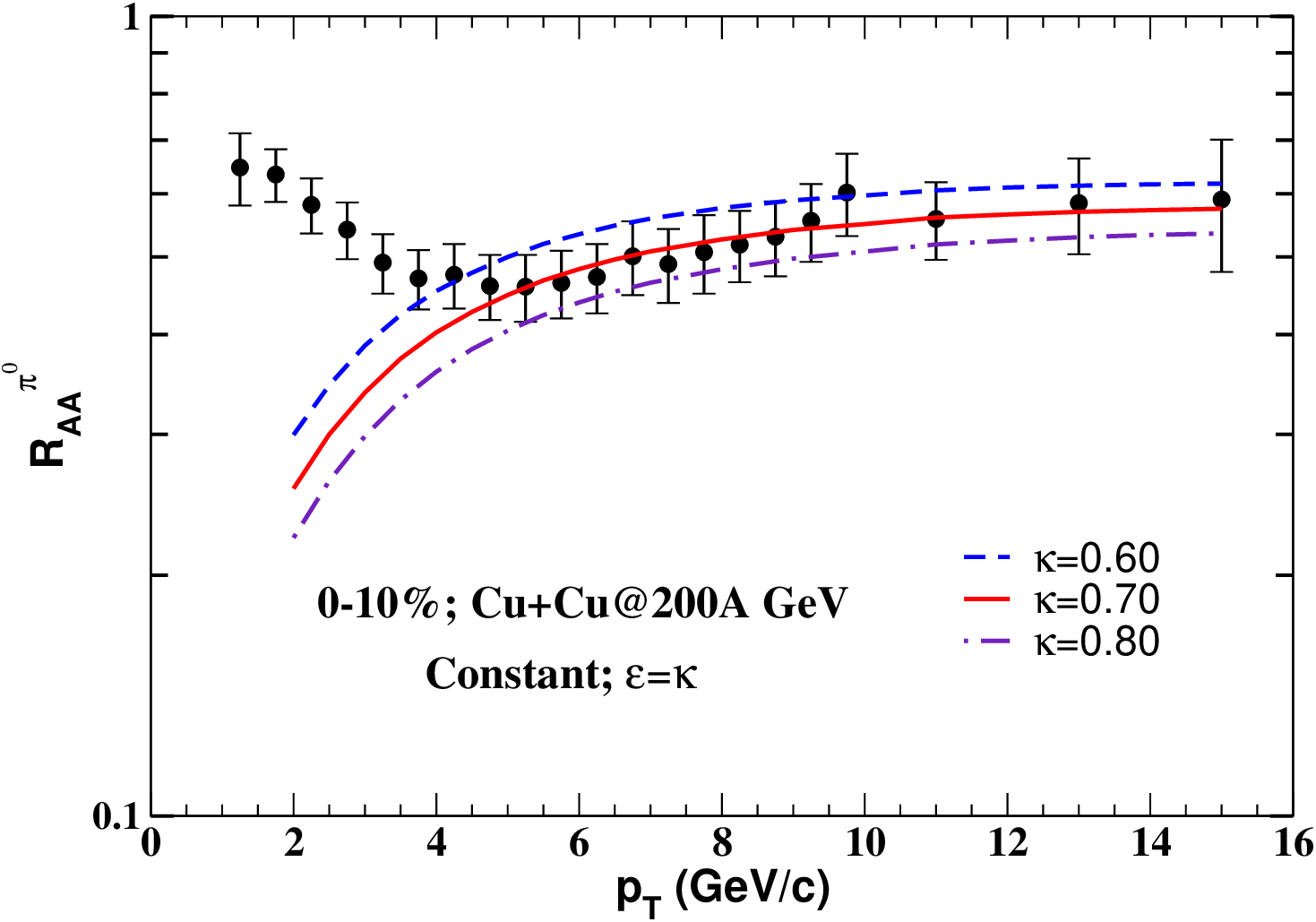}
\includegraphics[width=7.5cm, clip=true]{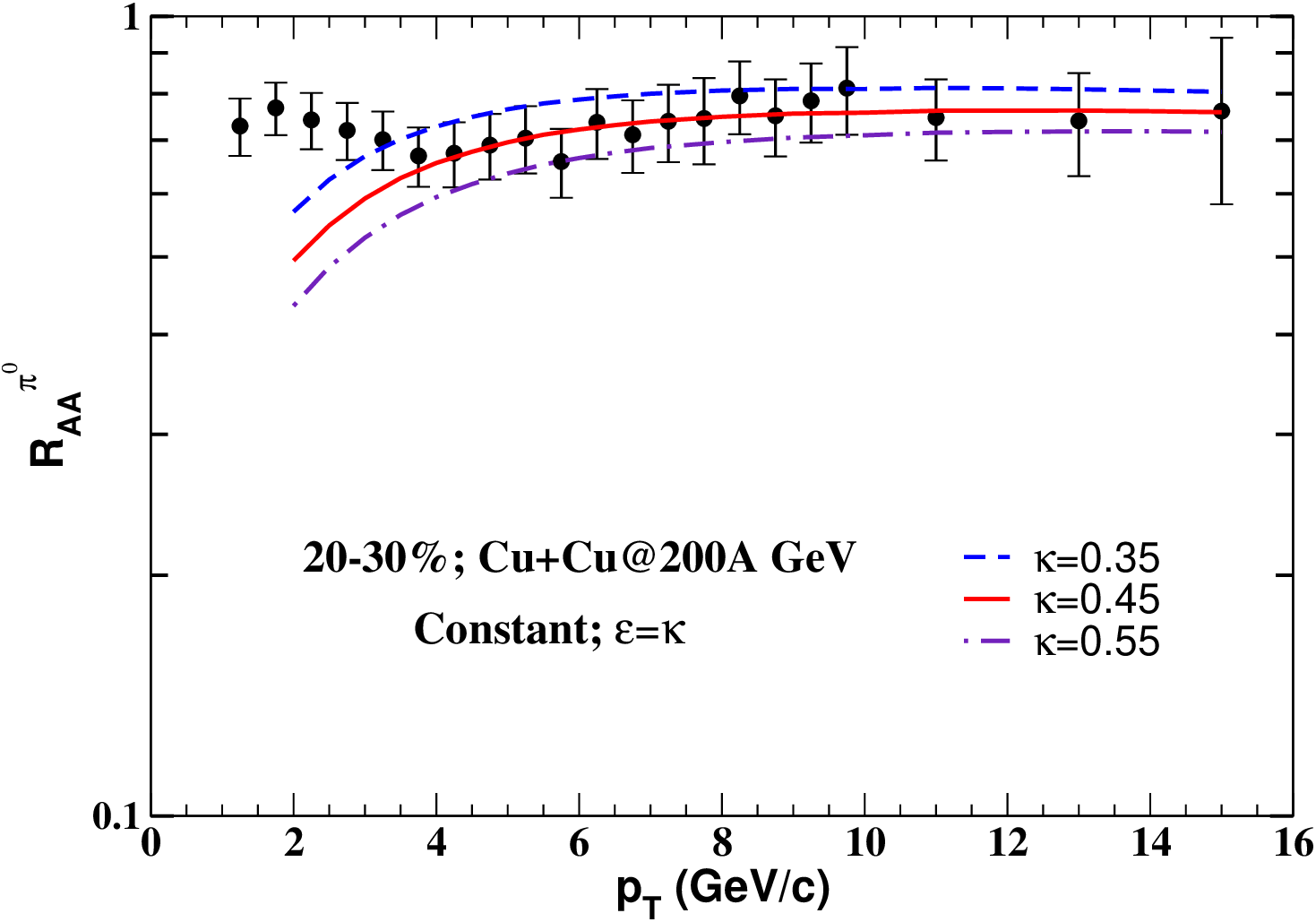}
\includegraphics[width=7.5cm, clip=true]{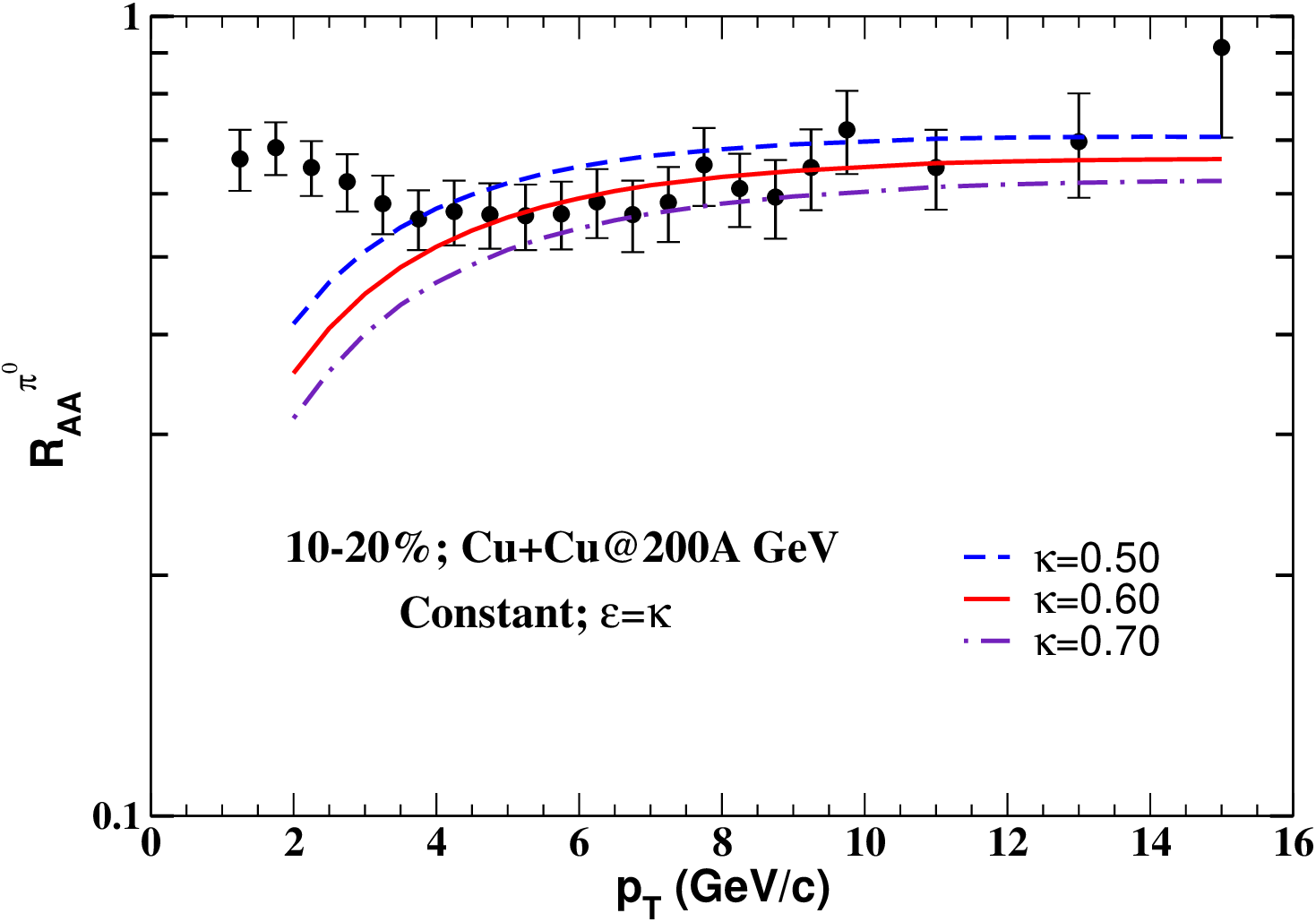}
\includegraphics[width=7.5cm, clip=true]{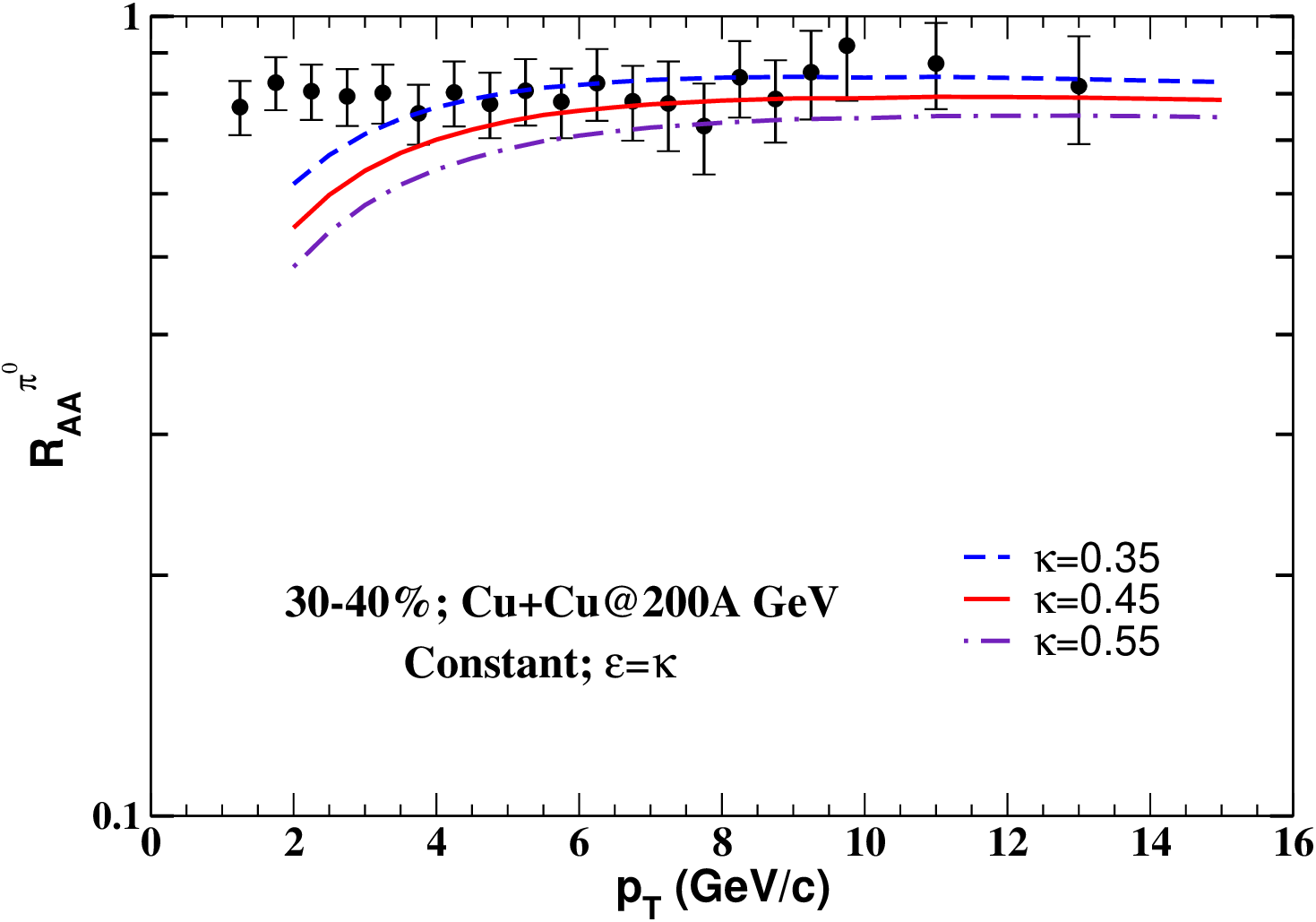}
\end{center}
\caption{(Colour on-line) 
Nuclear modification of neutral pion production for Cu+Cu collisions at $\sqrt{s_\textrm{NN}}$=200 GeV for constant energy loss
mechanism compared with PHENIX results~\cite{phenix_cu}.}
\label{fig9}
\end{figure}

Proceeding as before, we give our results for the nuclear modification factor $R_{AA}$ for  Cu+Cu collisions for 
centralities varying from 0--10\% to 30--40\%, in Figs.~\ref{fig7}--\ref{fig9}, using the BH, the LPM, and the constant energy loss
mechanisms and compare with the measurements by PHENIX collaboration~\cite{phenix_cu}. (See also Ref.~\cite{gale_cu}).

From Fig.~\ref{fig7}, we see that the energy loss mechanism suitable for the BH regime provides an accurate explanation of the
data for all the centralities under consideration for $p_T <$ 6 GeV/$c$. 

The energy loss mechanism denoted as LPM is seen to describe the nuclear modification function over the $p_T$ range of 
6--10 GeV/$c$ and even beyond for less central collisions~Fig.~\ref{fig8}.

And finally, the results for the mechanism admitting a constant energy loss per collision is seen to
work well for pions having $p_T >$ about 6--8 GeV/$c$ (see Fig.~\ref{fig9}).

We note that the energy loss co-efficient for the most central
collision of copper nuclei is close to that for the collision of gold nuclei for the 50--60\% centrality, for all the
three mechanisms. We also note the possible separation of the BH and LPM regimes around $p_T$ equal to 5--6 GeV/$c$, as before for gold
collisions, though the demarcation of the LPM and the constant energy loss mechanisms is not as sharp for the copper nuclei
as for the gold nuclei, especially for the less central collisions. This may have its origin in smaller $L$ for copper nuclei.

A figure similar to Fig.~\ref{fig5a} can also be prepared for the collisions
involving copper nuclei. We have not given it here, for the want of space.

\subsection{Centrality dependence of $dE/dx$}

\begin{figure}[ht]
\begin{center}
\includegraphics[width=11.5cm, clip=true]{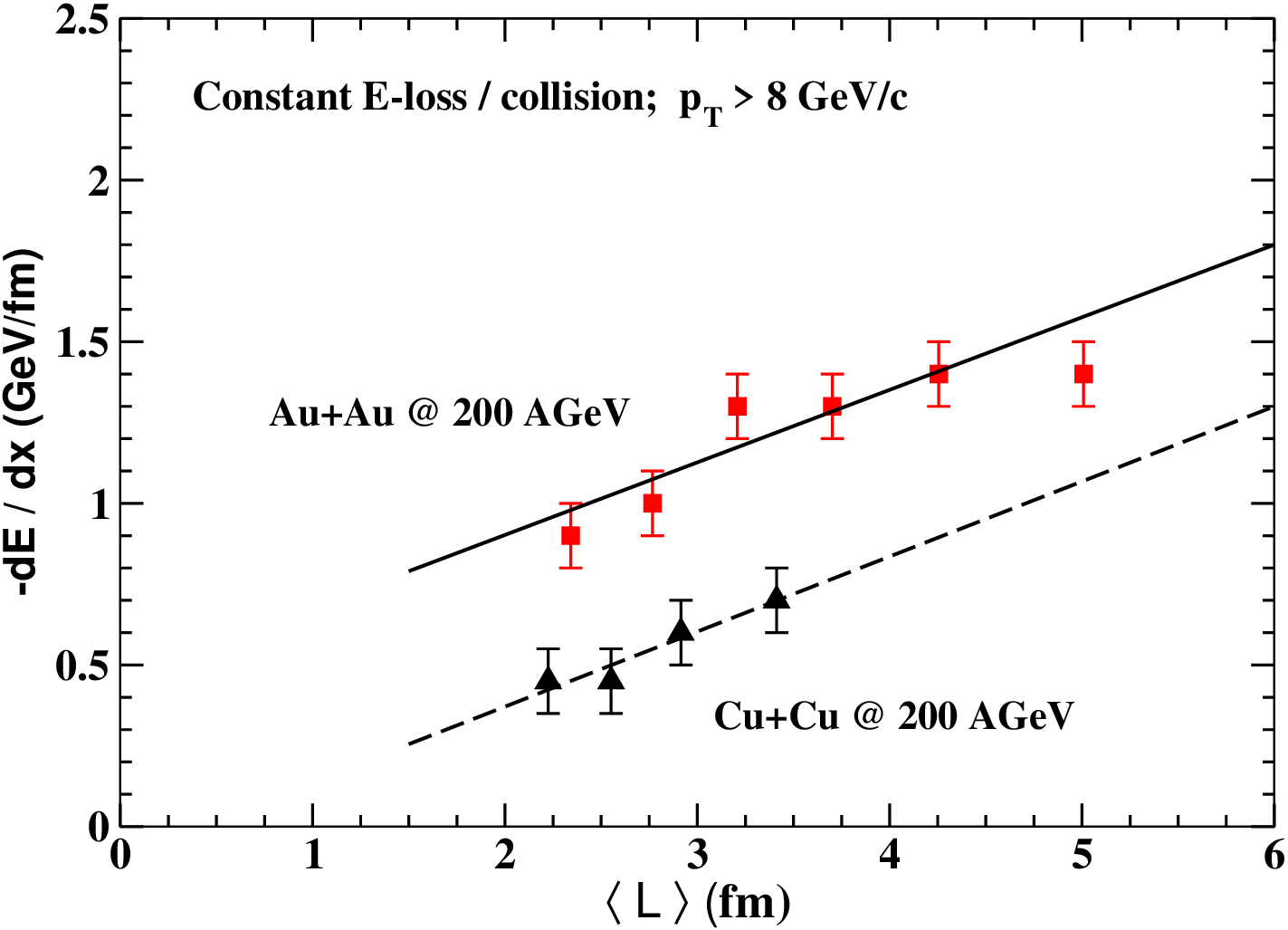}
\end{center}
\caption{(Colour on-line) Variation of $dE/dx$ with $\langle L \rangle$ for Cu+Cu and Au+Au collisions at 200 AGeV for the constant
energy loss per collision regime obtained in the present work.} 
\label{fig10}
\end{figure}

The results for the largest values of transverse momenta are summarized
 in Fig.~\ref{fig10}. Taking the case of constant energy loss
per collision, $-dE/dx=\varepsilon/\lambda$, we see that the rate of energy-loss necessary to explain the suppressed production of
hadrons at large $p_T$ varies as $\langle L \rangle$, the average path length of the partons in the medium, both for Cu+Cu and Au+Au collisions at the 
centre of mass energy of 200 GeV/nucleon. This empirical result
 confirms the conviction 
that for the range of transverse momenta covered, $p_T \geq$ 8--10
GeV/$c$, the concerned partons interact with the medium as a whole, 
and the energy lost by the partons, $\Delta E \propto L^2$.
This is different from the AdS/CFT description where the radiated parton cloud is brought on-shell by the drag of the
strongly coupled plasma and $\Delta E\propto L^3$~\cite{ads/CFT}, though
some moderation of the $L$ dependence is expected for an expanding plasma.

 We also see that even though the slopes of the results for Cu+Cu and Au+Au
collisions are similar, the energy loss for a given $\langle L \rangle$ for Au+Au collisions is about 40--60\% larger than that for Cu+Cu collisions.

We can use the Eq.~\ref{qhat} to obtain the value of the momentum transport
coefficient, $\widehat{q}$ from the above. We find that the $\widehat{q}$ varies from
0.25 GeV$^2$/fm for the collisions having  0--10\% centrality for Au+Au system at 200 AGeV to
0.39 GeV$^2$/fm for the 50--60\% centrality. The decrease for more central
collisions can be understood by noting that the systems 
produced in such cases would be at higher temperatures and thus much more dense
than for peripheral collisions. This would then lead to a smaller suppression of the
radiation due to the LPM effect. The faster rise with the path-length ($L^2$)
more than compensates this decrease, finally giving a large energy loss and larger
suppression of high momentum particles for more central collisions.

On the other hand, for the case of Cu+Cu system, 
we get $\widehat{q}\approx$  0.18 GeV$^2$/fm  for all the centralities 
considered by us. We note that it is about half of what we got for Au+Au
system. The near identity of $\widehat{q}$ for all the centralities 
for Cu+Cu system is related to a smaller variation in the path-length for
them. We also note that the Au+Au system is more efficient in degrading
the jet-energy.

We add that various authors have reported widely 
differing values of $\widehat{q}$. Thus, Wiedemann and Salgado~\cite{jet_rev}
have reported a value of 5--10 GeV$^2$/fm, while Gyulassy, Levai, 
and Vitev~\cite{jet_rev} report a value in the range of 0.35--0.85 GeV$^2$/fm.
Arnold, Moore, and Yaffe~\cite{jet_rev}, on the other hand, suggest a
value of about 2 GeV$^2$/fm, in contrast to a value of 3--4 GeV$^2$/fm reported
by Majumder~\cite{am,bass_jet,t.renk_jet}. It is felt that a part of
 this difference may be due to different physical attributes of the evolving
 system over which the average  is taken.

\section{Azimuthal anisotropy of high $p_T$ neutral pions}

We have seen that the phenomenological model of parton energy loss, which accounts for only the induced gluon radiation of the fast
partons, successfully describes the suppression of neutral pion spectra at large $p_T$ at different centralities for Au+Au and Cu+Cu collisions at
$\sqrt{s_\textrm{NN}}=200$ GeV. 
In case of non-central collisions, the path-length will have an azimuthal variation over the transverse plane (see Fig.~\ref{fig2}).
In what follows, we study the azimuthal dependence of the suppression of production of hadrons in terms of the azimuthal anisotropy
of their transverse momenta~\cite{jet_v2}.

We re-emphasize that this anisotropy is different from the elliptic flow seen for hadrons at lower $p_T$, which is explained using
hydrodynamics which converts the  spatial anisotropy of the initial
 state into the azimuthal anisotropy of the transverse momenta of the hadrons~\cite{flow_theo}.
The extent of validity of hydrodynamics for larger $p_T$ is not very clear. The success of the
 recombination model~\cite{recomb} suggests that the fragmentation
as a mode of hadronization may be valid  for $p_T \gg $ 3--5 GeV/$c$.  Thus the following results should be taken 
as indicative of anisotropy which can arise due to medium modification of fragmentation function due to energy loss of partons.

The differential azimuthal anisotropy is measured in terms of the parameter $v_2(p_T)$, which is the second Fourier coefficient of the 
azimuthal distribution of hadrons in the reaction plane:
\begin{equation}
v_2(p_T) =\frac{\int_{0}^{2\pi}\,d\phi\, \cos(2\phi) dN/d^2p_Tdy}
                        {\int_{0}^{2\pi}\,d\phi\, dN/d^2p_Tdy}~.
\end{equation}

We have calculated the azimuthal anisotropy coefficient $v_2(p_T)$ of neutral pions for $p_T>$ 2 GeV/c 
for Au+Au collisions  at $\sqrt{s_\textrm{NN}}$=200 GeV
for the six centralities mentioned earlier. 
The energy loss per collision is taken from the earlier analysis and $\phi$ dependent distribution of the pions is calculated
by incorporating the $\phi$ dependence of the path-length $L$.

\begin{figure}
\begin{center}
\includegraphics[width=7.5cm, clip=true]{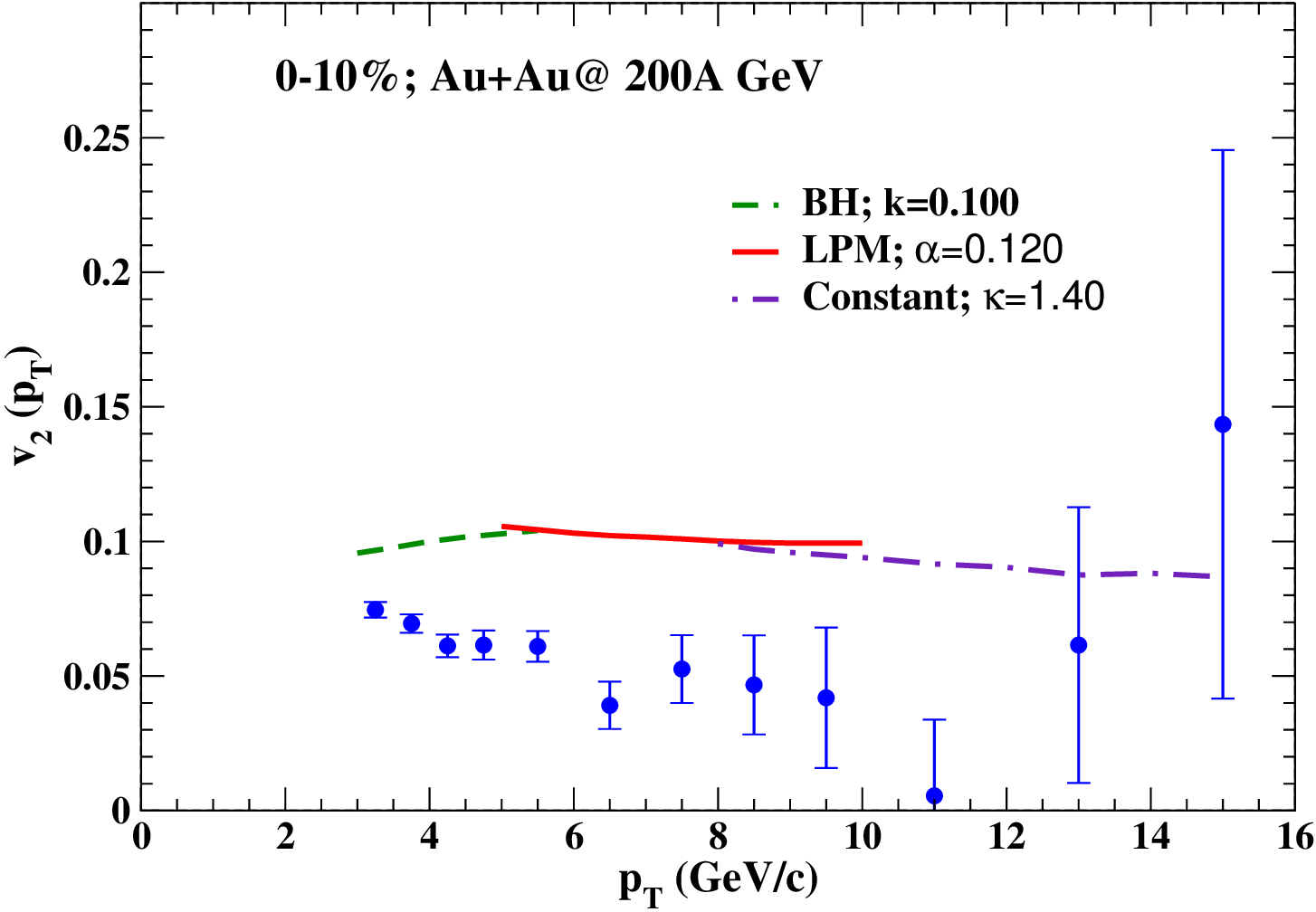}
\includegraphics[width=7.5cm, clip=true]{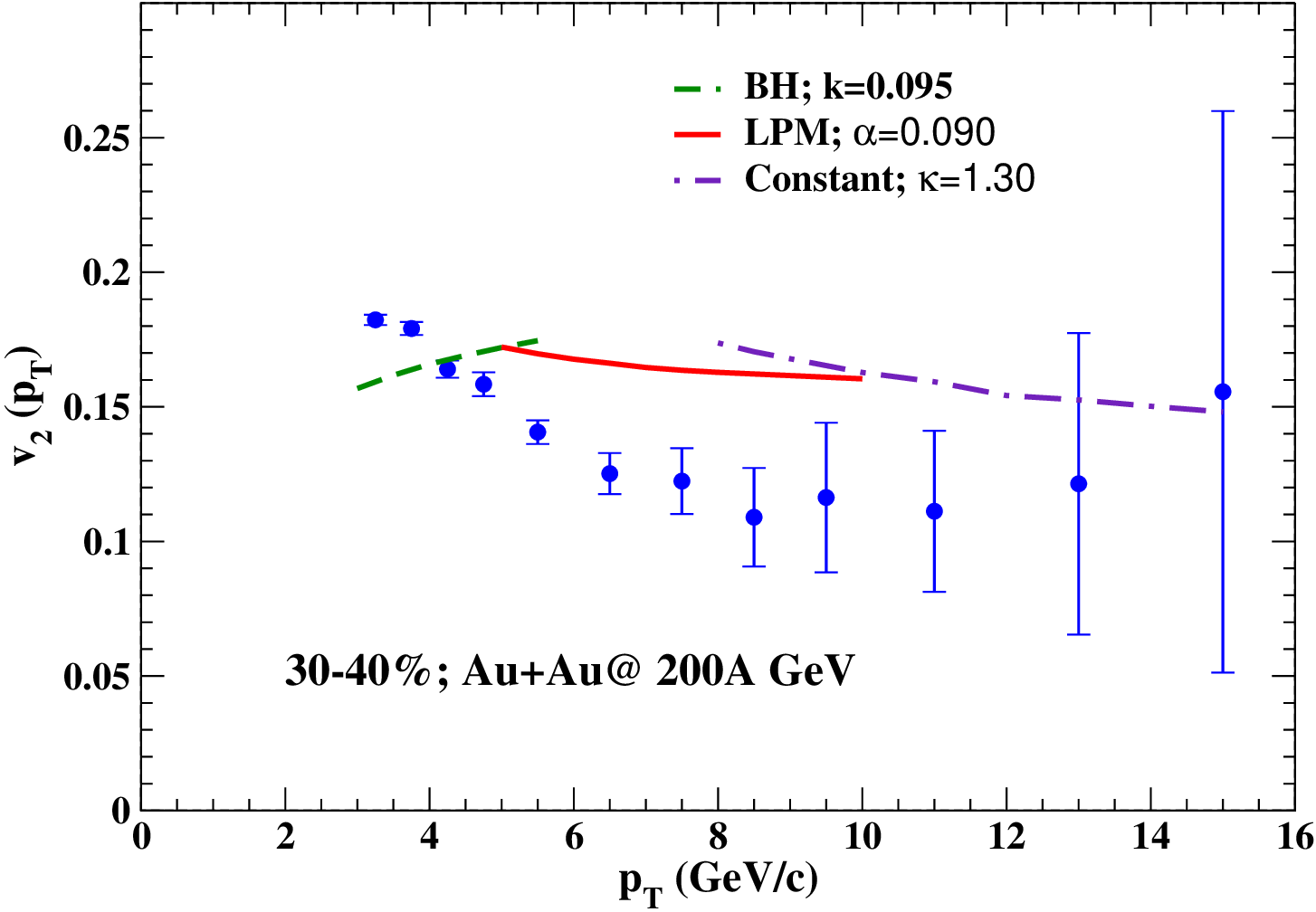}
\includegraphics[width=7.5cm, clip=true]{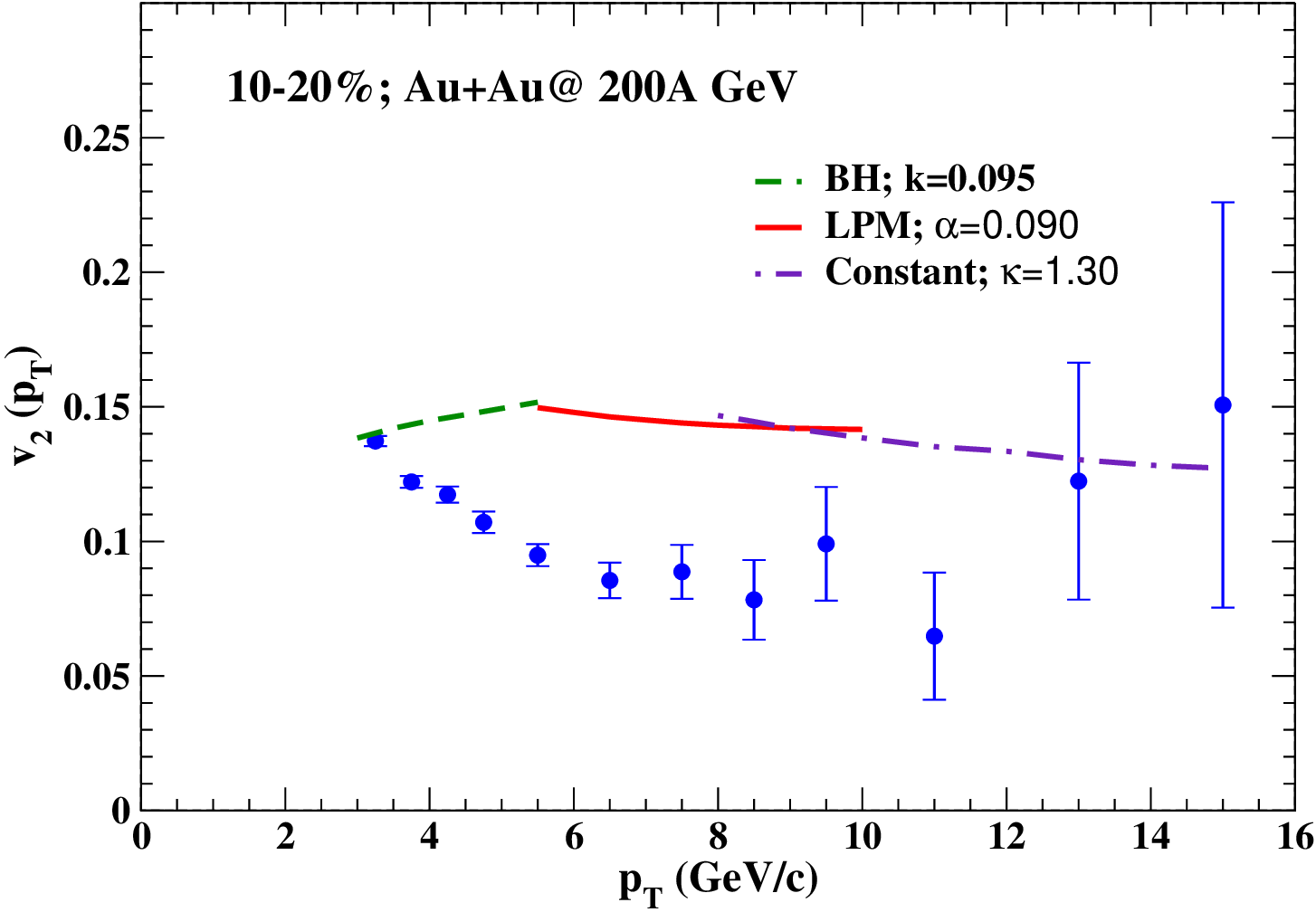}
\includegraphics[width=7.5cm, clip=true]{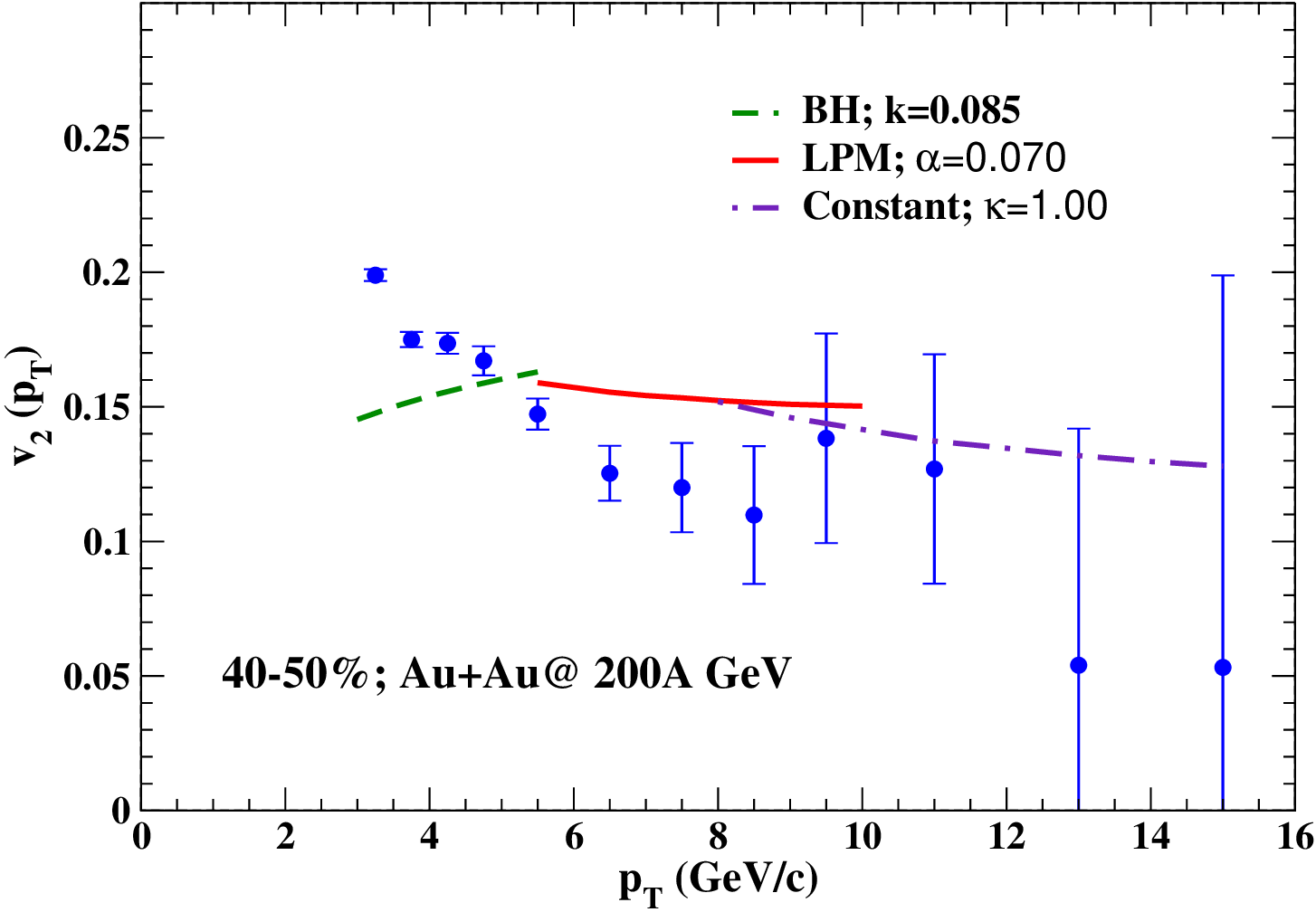}
\includegraphics[width=7.5cm, clip=true]{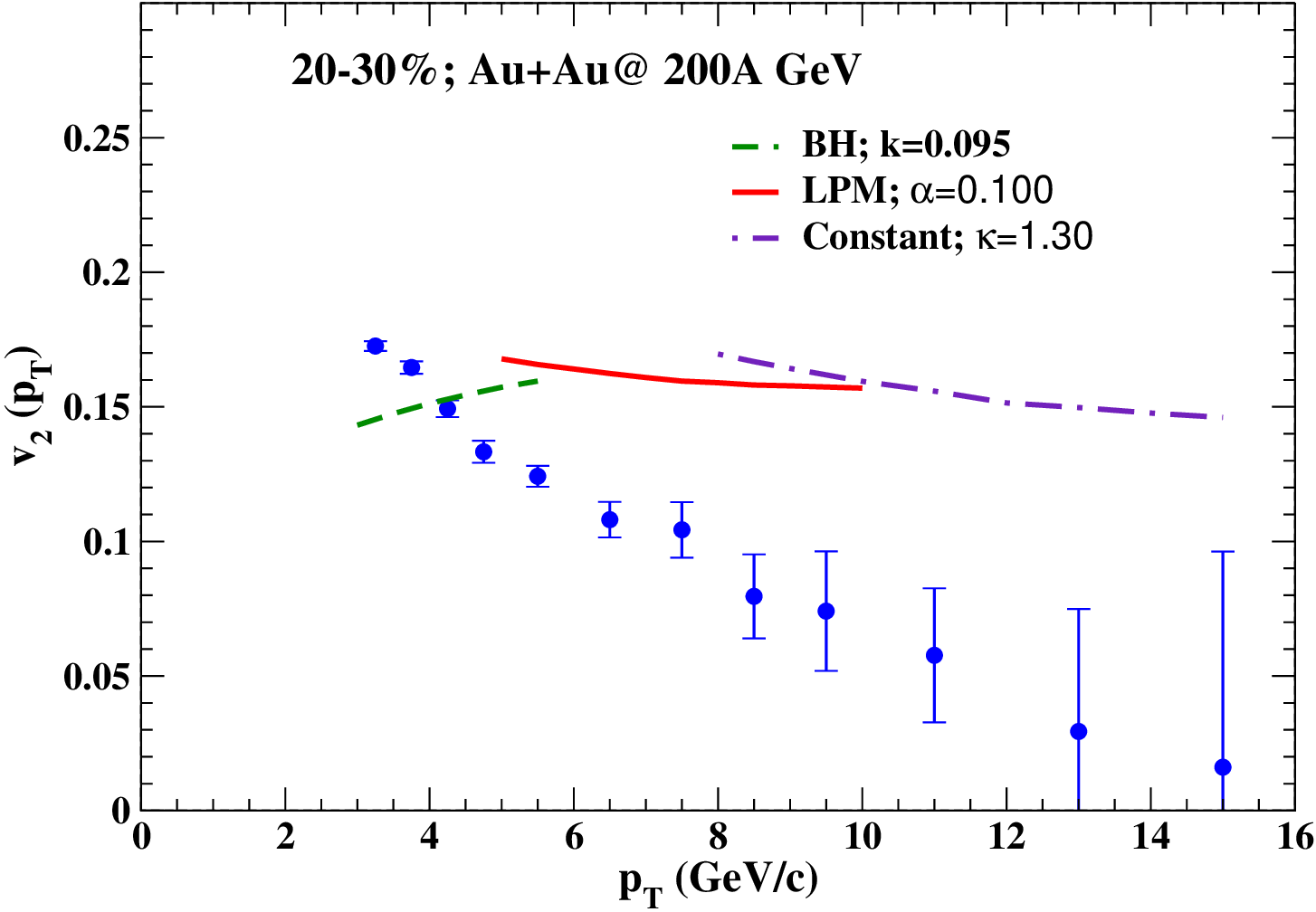}
\includegraphics[width=7.5cm, clip=true]{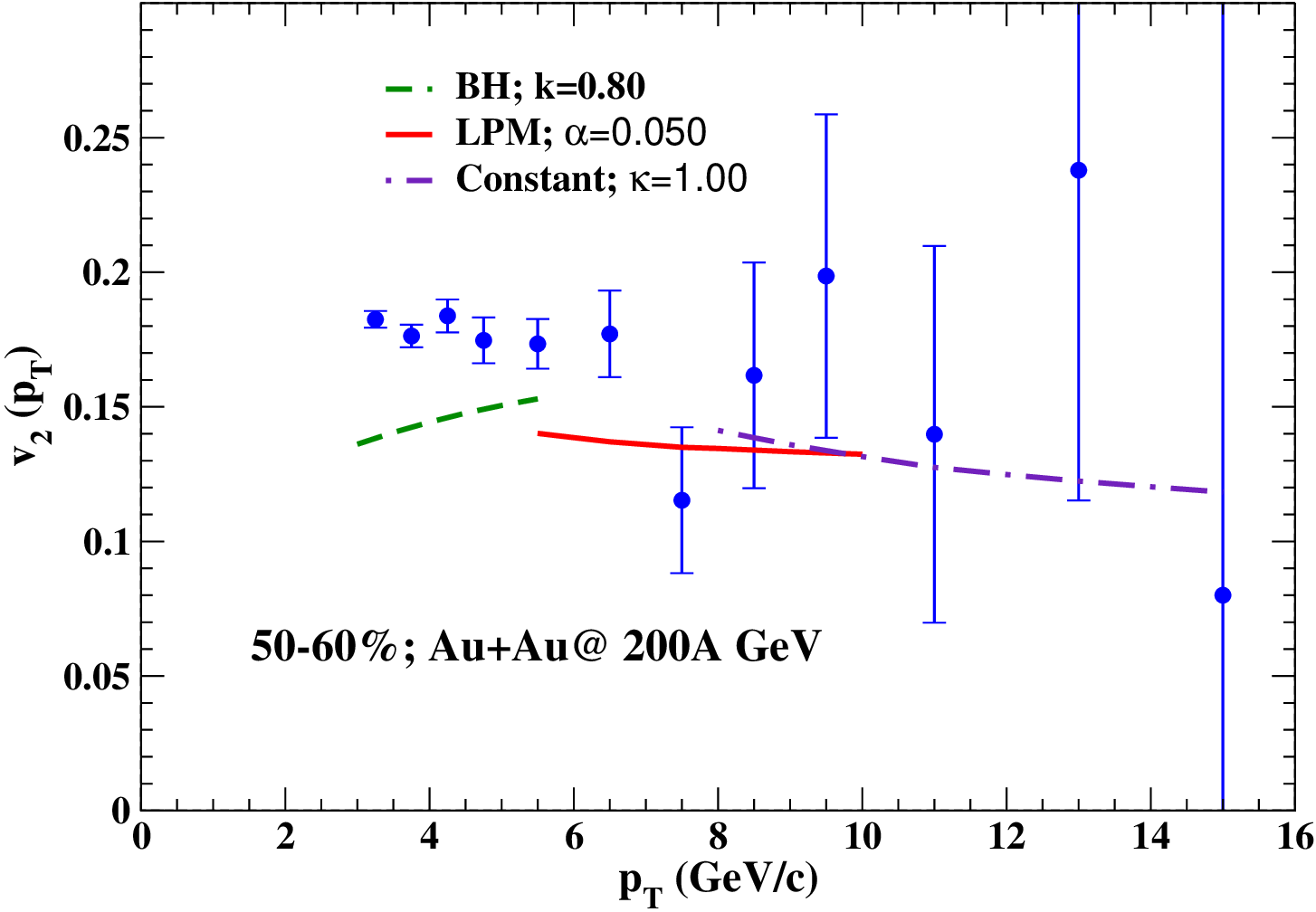}
\end{center}
\caption{(Colour on-line) 
The differential azimuthal anisotropy coefficient $v_2$ of neutral pion  calculated using BH, LPM, and constant energy loss mechanisms
for Au+Au collisions at $\sqrt{s_\textrm{NN}}$=200 GeV. The experimental data are
from the PHENIX collaboration~\cite{pi0v2_phenix}.}
\label{fig11}
\end{figure}

 The results  of our calculations are displayed in Fig.~\ref{fig11} along with the data available from PHENIX
collaboration~\cite{pi0v2_phenix}. We show the theoretical results
only over the $p_T$ window where the corresponding mechanism was found to
describe the nuclear modification data, earlier (see Fig.~\ref{fig5a}). 

Recalling that the BH mechanism was found to provide a good description of the nuclear suppression, up to $p_T$
equal to 5--6 GeV/$c$, while the mechanisms for the LPM and the constant energy loss per collision provided description for
higher $p_T$, the results are quite interesting. First of all, even for $p_T \leq $ 6 GeV/$c$ while the BH mechanism leads
to a $v_2(p_T)$ increasing with $p_T$, the experimental results show a contrary behaviour. It could be that a hydrodynamic
expansion of the system under viscosity affects the hadrons having $p_T$ up to 
5--6 GeV/$c$~\cite{h.song}.
 The other two mechanisms, though,
give a behaviour similar to the experimental data, in the region of transverse momenta where they are applicable. 
The theoretical values are larger by up to a factor of two in the worst cases of more central collisions, though 
they agree reasonably well with the experimental values for the 
least central collision considered here.

\begin{figure}[ht]
\begin{center}
\includegraphics[width=11.5cm, clip=true]{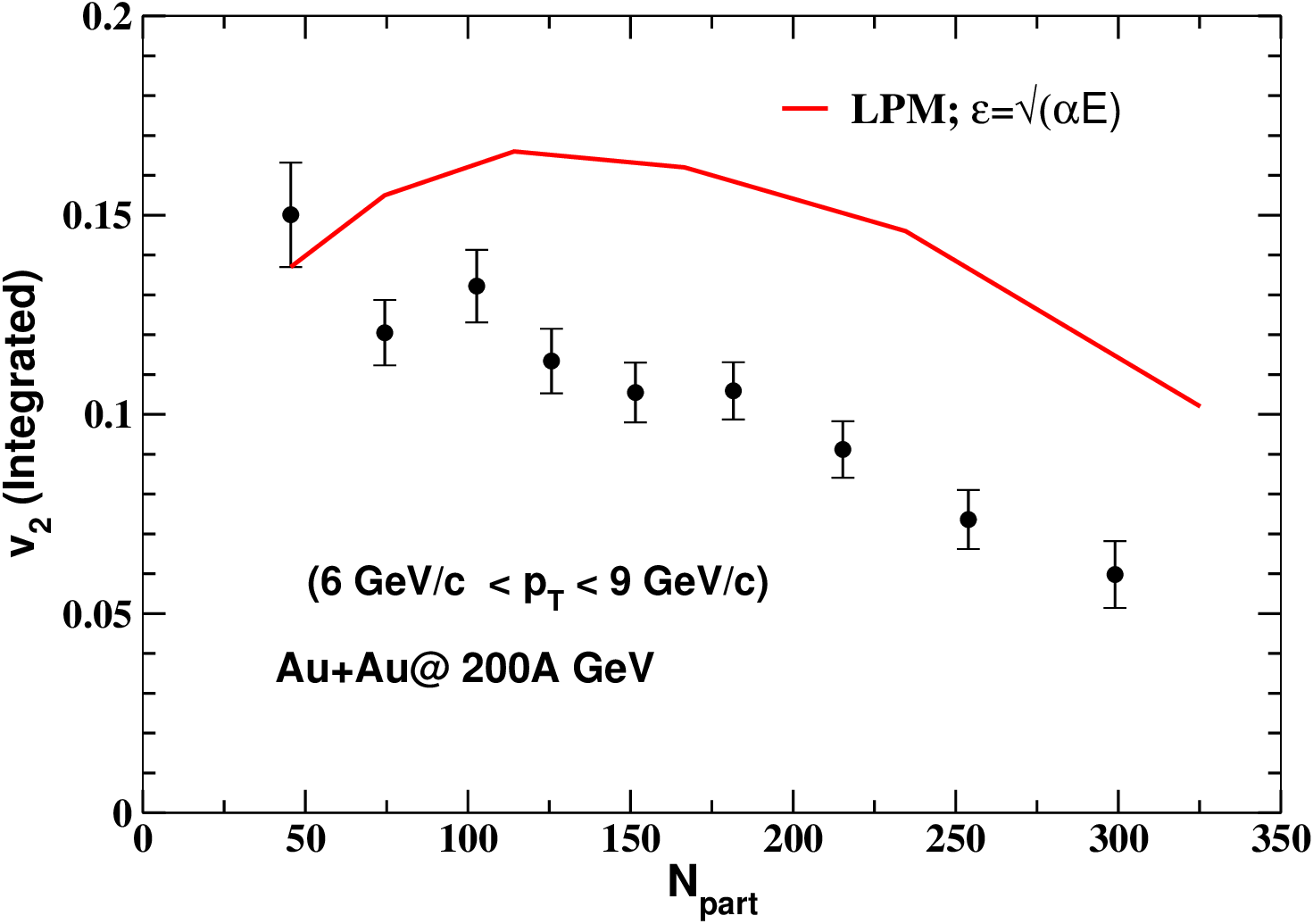}
\end{center}
\caption{(Colour on-line) The centrality dependence of integrated $v_2$ of neutral pions calculated for LPM mechanism of energy loss
 for Au+Au collisions $\sqrt{s_\textrm{NN}}$ = 200 GeV. The data points are taken from PHENIX collaboration~\cite{pi0v2_phenix}.}
\label{fig12}
\end{figure}

It is interesting to see the centrality dependence of the  integrated azimuthal anisotropy coefficient in the $p_T$ range of 6--9 GeV/c.
For this purpose, we have used the differential azimuthal anisotropy coefficient $v_2(p_T)$ for the LPM mechanism obtained 
earlier (see Fig.\ref{fig12}).
We see that in general our calculations reproduce the trend of the variation of $v_2$ with the number of participants, though 
the theoretical values are larger by about a factor of two (see Ref.~\cite{b.muller}, for similar results). 
One short-coming of our calculation immediately comes to mind- the description 
of the nuclei as having  a uniform density.
A Woods-Saxon density profile for the colliding nuclei would reduce 
the difference in the path-lengths for the
partons travelling in the reaction plane and travelling in a plane
 perpendicular to it, for example, and thus reduce $v_2$.
We postpone this study to a future work where additionally, the evolution of the medium will be taken into account.

\begin{figure}
\begin{center}
\includegraphics[width=7.5cm, clip=true]{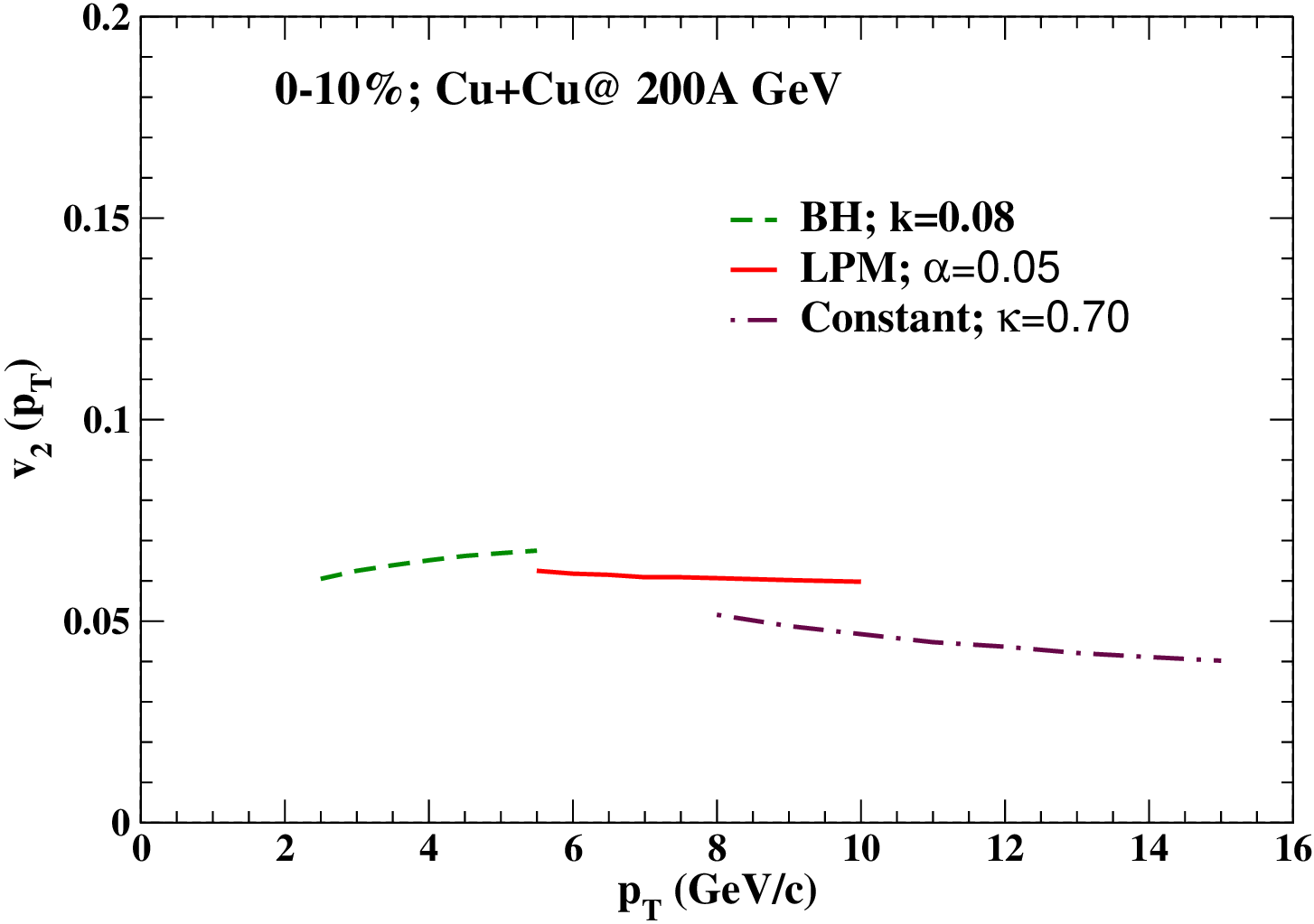}
\includegraphics[width=7.5cm, clip=true]{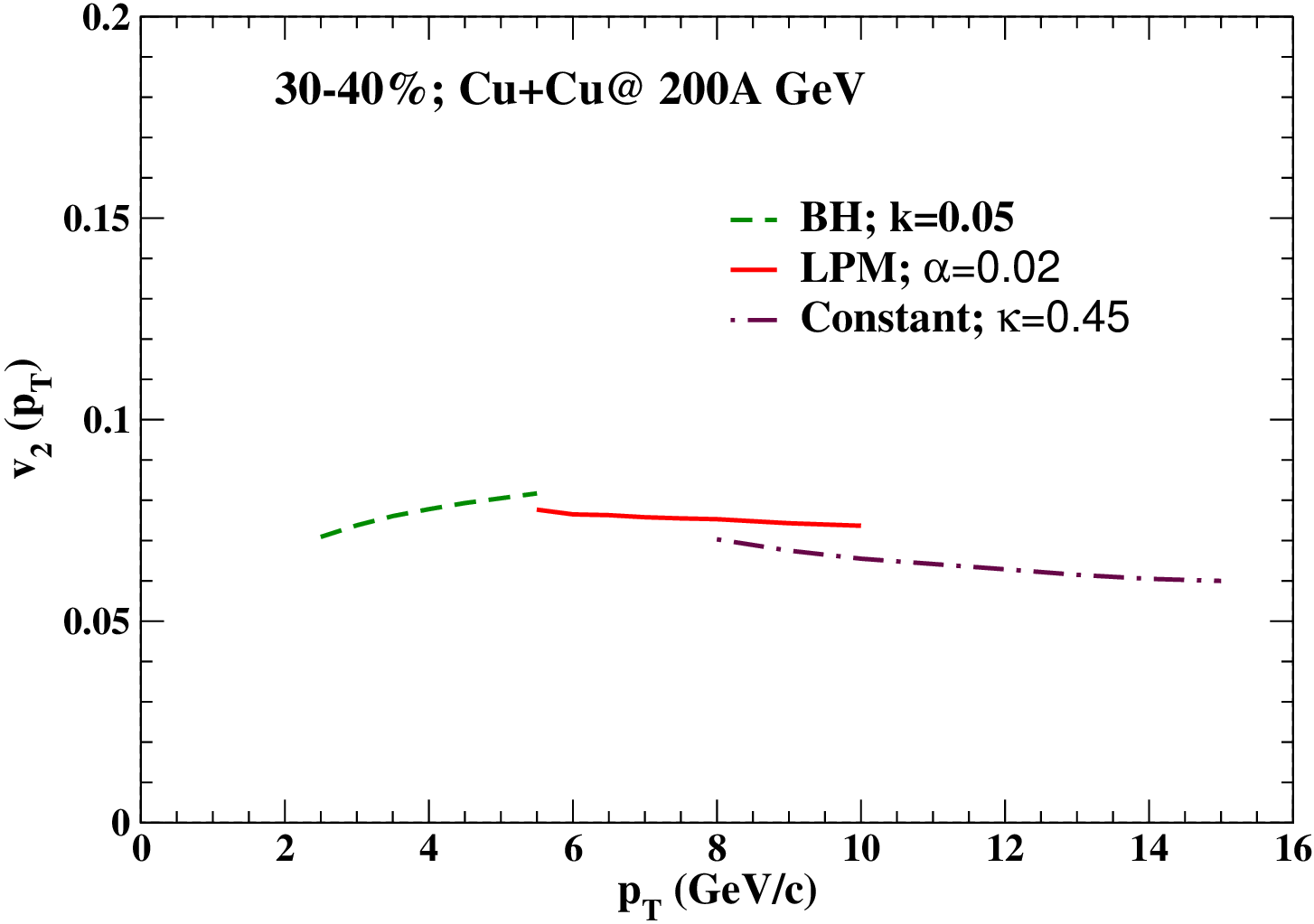}
\end{center}
\caption{(Colour on-line) 
The differential azimuthal anisotropy coefficient $v_2(p_T)$ of neutral pion calculated using BH, LPM, and constant energy loss mechanisms
for Cu+Cu collisions at $\sqrt{s_\textrm{NN}}$=200 GeV.}
\label{fig13}
\end{figure}

We also give our prediction for $v_2(p_T)$ for Cu+Cu collisions at 200A GeV for two 
typical centralities (Fig.~\ref{fig13}). A behaviour similar to that for Au+Au collisions is seen.

\section{Nuclear suppression and azimuthal asymmetry of prompt photons}

Prompt photons, as indicated earlier, originate from (a) quark-gluon Compton scattering 
($q+g\rightarrow q+\gamma$), (b) quark-anti-quark annihilation ($q+\bar{q}\rightarrow g+\gamma$), and (c) collinear fragmentation
of a high energy quarks ($q\rightarrow q+\gamma$) following a hard collision.
Of these the third process is affected by the energy loss suffered by the quark before its fragmentation~\cite{jeon,arleo}.
It will also lead to an azimuthal anisotropy of the momentum distribution of hard photons~\cite{v2_phot_highpt}. 

\begin{figure}[ht]
\begin{center}
\includegraphics[width=11.5cm, clip=true]{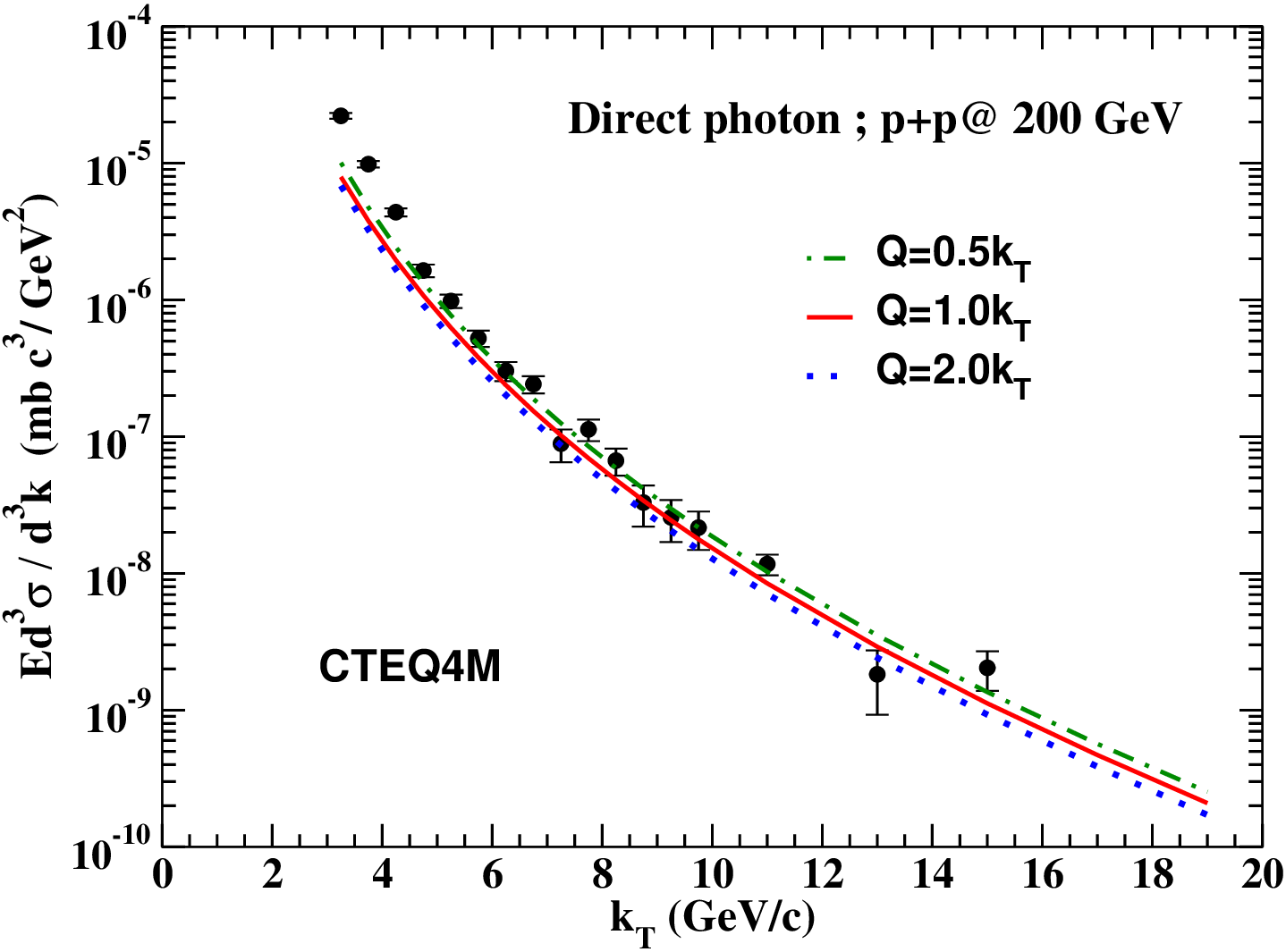}
\end{center}
\caption{(Colour on-line) A comparison of production of prompt photons in p+p collision
measured by the PHENIX collaboration~\cite{phot_pp} at $\sqrt{s_\textrm{NN}}$ = 200 GeV
with NLO pQCD calculations.}
\label{fig14}
\end{figure}

Once again, we use the NLO pQCD code of Aurenche et al~\cite{P.aur2}, suitably modified to account for the energy loss
suffered by the quarks before they fragment into photons.
As a first step we show our results for proton-proton collisions at 200 GeV (Fig.~\ref{fig14}) using the fragmentation, factorization,
and the renormalization scales as equal, with   $Q=$ $k_T$/2, $k_T$ and 2$k_T$.
We have used  the \emph{CTEQ4M} parton distribution function and the BKK 
fragmentation function for these calculations as well.
We see that our results are consistent with the data from PHENIX~\cite{phot_pp} and also
with earlier calculations (see Ref.~\cite{phot_pp}) along similar lines.
For the rest of the calculations we have used a  common scale $Q=k_T$.

While performing the calculations for the nucleus-nucleus collisions we include
nuclear shadowing and energy loss suffered by the quarks as before.
 We also properly account for the isospin
of the participating nucleons, which is essential while performing calculations for photons. The nuclear modification
factor $R_{\textrm{AA}}^{\gamma}$ is then obtained in a manner similar to Eq.~\ref{raa}.

\begin{figure}[ht]
\begin{center}
\includegraphics[width=11.5cm, clip=true]{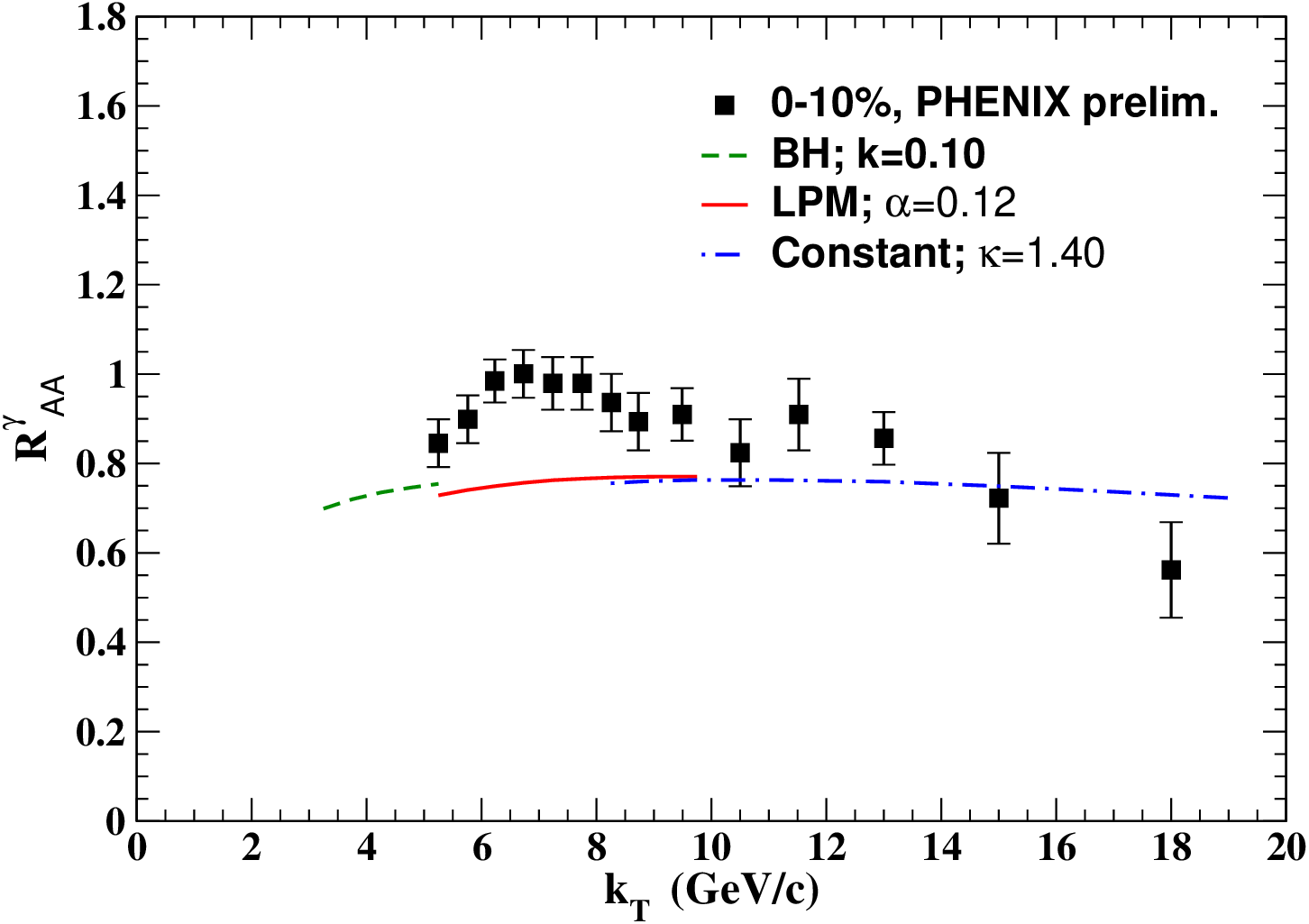}
\end{center}
\caption{(Colour on-line) Nuclear suppression of hard  photons calculated using BH, LPM, and constant energy 
loss per collision 
for Au+Au (0-10\%) collisions at $\sqrt{s_\textrm{NN}}$=200 GeV. The data points are taken from Ref.~\cite{phenix_prel}}
\label{fig15}
\end{figure}

We show our results for the nuclear modification of hard photon production using the energy loss parameters obtained earlier along
with the experimental results for collision of gold nuclei at 200A GeV in Fig.~\ref{fig15}.
We give the theoretical curves over the regions where the nuclear modification 
for pion production is in agreement with the experimental data (see Fig.~\ref{fig5a}).

We see that our results are in fair agreement with the data for $k_T$ beyond 10 GeV/$c$ 
where the mechanism of prompt photon production
included in the present work is expected to dominate~\cite{arleo}.
 We do realize that for lower $k_T$ other mechanisms like 
jet-conversion~\cite{fms}, induced bremsstrahlung~\cite{zakharov}
 and thermal production~\cite{phot_theory} will contribute.

Next we calculate the
the differential azimuthal anisotropy coefficient $v^{\gamma}_2(k_T)$ for hard photons for Au+Au collisions at 200A GeV.
Recall that for lower  $k_T$ ($\leq $  5--6 GeV/$c$), the  azimuthal anisotropy of the
 direct photons carries valuable
information about the development of the elliptic flow of the plasma and momentum anisotropy
of the thermalized partons~\cite{rupa}.

\begin{figure}
\begin{center}
\includegraphics[width=7.5cm, clip=true]{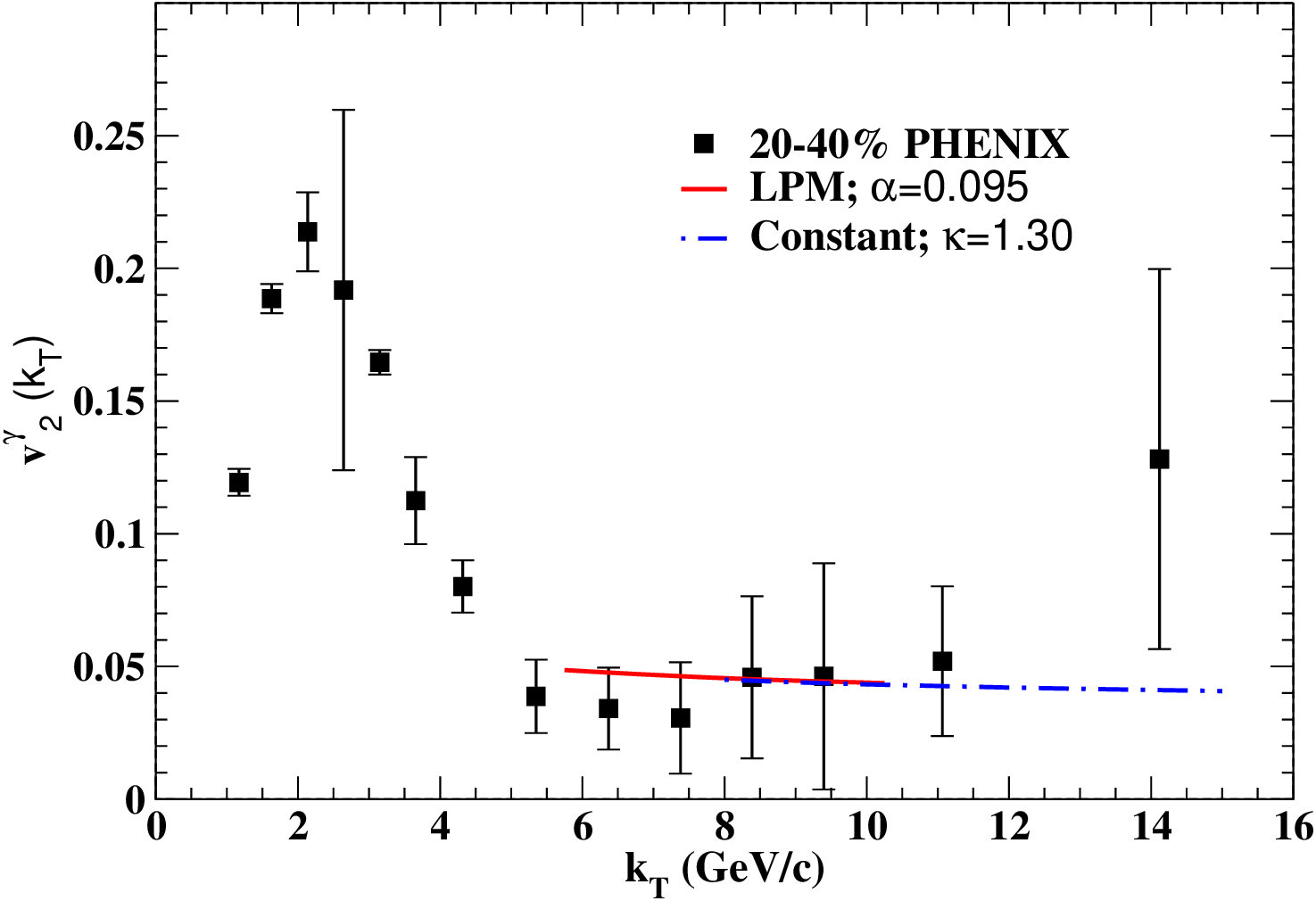}
\includegraphics[width=7.5cm, clip=true]{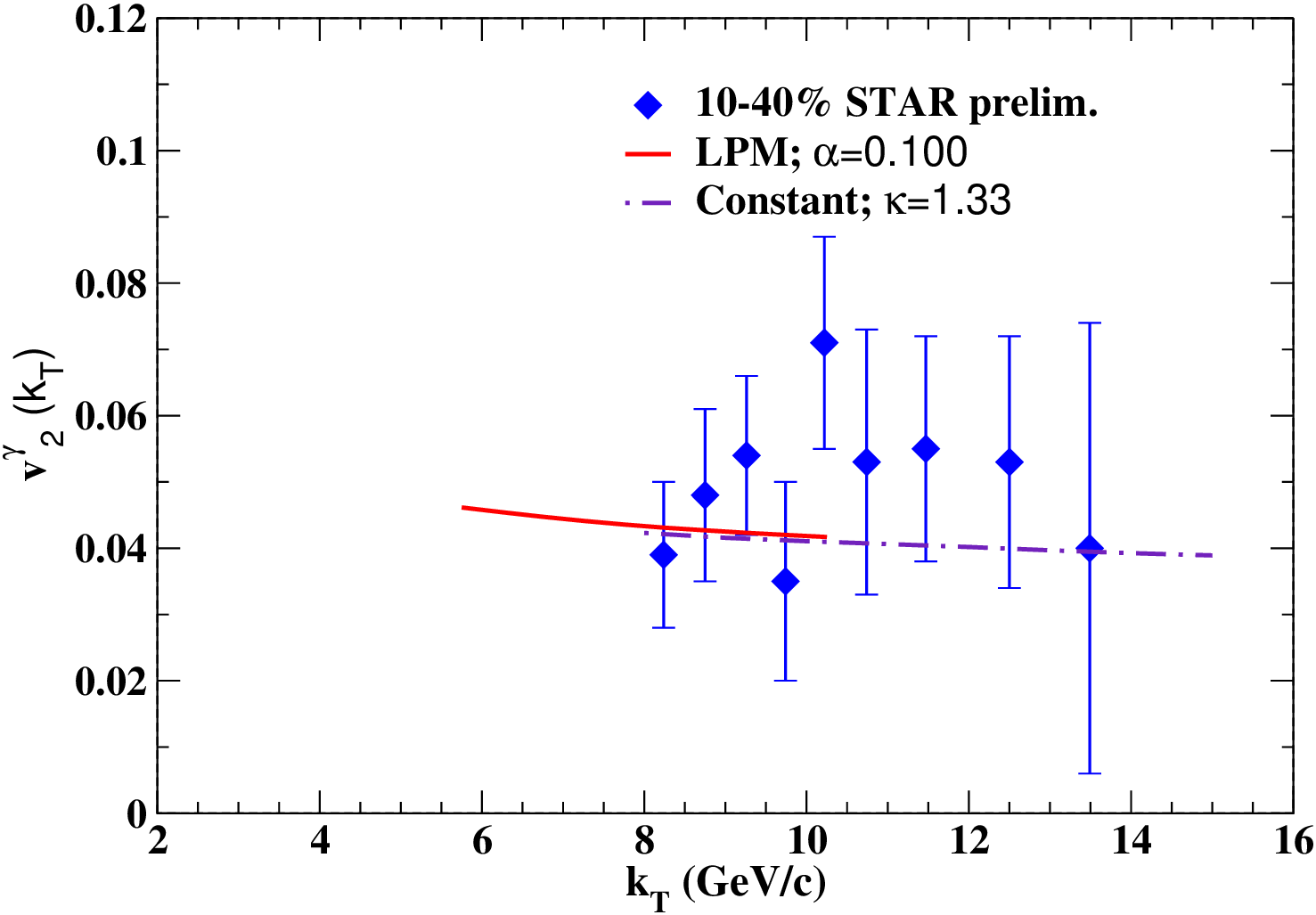}
\end{center}
\caption{(Colour on-line) 
The differential azimuthal anisotropy coefficient $v_2^{\gamma}$ of direct photons calculated using LPM scheme of parton
energy loss for Au+Au collisions at $\sqrt{s_\textrm{NN}}$=200 GeV along with experimental results
from the PHENIX~\cite{v2gama_phenix} (left panel) and the  
 STAR~\cite{v2gama_star} (right panel) collaborations.}
\label{fig16}
\end{figure}

 In Fig.~\ref{fig16} we have shown the results for $v^{\gamma}_2 (k_T)$ calculated using LPM
and the constant energy loss per collision  mechanisms of parton energy loss for two centrality
classes (20-40\% and 10-40\%) and compared them with the recent results from the PHENIX~\cite{v2gama_phenix} and the
STAR~\cite{v2gama_star} collaborations. 
We make no further adjustments of
 the energy loss parameters obtained earlier.

We see that the azimuthal anisotropy of direct photons  for $k_T >$ 6 GeV/$c$ is reproduced reasonably well from our
calculations. The results of Figs.~\ref{fig15} and~ \ref{fig16} come a pleasant
 surprise as we recall that while calculating the
energy loss of the partons and the consequent suppression of hadrons at large $p_T$, 
we did not distinguish between quarks and gluons.
The photon production, however, is sensitive only to the energy loss suffered by quarks. One possible reason for this could be the
dominating contribution of quarks to hadrons as well photons having large transverse momenta, which is then rightly sampled by our
procedure. One could use this procedure to first fix the energy loss per collision for
 quarks and then proceed to determine the
energy loss per collision for gluons by fitting the hadronic spectra. This would, however, require a 
much more accurate and extensive large transverse momentum data
for photons. A further availability of isolated photon spectra
along with the (total) single photon spectra will help us to
subtract out the Compton plus
annihilation contributions and increase the sensitivity of these 
calculations to energy loss suffered by quarks.

\section{Summary and Discussion}

We have used a simple and transparent model of multiple scattering and energy loss
 for partons passing through quark gluon plasma
produced in relativistic collisions of heavy nuclei, before they fragment into hadrons or photons. Taking the centrality dependence
of the average path-length into account and taking energy loss per collision as an adjustable parameter, we have obtained a description of 
hadronic suppression in Au+Au and Cu+Cu collisions at 200A GeV. The energy loss parameters are found to vary systematically with the
centrality and the $dE/dx$ for $p_T >$ 6--8 GeV/$c$ is found to vary linearly with the average path-length $\langle L \rangle$, in agreement with
the predictions of Baier et al.~\cite{baier}. 

The same parameters are then used without any further adjustments to obtain the azimuthal anisotropy for high $p_T$ pions. We find it to be about a factor of two larger than
the experimental values. 
Next, we have calculated the suppressed
production of hard photons and their azimuthal anisotropy, again
with out changing any of the parameters. These are
in reasonable agreement with the experimental findings. We have mentioned 
that in principal, the calculations
could be improved by incorporating diffused density distributions for the colliding
nuclei, instead of the uniform densities used in the present work. In that case,
it remains to be seen if the $v_2$ could be used to further constrain the energy loss 
parameters.

The success of the approach, considering that the dynamics of evolution,  the temperature and flavour dependence of some of
the parameters and a more detailed accounting of various interferences are not considered here, once again confirms the inadequacy of $R_\textrm{AA}$
in seriously distinguishing various details of the dynamics of the plasma~\cite{t.renk_jet}. 

Some improvements are definitely in order. The use of uniform density for the colliding nuclei can be easily improved upon. This,
as we argued, could improve the description of azimuthal anisotropy of hadron production at higher $p_T$. One may also take differing
values for the mean free paths for quarks and gluons and the energy loss per collision for them. Note, however, that if we 
argue that $dE/dx$ for gluons is twice that for quarks and also that their mean free path is half of that for quarks, then
the energy loss for quarks and gluons per collision will be identical. 

Why does this simple approach work so well? We realize that the $p_T$ distribution of jets produced in primary hard scatterings is
a rapidly falling function of the transverse momentum. Partons having a transverse momentum $p_T$ after undergoing multiple
collisions and losing a momentum $\Delta p_T$ will populate the momentum space occupied by partons having transverse momentum $p_T-\Delta p_T$
while the later will shift to still lower momenta. 
Thus if we make a good estimate of $\Delta p_T$, we are likely to get a 
 good description of the nuclear modification. 

But is this approach really so simple? NLO pQCD and accurate structure and 
fragmentation functions, along with the applicability of pQCD at larger transverse momenta ensure an accurate base-line for
p+p collisions. The influence of the dynamics of evolution is perhaps softened by the short life-time of the plasma at RHIC energies
and the fact that most of the energy is lost in collisions at early times.
 
Thus we feel that the present study, at the very minimum, 
gives a reliable average of the rate of energy loss of partons in quark gluon plasma.
 We also see that a simultaneous description of
nuclear modification of photon and pion production in nuclear collisions,
 holds out the hope of getting the energy loss suffered by quarks and gluons independently. 

It will definitely be of interest to see if the centrality dependence of 
$dE/dx$ ( $\propto \langle L \rangle$) 
 for partons having larger transverse momenta 
 at LHC energies (2.76 ATeV and 5.5 ATeV) continues to follow
the trend seen for the data at RHIC energies for Cu+Cu and Au+Au collisions,
 seen in Fig.~\ref{fig10}. If confirmed
at LHC energies as well, this will be a very valuable empirical affirmation of the suggestions of
 Baier et al~\cite{baier}. This aspect is under investigation.

\section*{Acknowledgments}
 SD would like to express his sincere gratitude to Department of Atomic Energy,
Government of India for financial support during the course of this work. We also thank 
Christine Aidala for drawing our attention to the final data for pp collisions 
(Fig.~\ref{fig1}).

\end{document}